\documentclass{elsart}

\usepackage{amssymb}

\usepackage{graphics}
\usepackage[dvips]{graphicx}
\usepackage{epsfig}

\begin{document}

\newcommand{\spd}{$sp$--$d$ }
\newcommand{\ef}{E_{\rm F}}
\newcommand{\Ang}{{\rm \AA}}

\newcommand{\be}{\begin{equation}}
\newcommand{\ee}{\end{equation}}
\newcommand{\ben}{\begin{eqnarray}}
\newcommand{\een}{\end{eqnarray}}
\newcommand{\beq}{\begin{equation}}
\newcommand{\eeq}{\end{equation}}
\newcommand{\B}{\mathrm{B}}
\newcommand{\NB}{\mathrm{NB}}
\newcommand{\RRe}{\mathrm{Re}}
\newcommand{\IIm}{\mathrm{Im}}

\begin{frontmatter}

\begin{flushright}
\begin{small}
{\it To appear in ``Physics of Quasicrystals'',}\\
{\it series "Handbook of Metal Physics",}\\
{\it Editors: T. Fujiwara, Y. Ishii  (Elsevier Science, 2007)}
\end{small}
\end{flushright}

\vspace{1cm}



\title{
%
%
%
Quantum transport  in quasicrystals and complex metallic alloys}

\author{Didier Mayou\corauthref{cor1}}
\ead{didier.mayou@grenoble.cnrs.fr}
\corauth[cor1]{Corresponding author}
\address{Institut N\'eel, CNRS and Universit\'e Joseph Fourier, B\^at D,\\
B.P. 166, 38042 Grenoble Cedex 9, France\\
Email: didier.mayou@grenoble.cnrs.fr
}

\author{Guy Trambly de Laissardi\`ere}
\ead{guy.trambly@u-cergy.fr}
\address{Laboratoire de Physique Th\'eorique et Mod\'elisation,
CNRS and Universit\'e de Cergy--Pontoise, St Martin, 95302 Cergy--Pontoise,
France\\ 
Email: guy.trambly@u-cergy.fr}

\begin{abstract}

The semi-classical Bloch-Boltzmann theory is at the heart of our understanding of conduction in solids, ranging from metals to semi-conductors. Physical systems that are beyond the range of applicability of this  theory  are thus of fundamental interest. This is the case  of disordered systems which present quantum  interferences in the diffusive regime, i.e. Anderson localization effects. This is also the case, for example, of systems that present  magnetic or electric breakdown when submitted to an electromagnetic field.   These exceptions, for which a full quantum transport theory must be developed have been intensively studied in the past and are now well known. 

It appears that in quasicrystals and related complex metallic alloys another type of breakdown of the semi-classical Bloch-Boltzmann theory operates. This  type of quantum transport is related to the specific  propagation mode of electrons in these  systems. Indeed in quasicrystals and related complex phases the quantum diffusion law deviates from the standard ballistic law characteristic of  perfect crystals in two possible ways. In a perfect quasicrystal the large time diffusion law is a power law  instead of a ballistic one in perfect crystals. In a complex crystal the diffusion law is always ballistic at large time but  it can deviate strongly from the ballistic law at sufficiently small times. We develop a theory of quantum transport that applies to a normal ballistic law but also to  these specific diffusion laws and we describe the behavior of conductivity that results from these specific laws. As we show phenomenological models based on this theory describe correctly the experimental transport properties. Ab-initio calculations performed on approximants confirm also the validity of this anomalous quantum diffusion scheme. This provides us with  the first ab-initio model of conductivity in approximants such as the $\alpha$-AlMnSi phase.

Although the present chapter focuses on electrons in quasicrystals and related complex metallic alloys the concept that are developed here can be useful for phonons in these systems. There is also a deep analogy between the type of quantum transport described here and the conduction properties of other systems where charge carriers are also slow, such as  some heavy fermions or polaronic systems.

\end{abstract}



\end{frontmatter}

\newpage
\tableofcontents



\newpage
\section{Introduction}
\label{SecIntro}

Immediately after the discovery by Shechtman et al. \cite{Shechtman84} of quasiperiodic intermetallics one major question was raised about the physical properties of  phases with this new type of order. In particular, one expected that the electronic and thermal properties could be deeply affected. Indeed the description of electrons or phonons in periodic phases rests on the Bloch theorem which cannot be applied to a quasiperiodic structure. Within a decade a series of new quasiperiodic phases and approximant were discovered and intensively studied. These investigations learned us that indeed the electrons and the phonons properties could be deeply affected by this new type of order.

The first quasiperiodic alloys AlMn where metastable and contained many structural defects. As a consequence they had conduction properties similar to those of amorphous metals with resistivities in the range 100--500\,$\mu\Omega$cm. In 1986 the first stable icosahedral phase was discovered in AlLiCu. This phase was still defective and although  its resistivity was higher (800\,$\mu\Omega$cm) it was still comparable to that of amorphous metals. The real breakthrough came with the discovery of the stable AlCuFe icosahedral phase, having a high structural order. The resistivity of these very well ordered systems where very high, of the order of 10\,000 $\mu\Omega$cm,  which gave a considerable interest in their conduction properties.  Within a few years several important electronic characteristics of these phases  were experimentally demonstrated.  The density of states in AlCuFe was smaller than in Al, about one third of that of pure Al, but still largely metallic. The conductivity presented a set of characteristics that were either  that of semi-conductors or that of normal metals. In particular  weak-localization effects were observed that are typical of amorphous metals. Yet the conductivity was increasing with the number of defects just as in semi-conductors.  Optical measurements showed that the Drude peak, characteristic of normal metals,  was absent. In 1993 another breakthrough was the discovery of  AlPdRe  which had resistivities in the range of  $10^6\,\mu\Omega$cm \cite{Pierce93_science,Berger93,Akiyama93}. 
This system gave the possibility of studying a metal-insulator transition in a quasiperiodic phase.  There are still many questions concerning electronic transport in AlPdRe phases. One difficulty concerns the homogeneity and the quality of samples which are  crucial for transport properties but are difficult to determine exactly.  

Since the discovery of  Shechtman et al. \cite{Shechtman84} our view of the role of quasiperiodic order has evolved. For  electronic or phonon properties of most known alloys  it appears that the medium range order, on one or a few nanometers, is the real length scale that determines properties. This observation has lead the scientific  community to adopt a larger point of view and consider quasicrystals as an example of a larger class. This new class of Complex Metallic Alloys contains  quasicrystals, approximants and alloys with large and complex unit cells with possibly hundreds of atoms in the unit cell. 

In this chapter we shall concentrate on ``the way electrons propagate'' 
in a quasicrystal or in a complex metallic alloy. 
The main objective is to show that the non standard  conduction properties of some quasicrystals and related complex metallic alloys result from purely quantum effects and cannot be interpreted through the semi-classical theory of transport. This is of great importance since the semi-classical Bloch-Boltzmann theory is at the heart of our understanding of conduction in solids, 
ranging from metals to semi-conductors.  
This new type of quantum transport is related to the specific  propagation mode of electrons in these  systems. Indeed in quasicrystals and related complex phases the quantum diffusion law deviates from the standard ballistic law characteristic of  perfect crystals in two possible ways. In a perfect quasicrystal the large time diffusion law is a power law  instead of a ballistic one in perfect crystals. In a complex crystal the diffusion law is always ballistic at large time but  it can deviate strongly from the ballistic law at sufficiently small times. It is this specific character that provides a basis for the interpretation of the strange conduction properties of AlCuFe, AlPdMn and probably also for those of AlPdRe.

This chapter is organized as follows. Part 2 is the most technical part but it can be skipped by readers not interested by mathematical aspects. Part 3 presents a detailed physical interpretation of anomalous diffusion and low frequency conductivity  laws in crystals and quasicrystals. Part 4 presents  evidence   of anomalous diffusion in experimental quasicrystalline and approximant phases.

In  part 2 we give some definitions and present  the mathematical relations that exist  between the low frequency conductivity, including the dc conductivity, and quantum diffusion.
We consider also the relaxation time approximation (RTA) that allows to treat the role of disorder on quantum diffusion and conductivity.  We demonstrate general formulas for quantum diffusion and low frequency conductivity (within the RTA) in  periodic and quasiperiodic models of potential.   

In part 3 we focus on the physical interpretation and consequences of the formulas derived in part 2. On a general ground we discuss  the limitations of the RTA and the possibility of a metal-insulator transition. We apply this to a general theory of low frequency conductivity and metal-insulator transition in crystals and quasicrystals.

In part 4 we present briefly  the experimental transport properties of phases such as AlMnSi, AlPdMn and AlCuFe or AlPdRe. These experimental transport properties indicate a conduction mode which is neither metallic nor semiconducting. For the $\alpha$-AlMnSi phase, recent ab-initio computations  are presented, which confirm the existence of an anomalous diffusion and allow for a semi-quantitative ab-initio computation of conductivity. Concerning AlCuFe and related quasiperiodic phases, which cannot be addressed by band structure calculations, we present a phenomenological model. This model  based on  anomalous quantum diffusion provides a coherent interpretation of the strange electronic transport of these systems.  

We conclude by a short summary. We  discuss also briefly the link with other problems such as phonons in quasiperiodic systems or electrons in heavy Fermions systems or in polaronic systems.



\section{Quantum formalism for electronic transport}

In this section we give some definitions and recall general properties of the conductivity. We present  the mathematical relations 
that exist  between the low frequency conductivity, including the dc conductivity, and quantum diffusion. 
We consider also the relaxation time approximation (RTA) \cite{Ashcroft76} 
that allows to treat the role of disorder on quantum diffusion and conductivity.  
We demonstrate general formulas for quantum diffusion and low frequency 
conductivity (within the RTA) in  periodic and quasiperiodic models of potential.

\subsection{Impulse response  and analytical properties of the conductivity}

Let us consider a system, at thermodynamical equilibrium, submitted to an impulse of electric field
\beq
 E(t)= E\delta (t)
\label{delta}
\eeq
where $\delta(t)$ is the Dirac function. 
The resulting current density is $J(t)$ ($J(t)=0$ for $t<0$) and the response $j(t)$  is defined by
\ben
j(t) = \frac{J(t)}{E}  ~~~~~~~~~~~~~~~~~ j(t) = 0 ~~~~{\rm for} ~~~t<0
\label{defj}
\een 
Then the complex conductivity ${\sigma}(\omega)$ and the response $j(t)=J(t)/E$ are related through 
\beq 
{\sigma}(\omega) = \int\limits_{0}^{\infty}e^{i\omega t} j(t)dt 
\label{sigma} 
\eeq 
From (\ref{defj}), (\ref{sigma}) one deduces that
\ben
{\RRe}\,\sigma(\omega) = {\RRe}\,\sigma(-\omega)
\label{symsigma}
\een
\ben
j(|t|) = \frac{1}{\pi} \int_{-\infty}^{+\infty} \e^{i \omega t} {\RRe}\,\sigma(\omega)
\d \omega
\label{j(sigma)}
\een
  
 In (\ref{sigma}) the integral over the  time $t$ runs over ($t>0$) only due to causality ($j(t) = 0~for ~t<0$). This implies that the conductivity $\sigma(\omega)$ is an analytical function of frequency $\omega$ in the upper half of the complex plane. From the analyticity of $\sigma(\omega)$ the Kramers-Kr$\rm \ddot o$nig relations, that relate the real part and the imaginary part of the conductivity,  can be deduced 
\beq
{\IIm\, \sigma}(\omega) = \frac{1}{\pi}~{\rm PP}~
\int\limits_{-\infty}^{\infty}
\frac{\RRe\, {\sigma(u)}}{\omega - u} \d u \label{KK1}
\eeq
\beq
{\RRe\, \sigma}(\omega) = 
-\frac{1}{\pi}~{\rm PP} \int_{-\infty}^{\infty}
\frac{\IIm\, {\sigma(u)}}{\omega - u} \d u \label{KK2}
\eeq
where $\rm PP$ means the principal part of the integral. This  implies also the following spectral decomposition for $z$ in the upper half complex plane: 
\ben 
\sigma(z) = \frac{i}{\pi} 
\int_{-\infty}^{+\infty} \frac{{\rm Re}~ \sigma(\omega')}{z - \omega '}
\d \omega '
\een
Finally we recall that the conductivity obeys sum rules. For example the response $j(t=0)$ is independent of the quantum character of electrons. It  depends only on their concentration $n$, mass $m$ and charge $e$ through
\ben
j(t=0) = \frac{ne^2}{m}
\label{j(t=0))}
\een
Combining with (\ref{symsigma}),(\ref{j(sigma)}) one get
\ben
\int_{0}^{+\infty} {\rm Re} ~\sigma(\omega)
\d \omega=\frac{\pi ne^2}{2 m}
\label{jsumrule)}
\een

\subsection{Relation between low frequency conductivity and quantum diffusion}

The quantum diffusion of states having an energy $ E$ is defined as $\Delta X^{2}(E,t)$:
\ben
\Delta X^{2}(E,t) =\Big<[X(t)-X(0)]^{2}\Big>_{E}
\een
where $<{\rm A}>_{E}$ means an average of the diagonal elements of the 
operator $\rm A$ over all states with energy $E$. $X(t)$ is the position operator 
along the axis $x$ expressed in the Heisenberg representation. 
The velocity operator is defined as $V_{x}(t)=\d X(t)/\d t$, 
its correlation function $C(E,t)$  is defined as
\ben
C(E,t) = \Big\langle V_x(t)V_x(0) + V_x(0)V_x(t) \Big\rangle_E
= 2\,{\rm Re}\, \Big\langle V_x(t)V_x(0) \Big\rangle_E 
\label{EqAutocorVit}
\een
and is related to quantum diffusion \cite{MayouPRL00} through
\ben
\frac{\d}{{\d} t} \Delta X^2(E,t) = \int_0^{t}C(E,t'){\d} t'.
\label{RelXC}
\een

As shown in \cite{MayouPRL00}, the real part of the low frequency conductivity is related to quantum diffusion. Indeed from the Kubo-Greenwood formula the real part of the conductivity is given by
\ben
\RRe\,\sigma(\omega)= \int_{\mu -\hbar\omega}^{\mu} \frac{{\d}E}{\hbar\omega}F(E,\omega).
\label{KG1}
\een
where $\mu$ is the chemical potential. 
In (\ref{KG1}) the Fermi-Dirac distribution function 
is taken equal to its zero temperature value.
This is valid provided that the electronic properties vary
smoothly on the thermal energy scale  $kT$. 
For finite  temperature, the effect of the   
Fermi-Dirac distribution function on the transport properties
has been studied in the 
literature \cite{Macia02,Macia04,Solbrig00,Haussler02}. 
But, theses analyzes could not
explain the unconventional conduction  of quasicrystals and 
related alloys 
(very high resistivity at low temperature, 
and conductivity that increases strongly 
when 
defects or temperature increases).
Therefore in the following,
the Fermi-Dirac distribution function 
is taken equal to its zero temperature value. 
But
the effect of defects and temperature  
(scattering by phonons\,...)
on the diffusivity is taken into account via
the relaxation time approximation (section \ref{PRTA}).
The central quantity $F(E,\omega)$ is given by
\ben
F(E,\omega)=\frac{2 \pi \hbar e^2}{\Omega}{\rm Tr} \Big\langle \delta (E-H)V_x \delta (E+\hbar \omega-H)V_x \Big\rangle
\label{KG2}
\een
where $\Omega$ is the volume of the system  and $Tr$ means the Trace of an operator. 
Expressing the operator $\delta (E-H)$ as the Fourier transform of the evolution 
operator $\e ^ {-iHt}$ one shows  that 
\ben
\frac{2F(E,\omega)}{e^2 n(E)} =
\int _{-\infty}^{\infty} \d t~ \e^{i\omega t}\Big\langle 
V_{x} (t)V_{x} (0)\Big\rangle_{E}
\label{KG3}
\een
and
\ben
\frac{2F(E-\hbar \omega,\omega)}{e^2 n(E)} =
\int _{-\infty}^{\infty} \d t~ \e^{i\omega t}\Big\langle 
V_{x} (0)V_{x} (t)\Big\rangle_{E}
\label{KG4}
\een
where $n(E)$ is density of states per unit volume 
(summed over up and down spins which are assumed to have the 
same transport properties here). Then one finds
\ben
2\, \RRe\,\tilde{\sigma}(E,\omega)=F(E,\omega)+F(E-\hbar \omega,\omega)
\label{KG5}
\een
where
\ben
\tilde{\sigma}(E,\omega) = e^2 \frac{n(E)}{2} \,\int_0^{\infty} 
\e^{i \omega t}  C(E,t) {\d} t
\label{defsigmatilde}
\een

Let us note that the function $\tilde{\sigma}(E,\omega)$ 
is analytical in the upper half of the complex plane. 
For large $\omega\,$: $\tilde{\sigma}(E,\omega)\propto1/\omega$ 
and the Kramers-Kr$\rm \ddot o$nig relations are valid. 
Finally the usual sum rule is valid
\ben
\int _{0}^{\infty} \RRe\,\tilde{\sigma}(E,\omega)\d\omega
=\frac{\pi e^2n(E)}{2}C(E,t=0)=\frac{\pi e^2n}{2m^*}
\label{Csumrule}
\een
where $m^*$ is the effective mass and $n$ the density of conduction electrons. 

If the variation of $F(E,\omega)$ with energy is small in the  interval 
$[\ef - \hbar \omega,\ef + \hbar \omega]$ of values of $E$, 
one deduces from the previous set of equations  that
\ben
{\rm Re} ~\sigma(\omega) \simeq e^2 \frac{n(\ef)}{2} ~{\rm Re}\,\int_0^{\infty} 
\e^{i \omega t} 
C(\ef,t) {\d} t
\label{RESC}
\een
(\ref{RESC}) is  valid at sufficiently small values of $\omega$. 
In particular at zero frequency the dc conductivity is given by 
the Einstein relation
\ben
\sigma(0) = e^2 n(\ef) D(\ef)
\een
with 
\ben
D(\ef) = \lim_{t \rightarrow \infty}  \frac{1}{2}\frac{\d}{{\d} t} \Delta X^2(\ef,t)  
\een

Finally there is a simple relation between the velocity correlation function at the Fermi energy and the impulse response $j(t)$. Indeed comparing (\ref{RESC}) and (\ref{sigma}) one deduces the following equivalence at large time
\ben
j(t) \simeq e^2 \frac{n(\ef)}{2}C(\ef,t)
\label{ReljC}
\een

\subsection{Relaxation time approximation (RTA)}
\label{PRTA}

 Within the relaxation time approximation one assumes that the response currents  respectively with disorder $j(t)$ and without disorder $ j_0(t)$ are related through

\ben
j(t) = j_0(t) \e^{-|t| / \tau}
\label{Relaxation}
\een

where $\tau$ is the relaxation time. So the relaxation time approximation (RTA) allows to treat the effect of disorder  on quantum diffusion and conductivity. We give the relations satisfied by conductivity and quantum diffusion in this approximation. The conditions of validity of the RTA are discussed in part 3.

Using (\ref{sigma}), and within the RTA, the conductivity with disorder $\sigma(\omega,\tau)$ and without disorder $\sigma_0(z)$ are related by
\ben
\sigma(\omega,\tau) =\sigma_0 \left(\omega+\frac{i}{\tau}\right)
\een

The real part of conductivities with defects ${\rm Re}~\sigma(\omega, \tau)$ and without defects ${\rm Re}~\sigma_0(\omega)$ are related simply. Using (\ref{sigma}), it is straightforward to get

\ben 
{\rm Re}~\sigma(\omega, \tau) = \frac{1}{\pi \tau} 
\int_{-\infty}^{+\infty} 
\frac{{\rm Re}~\sigma_0(\omega ')}{(\omega - \omega ')^2 +\frac{1}{\tau^2}}
\d \omega '
\label{convolution}
\een  

which allows to compute the real part of the conductivity with defects.

We discuss now the RTA from the point of view of quantum diffusion. In all cases we consider  that the influence of disorder is much stronger on the quantum diffusion than on the density of states. We thus neglect the variation of $n(E)$ with disorder. From (\ref{ReljC}) one deduces that, for not too large disorder i.e. for sufficiently large relaxation time  $\tau$ the RTA is equivalent to

\ben 
C(E,t) = C_0(E,t) \e^{-|t| / \tau}
\een

where $C(E,t)$ and $C_0(E,t)$ are respectively the velocity correlation functions with and without disorder.  After  equation (\ref{RelXC}) one deduces that the long time propagation is diffusive with a diffusion coefficient defined as

\ben
D(E) =\frac{1}{2} \int_{0}^{+\infty} C_0(E,t) \e^{-|t| / \tau}\d t
\label{DRTA}
\een

which is equivalent to
\ben
D(E) = \frac{1}{2}\frac{\d}{{\d} t} \Delta X^2(E,t)  ~~~~{\rm if}~~t \gg \tau
\een

At zero frequency the diffusivity can be written in the useful form

\ben
D(E_{\mathrm{F}},\tau)=\frac{L^2(E_{\mathrm{F}},\tau)}{2\tau}
\label{DLTAU}
\een

Using the $t=0$ conditions 
$\Delta X^2(E,t=0)=0$ and $\frac{\d}{\d t} \Delta X^2(E,t=0)=0$  
and performing two integrations by part one get
\ben
L^2(E_{\mathrm{F}},\tau) = 
\frac{\displaystyle{\int_0^{+\infty}}
\Delta X_0^2(E_{\mathrm{F}},t) \e^{-{t}/{\tau}} {\d}t}
{\displaystyle{\int_0^{+\infty}} 
\e^{-{t}/{\tau}} {\d}t}=\Big<\Delta X^2(E_{\mathrm{F}},t)\Big>_{\tau}
\label{LTAU}
\een
where $\Big<..\Big>_{\tau}$ is a time average on a time scale $\tau$. 
$\Delta X_{0}(E,t)$ is the spreading of states of energy $E$,  
in the perfect system i.e. without disorder.

More generally at low frequency, using ($\ref{RESC}$) one can 
define a frequency dependent diffusivity 
$D(E_{\mathrm{F}},\omega)$ such that
\ben
{\rm Re} ~\sigma(\omega) \simeq e^2 n(\ef) ~D(E_{\mathrm{F}},\omega)
\label{RESC2}
\een
and:
\ben
D(E_{\mathrm{F}},\omega)~=~\frac{1}{2}~{\rm Re}\,\int_0^{\infty}  \e^{i \omega t}  C(\ef,t) {\d} t
\label{RESC3}
\een
within the RTA (\ref{RESC3}) writes
\ben
D(E_{\mathrm{F}},\omega) = \frac{1}{2} \,  {\rm Re}\,
\int_0^{+\infty} \e^{(i\omega -1/\tau)t} C_0(E,t) {\d}t
\een

It can be convenient  to use the equivalent form which expresses the frequency dependent diffusivity $D(E_{\mathrm{F}},\omega)$  in terms of the quantum diffusion without disorder $\Delta X_0^2(E,t)$:
\ben
D(E_{\mathrm{F}},\omega) = \frac{1}{2} \,  {\rm Re}\,
\left\{ \left(\frac{1}{\tau} - i\omega \right)^2
\int_0^{+\infty} \e^{(i\omega -1/\tau)t} \Delta X_0^2(E,t) {\d}t
\right\}
\label{D(X)}
\een

\subsection{Application to periodic Hamiltonians}
   
In this section we analyze the quantum diffusion in a perfect crystal, then we derive formulas for the low frequency conductivity within the RTA.
   
{\it Quantum diffusion}
 
Due to the Bloch theorem an eigenstate of a periodic Hamiltonian is defined by its wave vector $\vec k$ and by its band index $n$. The diagonal element of the velocity correlation operator on a state $n \vec k$ can be decomposed as follows:
 
\begin{equation}
C(n \vec k,t) =
2 \sum_m \left| V_{n,m}(\vec k)\right|^2 {\rm cos}\left( \left(E_n(\vec k)-E_m(\vec k) \right)
\frac{t}{\hbar} \right).
\label{eq_Cnk_Vx2}
\end{equation}

where $V$ is the velocity operator (in the chosen direction X) and $V_{n,m}(\vec k)$ is the matrix element between states $n \vec k$  and $m \vec k$.

At small time, $t \ll {\hbar}/{W} $ where $W$ is a typical bandwidth,  using ($\ref{RelXC}$), the quantum diffusion is always ballistic with:

\ben
\Delta X^2 = V_{tot}^2 t^2 ~~~~ {\rm if}~~ t \ll  \frac{\hbar}{W} 
\een
and 
\ben
V_{tot}^2 = \Bigg\langle \sum_{m}
\Big| \langle n\vec k | V_x | m\vec k \rangle \Big|^2
\Bigg\rangle_{E_n=E_{\mathrm{F}}}
\een

But in general the relevant time scale for electronic conductivity, which is the scattering time, is much larger than ${\hbar}/{W}$. The following decomposition is important. 

Using ($\ref{RelXC}$) one shows that quite generally the quantum diffusion $\Delta X^2(E,t)$ can be decomposed in a  ballistic contribution and a bounded part:

\begin{equation}
\Delta X^2(E,t) = V_{B}(E)^2 t^2 + \Delta X_{\mathrm{NB}}^2(E,t).
\label{Eq_DeltaX2}
\end{equation}

The ballistic term $V_{B}(E)^2 t^2$ is due  to the diagonal elements of the velocity correlation function (intraband contribution) , whereas $\Delta X_{\mathrm{NB}}^2(E,t)$ is due to the off diagonal terms of the velocity correlation function (interband contribution). One has

\begin{equation}
V_{B}(E)^2 =  \bigg\langle 
\Big|\langle n\vec k | V_x | n\vec k \rangle \Big|^2
\bigg\rangle_{E_n=E}
\end{equation}
and
\ben
\Delta X_{\rm NB}^2(E,t) \leq 2 \Delta X_{\rm \infty NB}^2 (E)
\een
An important relation exists between $\Delta X_{\rm \infty NB}^2(E)$ 
and the square length of the unit cell in the chosen direction $L^2$ namely:
\ben
\Delta X_{\rm \infty NB}^2(E) \leq
\left(\frac{L}{2}\right)^2
\label{bound}
\een

Indeed one shows easily that 
\begin{equation}
\Delta X^2_{\mathrm{NB}}(E_{\mathrm{F}},t) = 2 \hbar^2
~\Bigg\langle
\sum_{m \,(m\neq n)}
\frac{1 - \cos\Big((E_n-E_m)\frac{t}{\hbar} \Big)}{(E_n-E_m)^2} 
\Big| \langle n\vec k | V_x | m\vec k \rangle \Big|^2
\Bigg\rangle_{E_n=E_{\mathrm{F}}}
\label{Calcul_DeltaX2}
\end{equation}
 
the above expression is bounded by considering that all cosines are equal 
to $-1$ and we define $\Delta X_{\rm \infty NB}^2(E)$ as
\ben
\Delta X_{\rm \infty NB}^2(E) = 2 \hbar^2
~\Bigg\langle \sum_{m \,(m\neq n)}
\frac{ \Big| \langle n\vec k | V_x | m\vec k \rangle \Big|^2 }
{(E_n-E_m)^2}
\Bigg\rangle_{E_n=E}
\een

From  the definition of the velocity operator
\ben
V_X = \frac{1}{i \hbar} \Big[X,H\Big]
\een
and by  considering the matrix elements of the position operator:
\ben
\langle n\vec k | X | m\vec k \rangle = 
\int_{Cell} u^*_{n\vec k}(\vec r) x u_{m\vec k}(\vec r) {\d} \vec r
\een
one get
\ben
\Delta X_{\rm \infty NB}^2(E) = 2~ \Bigg\langle \sum_{m \,(m\neq n)}
\Big| \langle n\vec k | X | m\vec k \rangle \Big|^2
\Bigg\rangle_{E_n=E}
\een

Let us define an operator $X^-$ that is constant in each unit cell and gives the average position of each unit cell along the chosen X direction. Since this operator is constant in each unit cell the orthogonality of ${m\vec k}$ and ${n\vec k}$  with ($n \neq m$), implies:
\ben
\int_{Cell} u^*_{n\vec k}(\vec r) x^- u_{m\vec k}(\vec r) {\d} \vec r = 0
~~~~ {\rm for} ~~n \neq m
\een 

Thus  the operator $ (X-X^-)$   has the same off diagonal elements as $X$  between ${m\vec k}$ and ${n\vec k}$  with ($n \neq m$)
\ben
\langle n\vec k | X | m\vec k \rangle = \langle n\vec k | X-X^- | m\vec k \rangle =
\int_{Cell} u^*_{n\vec k}(\vec r) (x -x^-) u_{m\vec k}(\vec r) {\d} \vec r
\een

The operator $ X_P = (X-X^-)$ has also well defined  diagonal elements $\langle n\vec k | X-X^- | n\vec k \rangle$ in the basis of Bloch states contrary to  the operator $X$ which  has not well defined diagonal elements in this basis.

Thus one can write
\ben
\Delta X_{\rm \infty NB}^2(E) = 2~ \Bigg\langle \sum_{m \,(m\neq n)}
\Big| \langle n\vec k | X_P | m\vec k \rangle \Big|^2
\Bigg\rangle_{E_n=E_{\mathrm{F}}}
\een
and
\ben
\Delta X_{\rm \infty NB}^2(E) = \Bigg\langle
\langle n\vec k | X_P^2| n\vec k \rangle
-  \langle n\vec k | X_P | n\vec k \rangle^2
\Bigg\rangle_{E_n=E_{\mathrm{F}}}
\een

\ben
\Delta X_{\rm \infty NB}^2(E) = \Bigg\langle
\langle n\vec k | (X_P -  \langle n\vec k | X_P | n\vec k \rangle)^2| n\vec k \rangle\Bigg\rangle_{E_n=E_{\mathrm{F}}}
\een

The above expression depends only on the density probability of the Bloch wavefunction $ |  \Psi_{n\vec k} (\vec r) |  ^2$. A bound of the above expression is easily established and according to the  equation announced in (\ref{bound}) writes $\Delta X_{\rm \infty NB}^2(E) \leq \Big(\frac{L}{2}\Big)^2$.

{\it Low frequency conductivity in the RTA}

Let us consider  now the low frequency conductivity and diffusivity of a crystal within the RTA. Using (\ref{Eq_DeltaX2}) and (\ref{D(X)}) one has
\ben
D(E,\tau) &=& V_{\B}(E)^2 \tau + \frac{L_{\NB}^2(E,\tau)}{2\tau}
\label{Eq_diffusivite}
\een
with
\begin{equation}
V_{\B}(E)^2 =  \bigg\langle 
|\langle n\vec k | V_x | n\vec k \rangle \Big|^2
\bigg\rangle_{E_n=E}
\end{equation}
and
\ben
L_{\NB}^2(E,\tau) = \frac{\displaystyle{ \int_0^{+\infty}} 
\Delta X_{\NB}^2(t) \e^{-{t}/{\tau}} {\d}t}
{\displaystyle{ \int_0^{+\infty}} \e^{-{t}/{\tau}} {\d}t}
\label{LXNB}
\een

At low frequency, neglecting any variation of density of states with energy on a scale $\hbar\omega$,  one can still write
\ben
\RRe\,\sigma(E_{\mathrm{F}},\omega)=e^2n(E_{\mathrm{F}})D(E_{\mathrm{F}},\omega)
\een
where the frequency dependent diffusivity can here also be decomposed in a Boltzmann and a Non Boltzmann contributions:
\ben
D(E_{\mathrm{F}},\omega)=D_{\B}(E_{\mathrm{F}},\omega)+D_{\NB}(E_{\mathrm{F}},\omega)
\label{DDD}
\een
with the Boltzmann contribution $D_{B}(E_{\mathrm{F}},\omega)$ and the Non Boltzmann contribution $D_{\NB}(E_{\mathrm{F}},\omega)$  given by
\ben
D_{\B}(E_{\mathrm{F}},\omega) = \frac{1}{2} \,  {\rm Re}\,
\left\{ \left(\frac{1}{\tau} - i\omega \right)^2
\int_0^{+\infty} \e^{(i\omega -1/\tau)t} \Delta X_{\B}^2(E_{\mathrm{F}},t) {\d}t
\right\}
\label{DB2}
\een
or
\ben
D_{\B}(E_{\mathrm{F}},\omega) = \frac{V^2 \tau}{1+\omega^2\tau^2} 
\label{DB3}
\een
and 

\ben
D_{\NB}(E_{\mathrm{F}},\omega) = \frac{1}{2} \,  {\rm Re}\,
\left\{ \left(\frac{1}{\tau} - i\omega\right)^2
\int_0^{+\infty} \e^{(i\omega -1/\tau)t} \Delta X_{\NB}^2(E_{\mathrm{F}},t) 
{\d}t
\right\}
\label{DNB2}
\een
or equivalently
\ben
D_{\NB}(E_{\mathrm{F}},\omega) = \frac{1}{2} \,  {\rm Re}\,
\left\{ \int_0^{+\infty} \e^{(i\omega -1/\tau)t} C_{\NB}(E_{\mathrm{F}},t) 
{\d}t
\right\}
\label{DNB3}
\een
where $C_{\NB}(E_{\mathrm{F}},t)$ is the Non Boltzmann contribution to the velocity correlation function
i.e. the off diagonal contributions $n\neq m$ in (\ref{eq_Cnk_Vx2}).

\subsection{Application to quasiperiodic Hamiltonians}
\label{QPSection}

In this section we describe briefly the nature of states 
in a perfect quasiperiodic system and the associated anomalous diffusion mode. Then we derive expressions of the conductivity within the RTA.

{\it Critical eigenstates in quasiperiodic Hamiltonians}

A first point is to define what a quasiperiodic Hamiltonian is 
\cite{IntroQC,FujiwaraSofArt}. 
Here we refer  to the construction by the method of the acceptance zone in a crystal 
of higher dimension with a cut by a space with irrational slopes. 
We assume also that the basic pattern, in the crystal of higher dimension, 
does not connect different unit cells as it is the case for some incommensurate phases. 
An important consequence is that a local environment of size $L$ 
must be repeated exactly within a distance of order $L$.

Consider a general one dimensional tight-binding Hamiltonian $H$ of the form
\begin{eqnarray}
H = \sum_{\langle i,j\rangle} t |i\rangle\langle j | +\sum_i \epsilon_i  |i\rangle\langle i |
\end{eqnarray}
where the first part corresponds to hopping between nearest neighbors $\langle i,j\rangle$ 
and the second part to on-site energies $\epsilon_i$. 
For the Fibonacci chain:
\begin{eqnarray}
\epsilon_i &=& V[i\varphi] \\
V(x) &=& V_0 {\rm ~~for~~} -\varphi < x \leq \varphi^3  \\
V(x) &=& V_1 {\rm ~~for~~} -\varphi^3 < x \leq \varphi^2
\end{eqnarray}
where $\varphi$  is the Golden mean. 
For this Hamiltonian all states are critical. 
A critical state is intermediate between spatially extended 
and exponentially localized. 
Its envelopp  presents an algebraic decay with distance 
\cite{Kohmoto83,Tokihiro88,FujiwaraSofArt,Roche96,Bellissard03}.

It is interesting to compare the eigenstates of the Fibonacci chain with those 
of two classical models: the 1D Anderson Hamiltonian and the Harper Hamiltonian.
For the Anderson Hamiltonian the on-site energies $\epsilon_i$  
are chosen randomly between two values $[-W,+W]$. 
In that case all states are exponentially localized with distance from some
point. The localization length $\mathcal{L}$ depends on the energy and on disorder 
(through the parameter $W/t$) and decreases when $W/t$ increases.
For the Harper Hamiltonian the on-site energies are of the form
\begin{eqnarray}
\epsilon_i = \lambda \cos(2\pi \omega i)
\end{eqnarray} 
For $\omega$ rational the structure is periodic and the eigenstates are Bloch states. 
For $\omega$ irrational the structure is aperiodic. 
Depending on the parameter $\lambda$ the states are either extended ($\lambda < 2$), 
critical ($\lambda = 2$) or exponentially localized ($\lambda > 2$).

A remarkable difference between the Harper and the Fibonacci model is that in the second model 
the states are critical for any values of the parameters. 
It is clear that the exact repetition of a given environment has a strong influence 
on the long range correlations of eigenstates. 
Indeed in the Fibonacci model a given sequence of length $L$ is repeated exactly within 
a distance of order $L$ whereas in the Harper model (or in the Anderson model) this 
exact repetition does not exist.

\begin{figure}[]
\begin{center}
\rotatebox[origin=c]{90}{
\includegraphics[width=8cm]{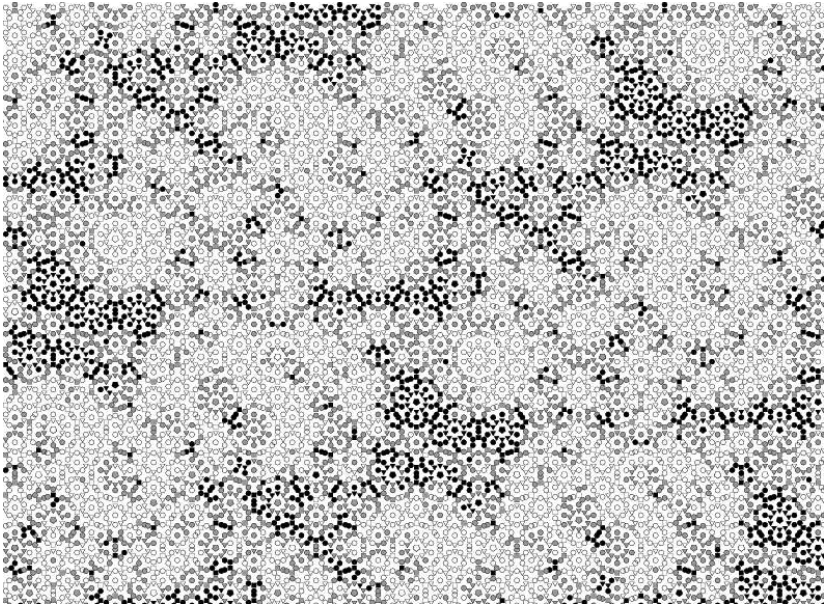}}
\vskip -1.5cm
\caption{Critical eigenstate in a decagonal model. The wavefunction density probability is small on white spots and strong on black spots.
From T. Fujiwara \cite{Fujiwara96}.
\label{FujiwaraDeca2}}
\end{center}
\end{figure}

A major question is whether the critical states persist in higher dimension. 
Although there could be exceptions it seems that in general states
are critical in 2 or 3 dimensions. There have been numerical studies of spectra in more than one dimension and  figure  \ref{FujiwaraDeca2}
presents a state calculated for a large 2-dimensional model by 
T. Fujiwara et al. \cite{Fujiwara96}. 

An argument proposed by C. Sire \cite{Sire94} is the following. 
If a wavefunction $\Psi_{L_0}$  is mainly localized in a region of extension $L_0$  
then it should live also on any similar environment. 
After Conway theorem, identical environments are 
located at a distance $2 L_0$  at most. Introducing a tunneling factor $z$ between 
the two local environments we write
\begin{eqnarray}
\Psi_{2 L_0} = z \Psi_{L_0} {\rm ~~and~~}
\Psi_{2^n L_0} = z^n \Psi_{L_0}
\end{eqnarray} 
Introducing $L=2^n L_0$, we can write equivalently
\begin{eqnarray}
\Psi_{L} = \left(\frac{L_0}{L}\right)^{\alpha} \Psi_{L_0} {\rm ~~with~~}
\alpha = \frac{\log z}{\log 2}
\end{eqnarray}
This qualitative argument points out the importance of the exact repetition of an environment 
within a distance comparable to the size of this environment, which is typical of a quasicrystal.

{\it Band scaling and anomalous diffusion}

The critical states are associated to a fractal energy spectrum in quasicrystal. 
The nature of this spectrum can be understood by considering a series of approximants 
with unit cell of increasing size $L$. 
When going from one approximant to the following, the bands are broken into smaller 
pieces which gives a fractal spectrum for the quasiperiodic system.

Typically the width $\Delta E$ of a band varies like
\begin{eqnarray}
\Delta E \simeq \frac{B}{L^{\gamma}}
\end{eqnarray} 

where $L$ is the length of the  unit cell of the system. 
The exponent $\gamma$ is greater than one due to the effect of the quasiperiodic potential. $\gamma$ depends on the energy of the wavepacket and of course on the parameters of the Hamiltonian.
This scaling law reflects also the critical character of eigenstates and their algebraic 
decrease with distance. 

The band scaling  has a direct consequence for the propagation of waves in quasiperiodic media. 
The spreading $L(t)$ of a wavepacket is neither ballistic (i.e. proportional to time $t$) 
as in crystals nor diffusive (i.e. proportional to the square root of time) 
as in disordered metals. 
In general it follows a power law and one can write after a time $t$: 
\begin{eqnarray}
L(t) 	\simeq A t^{\beta} \label{EqLAnomalusDif}
\end{eqnarray}
$\beta$ depends on the hamiltonian and on the energy of the wavepacket \cite{Piechon96}.
Indeed for an approximant with unit cell of size $L$ the characteristic velocity $v(L)$ 
is given by
\begin{eqnarray}
v(L) \propto \frac{1}{\hbar}\frac{\d E(\vec k)}{ \d \vec k} \propto \frac{L \Delta E}{\hbar} \propto \frac{B}{L^{\gamma - 1}}
\label{EqVitesseAnomalusDif}
\end{eqnarray} 

$\vec k$ is a wave vector in reciprocal space. 
When the magnitude of $\Delta \vec k$   is of order $1/L$ (i.e. the size of the Brillouin zone) 
the magnitude of the energy variation is of order $\Delta E$ 
(i.e. the width of a band). 
The last equality in (\ref{EqVitesseAnomalusDif}) is then obtained through equation 
(\ref{EqLAnomalusDif}). 
Since $\gamma > 1$ the typical velocity tends to zero when $L$ increases. 

Using the relation between the spreading of the wavepacket and the velocity at length scale $L$:

\begin{eqnarray}
\frac{\d L}{\d t} = v(L) \propto \frac{B}{L^{\gamma -1}}
\end{eqnarray}
One integrates straightforwardly the above differential equation and obtains:
\begin{eqnarray}
L(t) \simeq A t^{\beta} {\rm ~~with~~} \beta = \frac{1}{\gamma}<1
\end{eqnarray}

Thus in a quasiperiodic system  the asymptotic diffusion law  is at sufficiently large $t$:  

\ben
\Delta X_0^2(t) \simeq  At^{2\beta}
\een 

but let us recall that at small time the propagation is necessarily ballistic. Indeed after (\ref{RelXC}) and (\ref{EqAutocorVit}) one has

\ben
\Delta  X^2(E,t)\simeq  \Big\langle V_x(t=0)^2\Big\rangle_E t^2
\label{lowtX}
\een

\begin{figure}[]
\centerline{
\includegraphics[height=7.2cm]{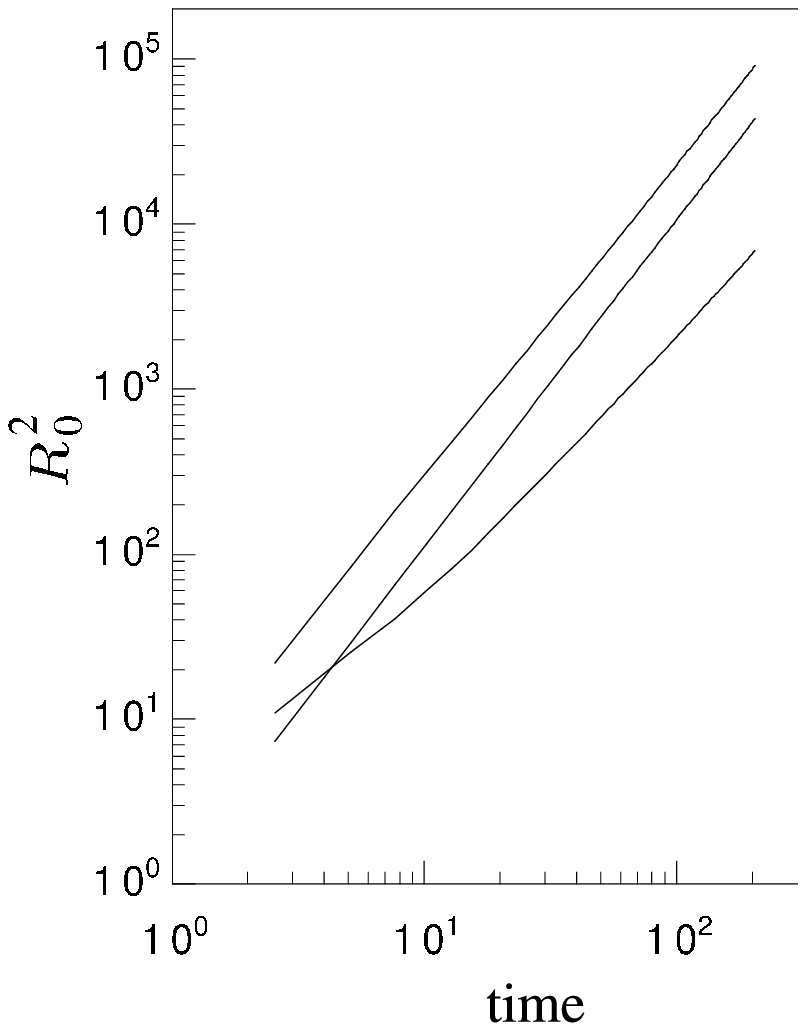}
\hspace{-.2cm}\includegraphics[height=7.2cm]{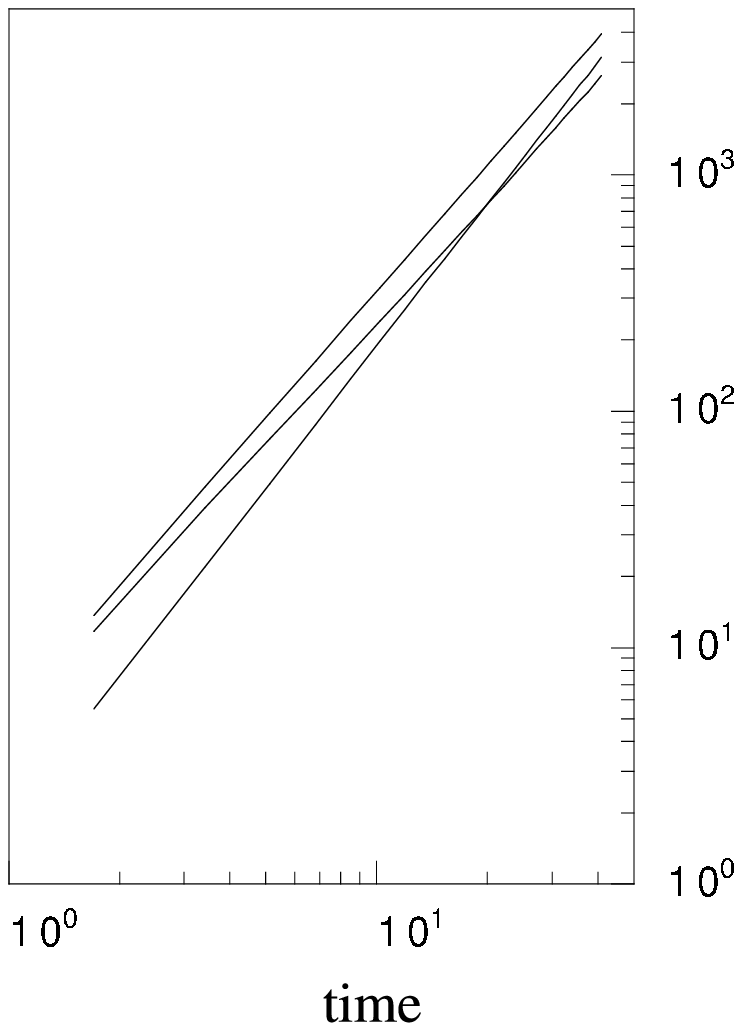}
}
\caption{Behaviour of $R_{0}^{2}(E,t)= \Delta X^{2}(E,t) +\Delta Y^{2}(E,t) + \Delta Z^{2}(E,t) $ for three energies  in the 2- (left) and the 3-dimensional (right) generalized Rauzy tiling. From \cite{Triozon02}.
\label{DIFFQUANT}}
\vspace{.5cm}
\centerline{
~~~~~~~\includegraphics[height=7cm]{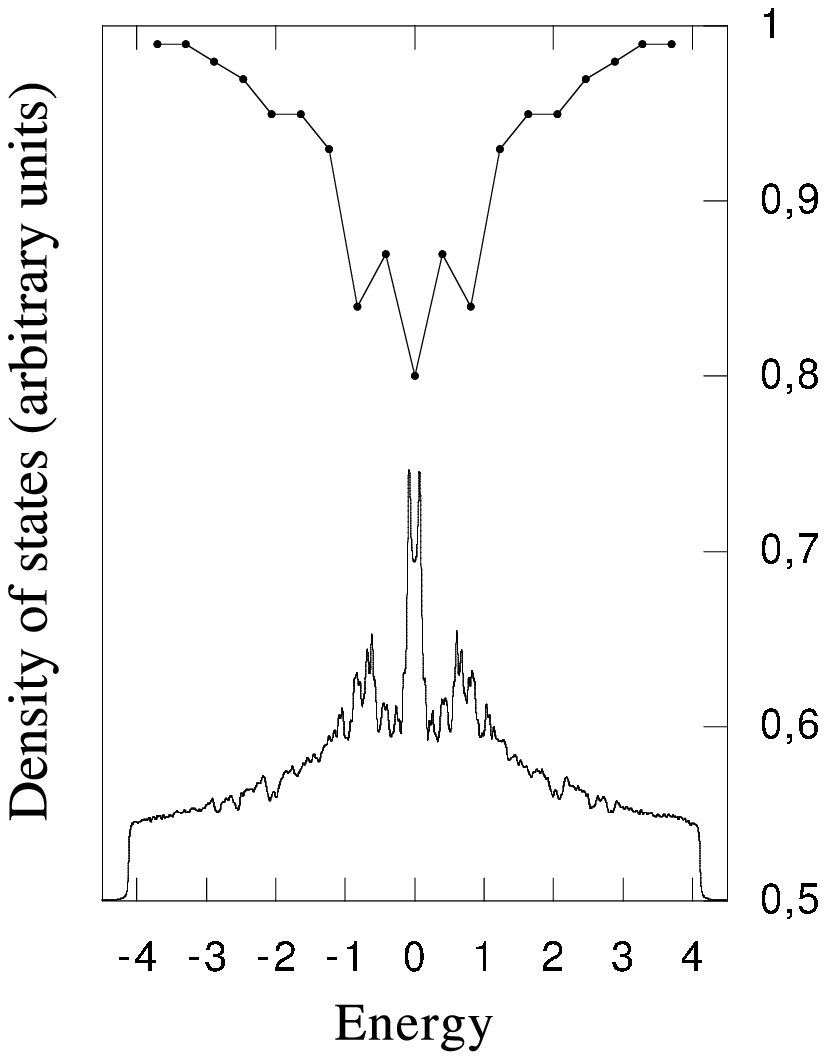}
\includegraphics[height=7cm]{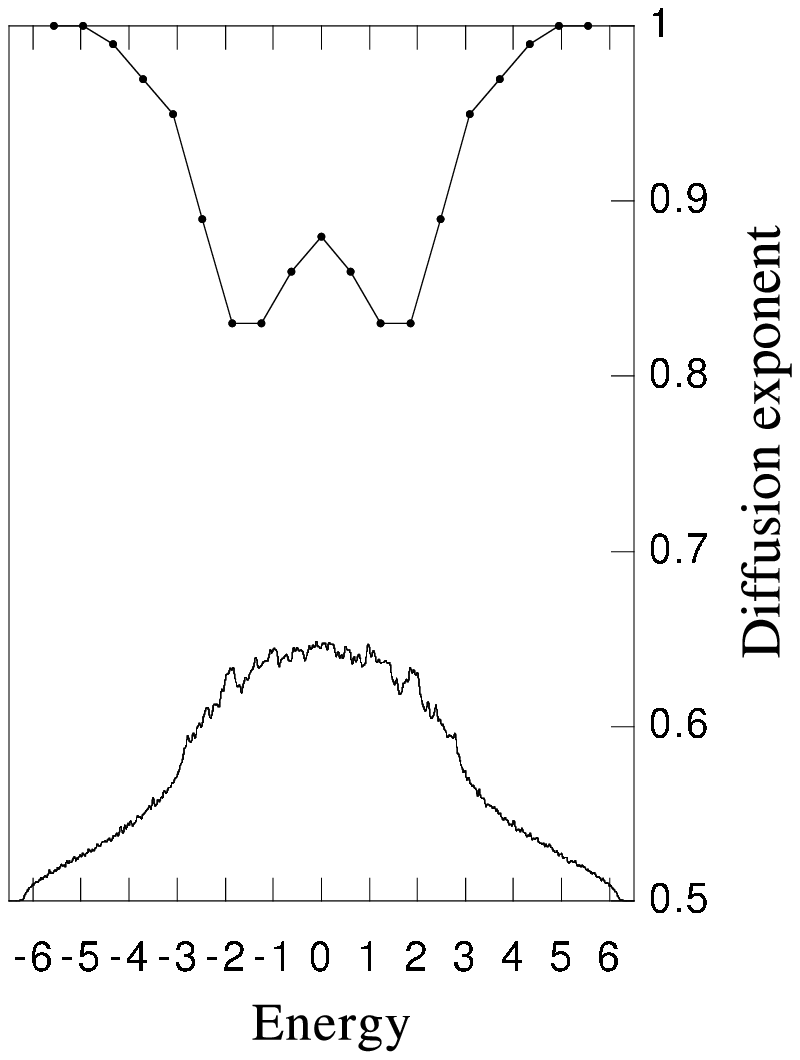}
}
\caption{Normalized density of states (lower curves) and diffusion exponents 
(upper curve) of 
the 2- (left) and the 3-dimensional (right) generalized Rauzy tiling. 
From \cite{Triozon02}.
\label{DOSBETA}}
\end{figure}

There is numerical evidence of anomalous diffusion in 1- or 2-dimensional systems
(see figures  \ref{DIFFQUANT}, \ref{DOSBETA}). It should be noted that these law are numerically approximate because in practice there is always fluctuations.
As a rule, one expects that the fluctuations are less important when 
the dimensionality of the system increases.

Associated to this anomalous diffusion one observes also an anomalous transmission 
of waves through 
a finite part of a quasiperiodic tiling of length $L$. 
The transmission coefficient varies like a power law of the length $L$. 
There is also a power law variation of the resistance of a stripe of length $L$.

It is interesting to consider the propagation law $\Delta X_0^2(L,t)$ in an approximant of unit cell size $L$, in the limit of large $L$,  and to relate it to the propagation in the quasicrystal $\Delta X_0^2(t)$. 

In the periodic approximant of unit cell size $L$ after equation (\ref{Eq_DeltaX2}):

\ben
\Delta X_0^2(L,t) =V(L)^2t^2 +\Delta X_{\NB}^2(L,t) 
\een
or equivalently 
\ben
\Delta X_{\NB}^2(L,t)=\Delta X_0^2(L,t)-V(L)^2t^2
\een
Since  the velocity $V(L)$ tends to zero at large $L$ one gets
\ben
\lim_{L\to \infty}\Delta X_{\NB}^2(L,t)=\lim_{L\to \infty}\Delta X_0^2(L,t)=\Delta X_0^2(t)
\een

Thus at a given time $t$  the Non Boltzmann contribution to the spreading $\Delta X_{\NB}^2(L,t)$, for an approximant of unit cell size $L$, tends to the anomalous diffusion law of quasicrystals $\Delta X_0^2(t)$ in the limit of large $L$.

{\it Low frequency conductivity in the RTA}

According to the anomalous law derived above and exemplified by numerical simulations we assume that at sufficiently large time $t\to\infty$ the diffusion law for a perfect quasicrystals  is

\ben
\Delta X_0(E,t)^2 \simeq  A(E)t^{2\beta(E)}
\label{Anomalous diff}
\een 

Let us recall that at small time $t$ the diffusion law is ballistic (\ref{lowtX}). According to the general expression relating frequency dependent diffusivity and quantum diffusion (\ref{D(X)})  
one has in the RTA: 
\ben
\RRe\, \sigma (E_{\mathrm{F}},\omega) = \frac{e^2n(E_{\mathrm{F}})}{2} \,  {\rm Re}\,
\left\{ \left(\frac{1}{\tau} - i\omega\right)^2
\int_0^{+\infty} \e^{(i\omega -1/\tau)t} \Delta X_0^2(E,t) {\d}t
\right\}
\label{RTAQC}
\een

If the asymptotic law (\ref{Anomalous diff}) is applicable on the time scales $\tau$ and $1/ \omega$ (which means a limit of small disorder and small frequency) then one can write
\ben
\RRe\, \sigma (E_{\mathrm{F}},\omega) = \RRe\, \tilde{\sigma} (E_{\mathrm{F}},\omega)
\label{GD1} 
\een
with
\ben
\tilde{\sigma} (E_{\mathrm{F}},\omega) = \frac{e^2n(E_{\mathrm{F}})}{2} A \Gamma (2\beta+1) \left( \frac{\tau}{1-i\omega\tau}\right) ^{2\beta-1}
\label{GD2}
\een

where $\Gamma$ is the Euler Gamma function. The formula (\ref{GD1}), (\ref{GD2}) 
is a  generalized Drude formula. 
Indeed if one considers the case $\beta=1$ then one recovers exactly the Drude formula. 

The generalized Drude formula is established for a given $\beta$ in the limit 
of infinite scattering time $\tau\to \infty$ and low frequency $\omega\to 0$. 
Thus this  formula cannot be applied  for a given $\tau$ in the limit $\beta\to 0$.  
In order to treat the case of small $\beta$, for a fixed $\tau$,  
we proceed in the
following way. We start from  the low frequency form (\ref{Modelsigma1}) 
for the perfect system and treat the effect of disorder within the RTA. 
For the perfect system we assume the conductivity satisfies:   
\beq
\RRe\, \sigma_{0}(\omega) = \sigma_{0}
\left(\frac{|\omega|}{\omega_{1}}\right)^{1-2\beta}   
\qquad\hbox{for $|\omega| < \omega_{1}$}
\label{Modelsigma1}
\eeq

After (\ref{sigma}), (\ref{Relaxation}) the conductivity
${\sigma}(\omega,\tau)$ of the system with defects satisfies:
\beq 
{\sigma}(\omega,\tau) \simeq {\sigma}_{0}\left(\omega + \frac{i}{\tau}\right)
\label{R}
\eeq
\noindent 
thus ${\sigma}(\omega,\tau)$ is determined once $\sigma_{0}(z)$ is known for 
$z=\omega +i/\tau$. In the limit
of low frequency and long scattering time we need $\sigma_{0}(z)$ 
for small values of $|z|$. 

\noindent Consider the two functions 
that are analytical in the upper half of the complex plane:
\beq
\tilde{\sigma}(z)=
\frac{\sigma_{0}}{\sin(\beta\pi)}\left(\frac{z}{i\omega_{1}}\right)^{1-2\beta}
\label{GD}
\eeq
\beq
\sigma_{x}(z)=\frac{ix}{ix+z}\tilde{\sigma}(z)      
\qquad\hbox{for  $x>0$}\label{x}
\eeq
\noindent In (\ref{GD}), we choose the following determination for the power of a complex number.  For a complex number $X=|X|e^{i\theta}$
with $-\pi< \theta \leq \pi$ we define $X^{\alpha}=|X|^{\alpha} e^{i\alpha\theta}$. 
From (\ref{Modelsigma1},\ref{GD}),
$\RRe\,\sigma_{0}(\omega) = \RRe\,\tilde{\sigma}(\omega)$ 
for $|\omega| < \omega_{1}$. 

For a function $f(z)$ which is analytical in the upper half of the complex plane,
 and which modulus  tends to zero at large $z$ one has
\beq
f(z) = \frac{i}{\pi}
\int\limits_{-\infty}^{+\infty}\frac{\RRe\,f(\omega)}{z-\omega}\d\omega
\label{spectral}
\eeq
\noindent The spectral representation (\ref{spectral}) applies to both 
$\sigma_{0}(z)$ because it is the conductivity of a real system  and 
to $\sigma_{x}(z)$ because of the factor ${ix}/({ix+z})$ 
which guarantees that the function $\sigma_{x}(z)$ is analytical in the 
upper half complex plane ($x>0$) and decreases sufficiently quickly at large $z$. So
\beq
{\sigma}_{0}(z) = \sigma_{x}(z) + \frac{i}{\pi}
\int\limits_{-\infty}^{+\infty}
\frac{\RRe\,\sigma_{0}(\omega)-\RRe\,\sigma_{x}(\omega)}{z-\omega}\d\omega
\label{spectral2}
\eeq
\noindent Since $\RRe\,\sigma_{0}(\omega)$ and $\RRe\,\sigma_{x}(\omega)$ 
are pair functions of $\omega$ one
can add together contributions of $\omega$ and $-\omega$ in (\ref{spectral2}).
Using $\RRe\,\sigma_{0}(\omega) = \RRe\,\tilde{\sigma}(\omega)$ for $|\omega| < \omega_{1}$ and taking the limit $x\rightarrow +\infty$ one gets
\beq
{\sigma}_{0}(z) = \tilde{\sigma}(z) - \frac{2zi}{\pi}
\int\limits_{\omega_{1}}^{+\infty}\frac{\RRe\,\sigma_{0}(\omega)-
\RRe\,\tilde{\sigma}(\omega)}{\omega^{2}(1-z^{2}/\omega^{2})}\d\omega
\label{spectral3}
\eeq
Developping the integral in a power series of $z$ and keeping only the 
term proportionnal to $z$ one gets
\beq
\sigma_{0}(z) = 
\tilde{\sigma}(z) +\frac{iz\sigma_{0}}{\pi\omega_{1}\beta}-
\frac{2iz\alpha\sigma_{0}}{\pi\omega_{1}} + O\left(\left(\frac{z}{\omega_{1}}\right)^{3}\right)
\label{sigma00(z)} 
\eeq
\beq
 \alpha =\frac{\omega_{1}}{\sigma_{0}} \int\limits_{\omega_{1}}^{+\infty}\frac{\RRe\,\sigma_{0}(\omega)}{\omega^{2}}\d\omega
\label{defalpha}
\eeq
\noindent For a given $\beta \neq 0$ the first term in the right member of 
(\ref{sigma00(z)}) dominates in the limit $z=0$. In this limit, one recovers, as expected, 
the Generalized Drude formula (\ref{GD1}), (\ref{GD2}) and \cite{MayouPRL00}. 

Yet it is interesting to look at the limit of $\beta\to0$ for a given 
$\tau$ or given frequency $\omega$ i.e.  
for a given value of $z=\omega+i/\tau$.  
In this limit, not considered in \cite{MayouPRL00}, one gets from 
(\ref{sigma00(z)}), (\ref{GD1}) and (\ref{GD2}):
\beq
\sigma_{0}(z) = \frac{2\sigma_{1} z }{i\pi\omega_{1}}
\left[\alpha - \log\left(\frac{z}{i \omega_{1}}\right)\right] 
+ O \left((z/\omega_{1})^{3}\right)
\label{sigma0(z)}
\eeq
\noindent In (\ref{sigma0(z)}) $\log(X)=\log|X| + i\theta$  for $X=|X|\e^{i\theta}$
 with
$-\pi< \theta \leq \pi$. Introducing $A=2\sigma_{0}/\pi\omega_{1}$ 
the dissipative part of conductivity, up to terms of order $(z/\omega _1)^3$,  
is given by
\beq
{\RRe\,\sigma} \simeq \frac{A}{\tau}
\left[\alpha + \log\left(\frac{\omega_{1} \tau}{\sqrt{1+(\omega \tau)^{2}}}\right)
+ \omega \tau {\rm Arctg}(\omega \tau)\right]
\label{Resigma}
\eeq

Finally we note that the quantum diffusion associated to the conductivity law 
($\ref {Resigma}$) is logarithmic. Indeed the zero frequency diffusivity 
is given by $D\propto \frac{A}{\tau} (\alpha + \log (\omega_{1} \tau ))$ 
and after ($\ref {DLTAU}$) the mean free path is 
$L^{2}(\tau) \propto (\alpha + \log(\omega_{1} \tau))$. 
Physically the mean free 
path represents the diffusion length between two scattering events in the perfect system.

\section{Anomalous quantum diffusion  and conductivity in periodic and quasiperiodic 
systems}
\label{Anomalous quantum diffusion  and conductivity}
In this part 3  we discuss first the limitations of the RTA and the possibility 
of a metal-insulator transition. Then we focus on the physical interpretation 
and consequences of the formulas derived in part 2. We apply this to a general 
theory of low frequency conductivity and metal-insulator transition in crystals 
and quasicrystals. 
 
 Let us recall first the main formulas derived in part 2 to treat low frequency 
conductivity.

 In the impulse response formalism one considers the response of a system to an impulse of electric field $E(t)= E\delta (t)$ where $\delta(t)$ is the Dirac function. The resulting current density is $J(t)$ ($J(t)=0$ for $t<0$) and the response $j(t)$  is defined by  
$j(t) = {J(t)}/{E}$.
The complex conductivity ${\sigma}(\omega)$ and the response $j(t)$ are related through
\beq
{\sigma}(\omega) = \int\limits_{0}^{\infty}\e^{i\omega t} j(t)\d t
\label{sigmap}\eeq

The quantum diffusion of states of energy $E$  is measured by the quantity 
$\Delta X^{2}(E,t)$:
\ben
\Delta X^{2}(E,t) =\Big<[X(t)-X(0)]^{2}\Big>_{E}
\label{Xp}
\een

where $<A>_{E}$ means an average of the diagonal elements of the operator 
$\rm A$ over all states with energy $\rm E$. $X(t)$ is the position operator 
along the $x$ axis in the Heisenberg representation. 
The velocity operator $V_x(t)=\d X(t)/\d t$ has a correlation function $C(E,t)$  defined as

\ben
C(E,t) = \Big\langle V_x(t)V_x(0) + V_x(0)V_x(t) \Big\rangle_E
= 2\,{\rm Re}\, \Big\langle V_x(t)V_x(0) \Big\rangle_E 
\label{EqAutocorVitp}
\een

and is related to quantum diffusion through

\ben
\frac{\d}{{\d} t} \Delta X^2(E,t) = \int_0^{t}C(E,t'){\d} t'.
\label{RelXCp}
\een
 
 The low frequency conductivity satisfies
 
\ben
{\rm Re} ~\sigma(\omega) \simeq e^2 n(\ef) ~D(E_{\mathrm{F}},\omega)
\label{RESC2p}
\een

with $n(E)$ the total density of states and

\ben
D(E_{\mathrm{F}},\omega) = \frac{1}{2} \,  {\rm Re}\,
\int_0^{+\infty} \e^{(i\omega)t} C(E_{\mathrm{F}},t) {\d}t
\label{D(C)p}
\een

In the limit of large time $j(t)$ and $C(\ef,t)$ are related by

\ben
j(t) \simeq e^2 \frac{n(\ef)}{2}C(\ef,t)
\label{ReljCp}
\een
where $E_{\mathrm{F}}$ is the Fermi energy and $n(\ef)$ the density of states 
at the Fermi energy (summed over spin up and down). 
 
Within the relaxation time approximation (RTA) 
one assumes that $j(t)$ and $C(E,t)$ 
with disorder are related to  $ j_0(t)$  and   $C_0(E,t)$ without disorder  through

\ben
j(t) = j_0(t) \e^{-|t| / \tau}
\label{Relaxationp1}
\een
\ben
C(E,t) = C_0(E,t) \e^{-|t| / \tau}
\label{Relaxationp2}
\een

Here the Fermi-Dirac distribution function 
is taken equal to its zero temperature value.
This is valid provided that the electronic properties vary
smoothly on the thermal energy scale  $kT$. 
But in the RTA, the effect of defects and temperature  
(scattering by phonons\,...)
is taken into account through the relaxation time 
$\tau$.
The diffusivity is given by 
 \ben
D(E_{\mathrm{F}},\omega) = \frac{1}{2} \,  {\rm Re}\,
\int_0^{+\infty} \e^{(i\omega -1/\tau)t} C_0(E_{\mathrm{F}},t) {\d}t
\label{D(C0)p}
\een

It can be convenient  to use the following equivalent form which expresses 
the frequency dependent diffusivity $D(E_{\mathrm{F}},\omega)$ 
in term of the quantum diffusion in the systemwithout disorder $\Delta X_0^2(E,t) $
\ben
D(E_{\mathrm{F}},\omega) = \frac{1}{2} \,  {\rm Re}\,
\left\{ \left(\frac{1}{\tau} - i\omega \right)^2
\int_0^{+\infty} \e^{(i\omega -1/\tau)t} \Delta X_0^2(E_{\mathrm{F}},t) {\d}t
\right\}
\label{D(X)p}
\een

Note that ($\ref{D(C)p}$), ($\ref{D(X)p}$) are valid at sufficiently small frequency. 
The condition is that the conductivity varies only slowly with $\ef$ on the energy 
scale $\hbar\omega$. If this is not the case it is still possible to define an 
average of the conductivity over a range of values of the energy $\ef$. 
In that case the above formulas are applicable to the conductivity averaged 
on a scale of Fermi energies greater than $\hbar\omega$.

At zero frequency the above formula ($\ref{D(X)p}$) is exact within the RTA and can 
be written as
\ben
D=\frac{L^{2}(E_{\mathrm{F}},\tau)}{2\tau} 
\label{D(L)}
\een 
with the mean free path $L(E_{\mathrm{F}},\tau)$ given by
\ben
L^2(E_{\mathrm{F}},\tau) = \frac{1}{\tau}
\int_0^{+\infty} \e^{ -t/\tau} \Delta X_0^2(E_{\mathrm{F}},t) {\d}t
\label{L(X)p}
\een

\subsection{Validity of the RTA and Anderson transition}

\begin{figure}[]
\begin{center}
\includegraphics[width=7cm]{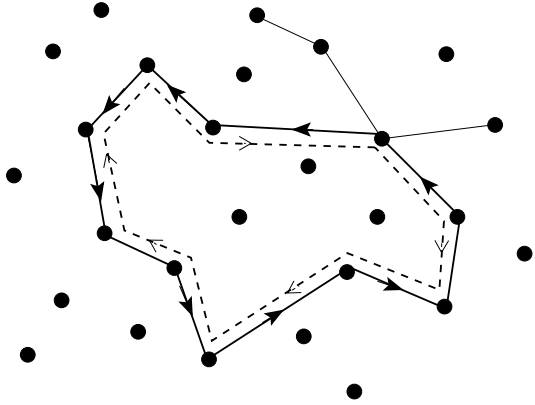}
\caption{Interference between a path (solid line) and its time reversal image (dashed line) for an electron which diffuses  through a static disordered potential. The interference that occurs between two different paths after several scattering events cannot be described by the RTA.
\label{Fig_WeakLocBoucle}}
\end{center}
\end{figure}

Let us discuss now the conditions of validity  of the RTA. We consider first the role of 
$\it{elastic~scattering}$ i.e. diffusion by a static potential. 
From the definition of the RTA it is clear that 
$j(t) = j_0(t) \exp\left(-|t| / \tau\right)$ is essentially zero beyond the the relation time $t>\tau$. However this is not the case in many disordered systems which present elastic scattering. For example in the case of free electrons scattered by static defects at random in a three dimensional system there are interferences between several scattering paths. This is represented in figure \ref{Fig_WeakLocBoucle}. As a consequence of these interferences the long time behavior of the response current $j(t)$, in the absence of a magnetic field,  is
\ben
j(t) = - A t^{-{3}/{2}}   ~~~~{\rm with}~~ A > 0 
\een

Thus interferences effects in the diffusive regime cannot be in general described properly by the RTA. Indeed in the RTA all quantum correlations are lost beyond the scattering time scale $\tau$. Figure \ref{Fig_WeakLocBoucle} shows a type of interference that occurs on a time scale greater than the scattering time $\tau$ and thus cannot be described in the RTA.

\begin{figure}[]
\begin{center}
\includegraphics[width=9cm]{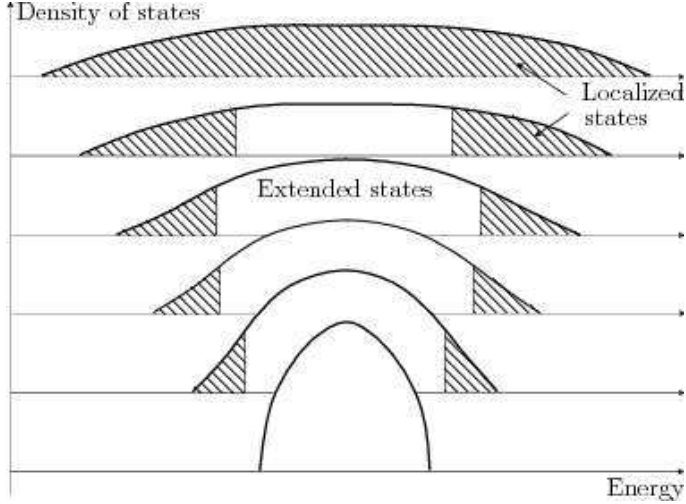}
\caption{Schematic band structure of an Anderson insulator, 
with increasing disorder from bottom to top. At zero temperature the system is metallic if the Fermi energy is in the extended states region. The system is insulating if the Fermi energy is in the localized states region.
\label{Fig_BandAndersonInsulator}}
\end{center}
\end{figure}

Ultimately these interferences can lead to a localization of the states, 
provided that the disorder is sufficiently strong. 
This Anderson localization is schematically represented 
in figure \ref{Fig_BandAndersonInsulator}. 
Of course  a correct theory of the Anderson localization is also 
beyond the  RTA.

For a 3-dimensional system the importance of quantum interferences depends on the ratio 
between the dc-conductivity of the system $\sigma_{dc}$ and the Mott value 
$\sigma_{\rm Mott}\, \simeq 600 \,{\rm (\Omega cm)^{-1}}/ \Lambda $ 
where $\Lambda$ 
is the mean-free path expressed in Angstr\"oms. 
If $R=\sigma_{dc}/\sigma_{Mott} \gg  1 $ the effect of 
the quantum interferences 
on $\sigma_{dc}$ is small. 
In that case the quantum interferences in the diffusive 
regime change only slightly the value of the conductivity and the RTA 
can describe the role of elastic scattering. If the  ratio R is closer to one then the RTA cannot be used to describe the role of elastic scattering.

The case of $\it{inelastic~scattering}$ is different. 
It is generally assumed that inelastic scattering with a scattering time  $\tau_{in}$ 
destroys quantum interferences on a time scale $t>\tau_{in}$. So  the relaxation time 
approximation is expected to be valid in the case of inelastic scattering. 
Yet a condition is that  the system contains a sufficiently large number of states so that  
the electron can find a state to scatter into close to the scattering event. 
 
 For this reason the hopping regime  between localized states 
(either short range hopping or variable long range hopping) is not described by the relaxation time approximation. Indeed  each hopping  process requires an exchange of energy with phonons that provides to the electron the difference in energy between the initial and final localized states. 
 
{\it General~case~according~to~the~scaling~theory}

We discuss now the case where there is both elastic scattering (characterized by $\tau_{el}$) and inelastic scattering scattering (characterized by $\tau_{in}$) and assume that $\tau_{in} >\tau_{el}$. This often happens since $\tau_{in}$ diverges at low temperature whereas $\tau_{el}$ is temperature independent at sufficiently low temperature. In the case where $\tau_{in} < \tau_{el}$ the elastic scattering is expected to have a minor effect on transport, and one is back to the case of pure inelastic scattering.

Let us define  $t(L)$  the time needed for a wavepacket to spread on a length scale $L$ and $D(L)={L^2}/{2t(L)}$  the diffusivity $D(L)$ at length scale $L$. Let us define also $L_{in}$ which is such that $\tau_{in}=t(L_{in})$.  

According to the scaling theory the diffusivity depends on the length scale due to the quantum interferences in the diffusive regime. One can distinguish three steps. 

1) Define the diffusivity at the length scale of the elastic mean free path. This can be done by using the RTA with a relaxation time equal to the elastic scattering time. This is valid because the quantum interferences do not act on length scales smaller than the elastic mean free path.

2)  Consider the conductance $g(L)$ of a cube with size $L$.

\begin{figure}[]
\begin{center}
~~~~~~~~~~~~~~~\includegraphics[width=9cm]{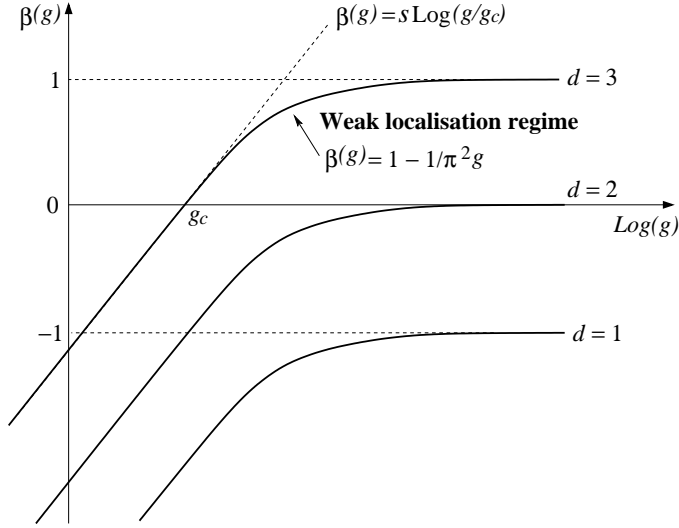}
\caption{Schematic representation of the $\beta(g)$ 
function for systems of dimension $d$.
\label{Fig_Beta}
}\vspace{.5cm}
\end{center}
\end{figure}

\ben
g(L)=e^2\frac{n(E)}{2}\frac{L^3}{t(L)}=e^{2}n(E)D(L)L
\label{defg}
\een
and apply the scaling relation on length scales $L$ such that 
$L_{in}>L>L_{e}$. 
\ben
\frac{\d \log(g)}{\d \log(L)}=\beta(g)
\label{scaling}
\een
$beta(g)$ is represented schematically in figure \ref{Fig_Beta}.
The relation (\ref{scaling}) defines entirely the quantum diffusion and thus the velocity correlation function as a function of the length scale $L$ or also as a function of time scale $t(L)$. Thus it allows to compute $C_{scaling}(E,t) $ where $C_{scaling}(E,t)$ is the velocity correlation function corresponding to the purely elastic scattering.

3) One stops the renormalization procedure at the length scale
 $L_{in}=L(\tau_{in})$, which means that the macroscopic diffusivity is simply 
$D=D(L_{in})$. Equivalently the role of inelastic scattering is simply to destroy the velocity correlation on the time scale $ \tau_{in}$ i.e. one has
\ben
C(E,t)=C_{scaling}(E,t) \e^{-{t}/{\tau_{in}}}
\label{Cscalein}
\een

The renormalization procedure determines entirely $C_{scaling}(E,t)$ and thus through 
(\ref{Cscalein}) determines entirely the quantum diffusion 
of the system with inelastic scattering.

\subsection{Phenomenon of backscattering}

\begin{figure}[]
\begin{center}
\includegraphics[width=9cm]{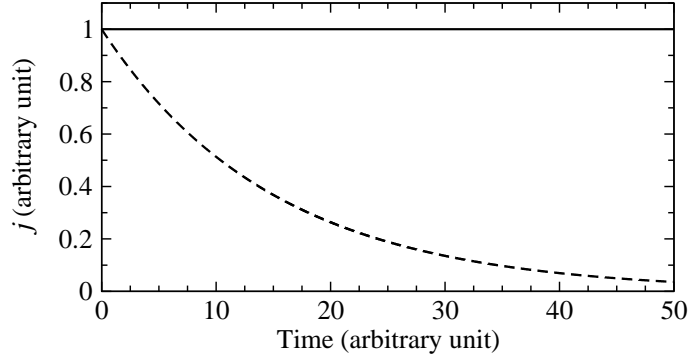}
\caption{Response current $j(t)$ within the Drude model. Without disorder (solid line) the response is constant $j=ne^2/m$. With disorder (dashed line) the response decays exponentially on the time scale $\tau$.
\label{drudepdf}}
\end{center}
\end{figure}

Let us consider the response $j(t)$ to an electric field $E \delta (t)$ applied to a system. During the impulsion the dynamics is dominated by the effect of the electric field. This implies that $j(0)$ is independent of the atomic forces applied to the electrons and is given by the classical response $j(0)={ne^2}/{m}$. In particular $j(0)$ is positive. In the classical Drude model  $j(t)=j(0)\exp\left(-t/\tau\right)$ tends to zero at larger times but is always positive. This is illustrated by figure \ref{drudepdf}.

 Quantum effects can lead to a counter intuitive situation where $j(t)$ becomes negative in some time interval. This  is the phenomenon of backscattering. This phenomenon occurs in a recurrent manner in the physics of quasicrystals and related complex intermetallics. So we give here an account of  its characteristics and its consequences on conductivity.

We illustrate first the phenomenon of backscattering in two cases concerning disordered systems:  the weak-localization and the strong localization regimes.

{\it Inelastic ~scattering ~in  ~the ~ weak~localization~regime}

\begin{figure}[]
\begin{center}
\includegraphics[width=9cm]{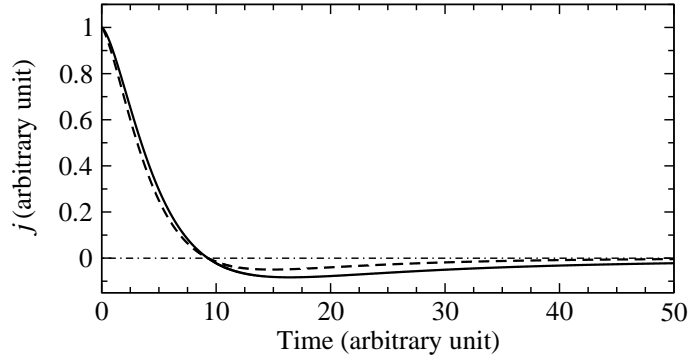}
\caption{Response function $j(t)$  in a disordered system with weak-localization effect, and thus backscattering at large time.  Without inelastic scattering (solid line)  and  with inelastic scattering (dashed line).
\label{WeakLocpdf}}
\end{center}
\end{figure}

We assume that the elastic scattering time $\tau_{e}$ is much shorter 
than the inelastic scattering time $\tau_{in}$. On the time scale 
$\tau_{e} <t<\tau_{in}$ the interferences are insensitive to inelastic 
scattering and  $j(t) = - A t^{-{3}/{2}}$ (for a 3-dimensional system), 
but on the time scale $t>\tau_{in}$ the interferences are destroyed. 
Thus a general expression of $j(t)$ for $t >\tau_{el}$ is (figure \ref{WeakLocpdf}):
\ben
j(t) = - A t^{-{3}/{2}} \e^{-{t}/{\tau_{in}}}  ~~~~{\rm with}~~ A > 0 
\een

In that case using equation ($\ref{sigmap}$) one finds for the dc conductivity:

\ben
\sigma_{dc} = \int\limits_{0}^{\infty} j(t)\d t = \sigma_{0} + \frac{2A\sqrt{\pi}}{\sqrt{\tau_{in}}}
\label{sdc}
\een

where $\sigma_{0}$ is the dc-conductivity in the absence of inelastic scattering. That is inelastic scattering tends to increase the conductivity. This paradoxal result is directly related to the backscattering phenomena that occurs with static disorder. Indeed in the integral ($\ref{sdc}$) with $j(t) = - A t^{-{3}/{2}} \e^{-{t}/{\tau_{in}}}$,  $\tau_{in}$ can be considered as a cut-off which supresses the negative contributions of $j(t) = - A t^{-{3}/{2}}$ at $t>\tau_{in}$.

The role of a small frequency  $\omega$ is comparable to that of inelastic scattering at a time scale $\tau_{in}\simeq {1}/{\omega}$. It acts also as a cut-off. For example in the case of the weak localization one has for $\tau_{in} \to \infty$:
\ben
\RRe\,\sigma(\omega)=\sigma_{0}+A\sqrt{2 \pi  \omega}
\een

This means that $\RRe\,\sigma(\omega,\tau)$ increases with frequency at small frequencies.  This is at the opposite of the standard behavior of metals, where the low frequency conductivity is dominated by the Drude peak.

The role of frequency can also be deduced from  the relation (\ref{convolution}) between the real parts of conductivity with scattering and without scattering. Indeed this relation (\ref{convolution}) shows that $\sigma_{dc}$ is the average of $\RRe\,\sigma_{0}(\omega)$ by a Lorentzian of width ${1}/{\tau}$ centered at zero frequency.   $\sigma_{dc}$ can increase with $1/\tau$ only if  $\RRe\,\sigma_{0}(\omega)$ increases with $\omega$.
 
{\it Inelastic ~scattering ~for ~ localized~ states~:~ the~Thouless~regime}

\begin{figure}[]
\begin{center}
\includegraphics[width=9cm]{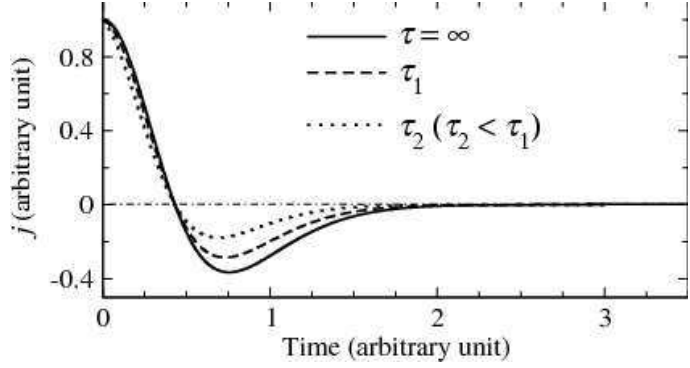}
\caption{Response function $j(t)$ for an insulator. Beyond a characteristic time ($T_{c}\simeq 2$ here)  the response $j_{0}(t)$ is essentially zero. Without inelastic scattering (solid line) the conductivity is zero  and the integral  $\sigma_{dc}=\int\limits_{0}^{\infty} j_{0}(t)\d t =0$. With inelastic scattering (dashed lines) the system is conducting and the integral $\sigma_{dc}=\int\limits_{0}^{\infty} j_{0}(t)\e^{-t/\tau_{in}}\d t >0$.
\vspace{0.3cm}
\label{jinsulator}}
\end{center}
\end{figure}

Let us consider a system that it is an insulator. In that case the zero frequency conductivity is zero:
\ben
{\sigma}(\omega=0) = \int\limits_{0}^{\infty} j(t)\d t=0 \label{sigma=0}
\een

Since the integral is zero, and since the small time response $j(t)>0$  it means that there are times interval where $j(t)<0$. Thus in an insulating system the backscattering phenomenon necessarily occurs. For Anderson localization one expects that  $j(t)>0$ at small times and $j(t)<0$ at large time. Beyond a characteristic time $T_c$,  the response $j(t)$ is essentially zero (see figure \ref{jinsulator}). 

Provided that the RTA is applicable one get 

\ben
{\sigma}(\omega=0,\tau) = \int\limits_{0}^{\infty} j_0(t)\e^{{-t}/{\tau}}\d t=0
\label{sigma=0p}
\een

If $\tau > T_c$ then the exponential differs significantly from $1$ only in the large  time region $t>T_c$. In that case one can replace  $\exp({{-t}/{\tau}})$ by $1-t/\tau$ in the range ($t<T_c$) where $j(t)\neq 0$. This gives 

\ben
{\sigma}(\omega=0,\tau) =-\frac{1}{\tau} \int_{0}^{\infty} j_0(t)t \d t
= 0
\label{sigma=0p0}
\een

Although $\int\limits_{0}^{\infty} j_0(t)\d t=0$  the integral 
$ \int\limits_{0}^{\infty} j_0(t)t\d t$ in ($\ref{sigma=0p0}$) is negative because $j_{0}(t)$ is negative at large times and positive at small times.

The regime described here correspond to the Thouless regime, that can be described by analyzing the quantum diffusion. In the Thouless regime, the physical picture is that of electrons spreading during a time $\tau_{in}$ between two inelastic scattering events and then loosing completely their phase at each {\it inelastic} scattering event.   According to the Thouless scenario, in the limit of large {\it inelastic} scattering time $\tau_{in}$, the spreading of the electron wavefunction between two inelastic scattering  events  is bounded by the localization length $\xi_E$.  Since the electron must wait till the next inelastic scattering event to loose phase memory and spread again the diffusivity is given by
\ben
D_{Thouless} (E)=\frac{\xi_E^2}{2\tau_{in}}
\label{DThouless}
\een
Within the RTA the diffusivity which is the square of the spreading during $\tau_{in}$ divided by $\tau_{in}$ is given by
\ben
D(E)=\frac{L^2(E,\tau_{in})}{2\tau_{in}}
\label{DLTAU2}
\een
with the mean free path $L^2(E,\tau_{in})$ given by ($\ref{L(X)p}$) which is equivalent to 
\ben
L^2(E,\tau_{in}) = \sum_{E'\neq E} \frac{(E-E')^2}
{(E-E')^2 + \left({\hbar}/{\tau_{in}}\right)^2}
X^2_{E,E'}
\een
with
\ben
X_{E,E'} = \langle E | X | E'\rangle
\een

At large inelastic scattering time $\tau_{in}$ one gets 
$L^2(E) = \lim_{\tau_{in} \rightarrow \infty} L^2(E,\tau_{in})$ 
which  can be written as
\ben
L^2(E) = \langle E | (X-\langle E | X | E\rangle)^2 | E\rangle = \xi_E^2
\een
where $\xi_E$ is the localization length of the state of energy $E$. After (\ref{DLTAU2}) 
the diffusivity is  $D(E)={\xi_E^2}/2\tau_{in}$,  in agreement with the argument of Thouless. 

One thus recovers the typical dependence on the scattering time $\tau$ i.e. $\sigma_{dc}\propto 1/\tau$ that was deduced from the analysis of the response $j(t)$ and the backscattering.

\subsection{Anomalous quantum diffusion and conductivity of periodic systems}

{\it Conductivity within the RTA}

The semi-classical theory of conduction in crystals  
is based on the concept  of a charge carrier wave-packet 
propagating at a velocity  $V = (1/\hbar) \partial E_n(k)/ \partial k$,
where ``$E_n(k)$'' is the dispersion relation for band $n$ and wavevector k. 
The validity of the wave-packet concept requires that  the extension 
$L_{\mathrm{wp}}$ of the wave-packet  of the charge carrier is smaller 
than the distance $V \tau $ of traveling on the  time scale  $\tau $. On the contrary, if   
$V \tau < L_{\mathrm{wp}}$, the semi-classical model breaks down. The  quantum
formalism presented here  allows  to treat on the same footing the  standard
 regime  where the semi-classical approach is valid and the small time regime  $V \tau < L_{\mathrm{wp}}$. As shown in part 2 the spreading of states with energy $E$ in crystals is given by

\ben
\Delta X_{0}^2(E_{\mathrm{F}},t)=V^2t^2 + \Delta X_{\NB}^2(E_{\mathrm{F}},t)
\label{XCP}
\een

The first term in the right hand side of (\ref{XCP}), 
$V^2t^2$, corresponds to the Boltzmann contribution. 
This term dominates at large times and describes the {\it intercell} ballistic propagation of wavepackets on long time scale in crystals. The physical origin of this term is the coupling between successive unit cell that allows the electron to travel in the whole crystal. 
The second term $\Delta X_{\NB}^2(E_{\mathrm{F}},t)$ is the Non-Boltzmann contribution. It describes the {\it intracell} spreading of the electron. Indeed as shown in part 2 this spreading is bounded by a term of the order of the square of the unit cell size. In a standard crystal the Boltzmann term dominates at the  time scale relevant for transport, i.e. the scattering time due to disorder
In approximants of quasicrystals the Non Boltzmann term can dominate.

Let us anticipate on the ab-initio calculations which show  that 
$\Delta X_{\NB}^2(E_{\mathrm{F}},t)$ is nearly constant  
$\Delta X_{\NB}^2(E_{\mathrm{F}},t)\simeq \Delta X_{\NB}^2$ except at  
very small times $t \ll \tau$. Then the equation (\ref{D(X)p}) leads to
\ben
D(E_{\mathrm{F}},\omega) ~ \simeq  ~\frac{V^2 \tau}{1+\omega^2\tau^2} ~+~ \frac{\Delta X_{\NB}^2}{2\tau}
\label{Dcrystal}
\een
Thus the frequency dependent  diffusivity $D(E_{\mathrm{F}},\omega)$ is the sum of a Drude like contribution (first term on the right hand side of 
(\ref{Dcrystal})) and a contribution independent of frequency which increases with disorder. 
As we show from ab-initio calculations (see below) the Drude like contribution can be small in some periodic approximants of quasicrystals. This explains why in these systems the optical conductivity presents no Drude peak and why the dc-conductivity increases with disorder.

We define $\tau^*$ as the time for which the Boltzmann and Non Boltzmann contributions to the dc-diffusivity are equal. Thus:

\ben
\tau^*=\frac{\Delta X_{\NB}}{V\sqrt{2}}
\een

\begin{figure}[]
\begin{center}
\includegraphics[width=7cm]{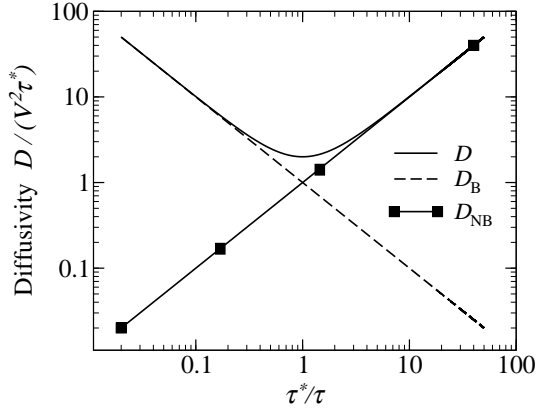}
\caption{Variation of the normalized zero frequency diffusivity $D/V^2\tau^*$ with scattering time for the expression ($\ref{Dcrystal}$). For $\tau=\tau^*$ the Boltzmann and non-Boltzmann contributions are equal.
\label{DCDIFF}}
\end{center}
\end{figure}

\begin{figure}[]
\begin{center}
\includegraphics[width=7cm]{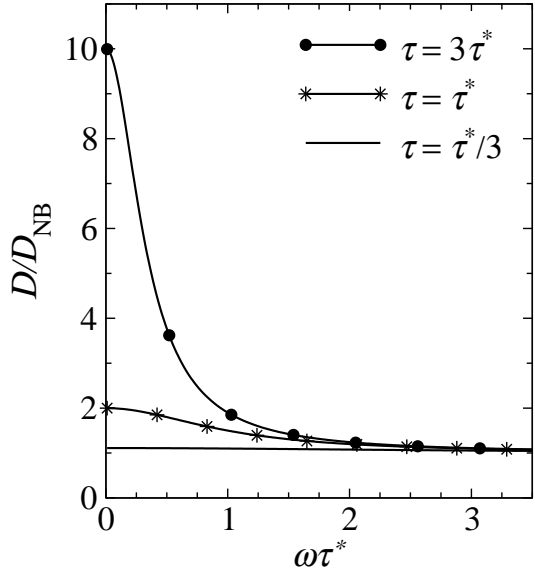}
\caption{Variation of the normalized frequency dependent diffusivity $D/D_{NB}$. For $\tau>\tau^*$ the transport is dominated by the Boltzmann term and the Drude peak is well defined. For  $\tau<\tau^*$ the transport is dominated by the Non-Boltzmann term and the Drude peak is absent.
\label{CondOptDrude}}
\end{center}
\end{figure}

In that case the dc-diffusivity can be represented as in figure \ref{DCDIFF}. The ac-diffusivity is represented  in figure \ref {CondOptDrude}.

{\it Backscattering}

The Non-Boltzmann contribution to the diffusivity 
$D_{\NB}={\Delta X_{\NB}^2}/({2\tau})$ 
is formally similar to the Thouless expression for localized states 
$D_{Thouless}={\xi_E^2}/({2\tau})$. 
Indeed in the Thouless regime the quantum diffusion between two inelastic scattering events is limited by the localization length $\xi_E$. For crystals it is the Non-Boltzmann contribution to quantum diffusion which tends to $\Delta X_{\NB}$. Let us recall that $\Delta X_{\NB}$ is itself limited by a term of the order of the unit cell size. 

As discussed previously (see section 3.2)  the increase of conductivity with disorder is a direct consequence of the backscattering.  
Indeed we shall find in part 4 that {ab-initio}  calculations prove the existence of backscattering. 

Finally we emphasize an important  difference with the Thouless regime. In the Thouless regime it is the inelastic scattering that destroys the localization produced by the elastic scattering. Here, provided that the RTA is valid,  it is either the elastic or the inelastic  scattering that destroys the localization induced by the periodic potential. 

{\it Metal-Insulator transition}

Let us discuss now the role of quantum interferences according to the scaling theory of localization. 
As explained in part 3.1, the central quantity is the conductance of a cube with a 
size equal to the elastic mean free path $L(E_{\mathrm{F}},\tau)$. 

\ben
g \simeq e^2 n(E_{\mathrm{F}}) D(E_{\mathrm{F}},\tau) L(E_{\mathrm{F}},\tau) 
\een

Since the quantum interferences effect have not the possibility to operate at smaller length scale than $L(E_{\mathrm{F}},\tau)$ then this quantity can be computed with the RTA according to ($\ref{L(X)p}$).

We still assume that $\Delta X_{\NB}(E_{\mathrm{F}},t)$ is nearly constant equal to $\Delta X_{\NB}$ except at the smallest time (see below the ab-inito results on an approximant of $\alpha$-AlMnSi). 

Thus the typical propagation length $L(E_{\mathrm{F}},\tau) $ on a time scale $\tau$, i.e. the mean-free path,  is
\ben 
L^2(E_{\mathrm{F}},\tau)= \Delta X_{\NB}^2 + 2 V^2 \tau^2
\een
Let us introduce $g_0$, which is characteristic of the {\it perfect crystal} and is defined by
\ben 
g_0 = e^2 n(E_{\mathrm{F}})  \Delta X_{\NB}^2 V
\label{g0}
\een

Let us introduce an adimensional  value $\tilde{\tau}$ of the scattering time $\tau$ defined by
\ben 
\tilde{\tau} = \frac{V \tau}{\Delta X_{\NB} } = \frac{\tau}{\sqrt{2} \tau^*}
\label{tildetau}
\een 

Let us define also  the function $f(x)$:
\ben
f(x)= \left(\frac{1}{2 x} + x \right) \sqrt{(1 +2 x^2)} 
\label{f(x)}
\een

Then one has
\ben
g= g_0 f(\tilde{\tau})
\een

\begin{figure}[]
\begin{center}
\includegraphics[width=10cm]{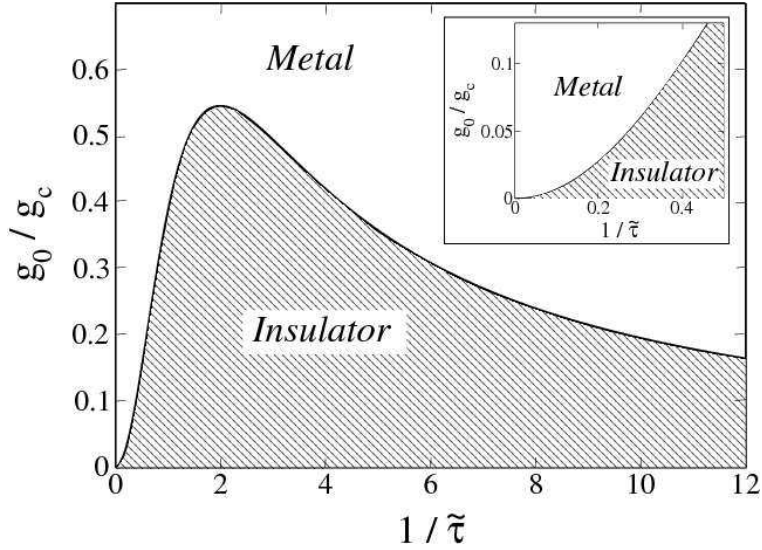}
\caption{Metal-Insulator phase diagram as a function of the two parameters 
$g_0/g_c$ and $1/\tilde{\tau}=\frac{\sqrt{2}\tau}{\tau^*}$. 
The insert represents the limit of a normal metal i.e. for  fixed 
$\tau$ and $V$ the limit of a small $\Delta X_{\NB}$. 
After ($\ref{g0}$) and ($\ref{tildetau}$) this limit is 
in the region of the phase diagram at small $g_0/g_c$ and small $1/\tilde{\tau}$. 
\label{g0_sur_gc_invX_b3}}
\end{center}
\end{figure}

After the scaling theory a three dimensional  system is insulating 
(metallic, respectively) 
if $g<g_c$ (resp. $g>g_c$) where $g_c$ is the value of the universal critical 
conductance in the scaling theory. Using $g= g_0 f(\tilde{\tau})$ it is 
equivalent to say that the system is insulator if 
$g_0/g_c<1/f(\tilde{\tau})$ and metallic if $g_0/g_c>1/f(\tilde{\tau})$. 
This is illustrated in figure  \ref{g0_sur_gc_invX_b3}.  
Note that $g_0/g_c$  is characteristic of the perfect crystal 
whereas $1/\tilde{\tau}$ measures the scattering 
rate $1/\tau$ in units of  $V/\Delta X_{\NB}$.

\begin{figure}[]
\begin{center}
\includegraphics[width=9cm]{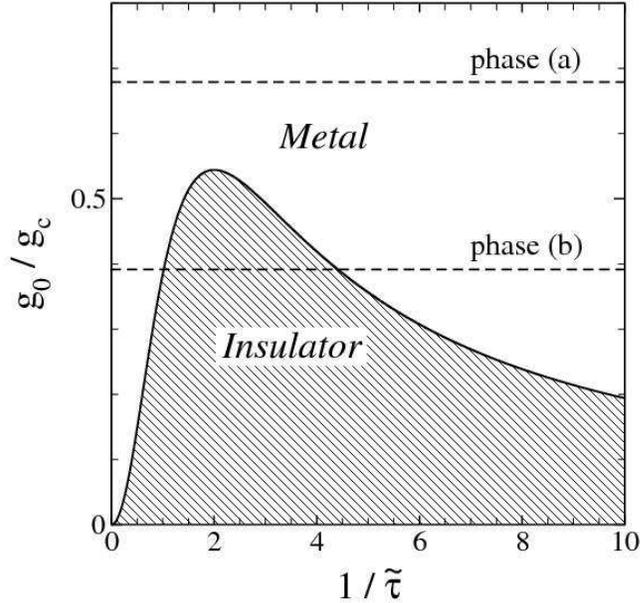}
\caption{Two types of systems can exist. Phases of type (a)  are always metallic whatever the value of the scattering time $\tau$. Phases of type (b) can be either metallic or insulating depending on the value of the scattering time. For phases of type (b) the Metal-Insulator transition that occurs at the highest value of $1/\tilde{\tau}$ is unconventional since the metallic state is obtained by increasing  $1/\tilde{\tau}$ i.e. by increasing disorder.
\label{g0_sur_gc_invX_alpha}}
\end{center}
\end{figure}

A first remarkable property of this phase diagram is that if $g_0>Rg_c$ 
with $R=(2/3)^{3/2}\simeq 0.54$ then the system is always metallic whatever 
the value of the scattering rate 
(phase (a) in figure \ref{g0_sur_gc_invX_alpha}). 
This is not the case for normal 
metals that always become insulating at sufficiently small scattering time 
$\tau$ (i.e. at  sufficiently large disorder). 

If  $g_0<Rg_c$ the system is metallic at large and small scattering rates 
and insulator in an intermediate zone
(phase (b) in figure \ref{g0_sur_gc_invX_alpha}).
This means that if the system is in the large $1/\tilde{\tau}$ metallic 
region it will become insulating by {\it decreasing} $1/\tilde{\tau}$ that is by 
{\it decreasing} disorder! This is just the opposite of the standard conditions 
for the occurrence of the Anderson localization transition. 
The other insulator-metal transition is normal in the sense that the metallic state is obtained by decreasing disorder. 

Note that the case of a normal metal corresponds to 
the limit $\Delta X_{\NB} \to 0$. 
In that case one uses the asymptotic form of the function $f(\tilde{\tau})$ 
for large $\tilde{\tau}$  namely $f(\tilde{\tau})\simeq \sqrt{2}\tilde{\tau}^{2}$.
One then recovers the standard criterion for free-like electrons.

\subsection{Anomalous quantum diffusion and conductivity of quasiperiodic systems}

{\it Conductivity within the RTA}

\begin{figure}[]
\begin{center}
\includegraphics[width=12cm]{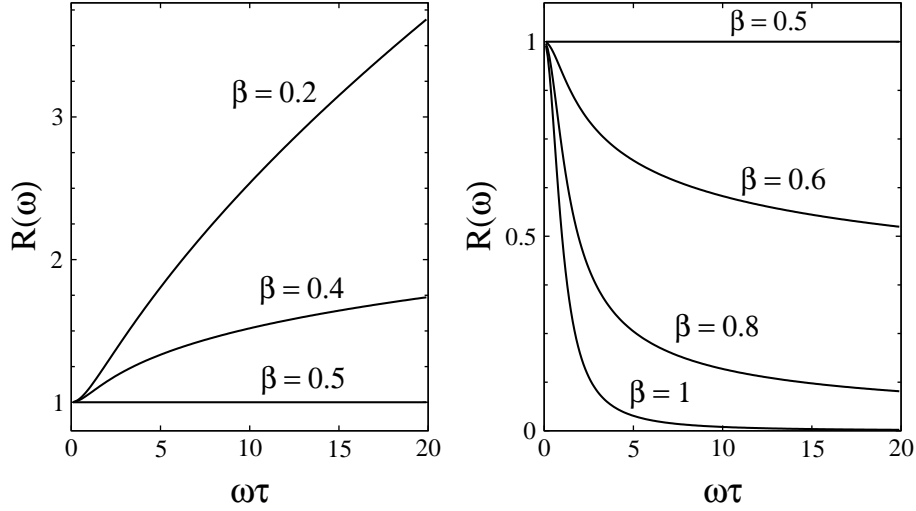}
\caption{Dissipative part of the conductivity $R(\omega)$ normalized by 
the zero frequency value ($R(\omega)=\RRe\,\sigma(\omega)/\sigma(\omega=0)$), 
as a function of $\omega\tau$ (frequency normalized by $1/\tau$). 
The left panel  shows that the conductivity  increases with frequency 
when $\beta <0.5$. The right panel shows that there is a low frequency 
peak if $\beta>0.5$. The case $\beta$ $=0.5$ is an intermediate case for 
which the conductivity is independent of frequency.
\label{CondDrudeG}}
\end{center}
\end{figure}

A generalized Drude formula ($\ref{GD1}$,$\ref{GD2}$) for the low frequency conductivity is derived in part 2.5. Yet it is interesting to derive it from simple physical arguments. One notes first that at zero frequency the dependence  on the scattering time is easy to establish. Indeed the diffusivity is

\ben
D(E_{\mathrm{F}},\tau)=\frac{L^2(E_{\mathrm{F}},\tau)}{2\tau}
\een

where $L^2(E_{\mathrm{F}},\tau)$ is the mean free path 
i.e. a typical distance of propagation in the perfect structure  
during the scattering time $\tau$. 
Assuming that in the perfect quasiperiodic structure the spreading 
of a wavepacket is $\Delta X^2(t)\simeq A t^{2\beta}$ one gets
\ben
L^2(E_{\mathrm{F}},\tau)\simeq A\tau^{2\beta}
\een
and 
\ben
D(\tau)\simeq \frac{L^2(E_{\mathrm{F}},\tau)}{2\tau}\simeq \frac{A}{2} \tau^{2\beta-1}
\een

One notes also that the conductivity  depends on scattering time and frequency 
only through the combination ${\tau}/({1-i\omega\tau})$. This stems directly from the RTA formula 
(\ref{D(X)p})  which is expressed as a Fourier-Laplace integral with  
$(\frac{1}{\tau} - i\omega)=({1-i\omega\tau})/{\tau}$.  Thus the frequency dependent diffusivity is
\ben
D(\omega,\tau) \simeq \frac{A}{2}~ \RRe\, \left( \frac{\tau}{1-i\omega\tau}\right) ^{2\beta-1}
\een
Except for a numerical factor $\Gamma (2\beta+1)$ this formula is equivalent 
to the Generalized Drude formula (\ref{GD1}), (\ref{GD2})  which we recall here:

\ben
\RRe\, \sigma (E_{\mathrm{F}},\omega) = \RRe\, \tilde{\sigma} (E_{\mathrm{F}},\omega)
\label{GD1p} 
\een
with
\ben
\tilde{\sigma} (E_{\mathrm{F}},\omega) = \frac{e^2n(E_{\mathrm{F}})}{2} A \Gamma (2\beta+1) \left( \frac{\tau}{1-i\omega\tau}\right) ^{2\beta-1}
\label{GD2p}
\een

The behavior of the conductivity depends on the value of $\beta$ compared to $0.5$. The frequency dependence is represented in figure \ref{CondDrudeG}. 
If $\beta >0.5$ the behavior is similar to that of a normal metal. 
The dc-conductivity decreases when disorder increases and the low frequency conductivity presents a peak at low frequency, somewhat similar to the Drude peak. 
If $\beta <0.5$ the behavior is not that of a metal. 
In the absence of disorder the system is insulating, and the dc-conductivity increases 
when disorder increases \cite{Roche97}. 
The real part of the conductivity increases when frequency increases. 
Instead of a Drude peak there is a dip.

One notes also that even in the absence of scattering i.e. for $\tau\to\infty$ the real part of the conductivity is non zero in the limit of small frequency. This means that there is absorption of electromagnetic energy by the system. This is not the case in a normal metallic crystal. Here the absorption of energy can be understood by considering approximants with large unit cell. In perfect approximants the absorption of energy is made through {\it interband transitions}. For a given frequency the absorption of energy by interband transition can occur with sufficiently large unit cell because the bands become very narrow and very close in energy allowing for interband transitions.

{\it Backscattering}

\begin{figure}[]
\begin{center}
\includegraphics[width=9cm]{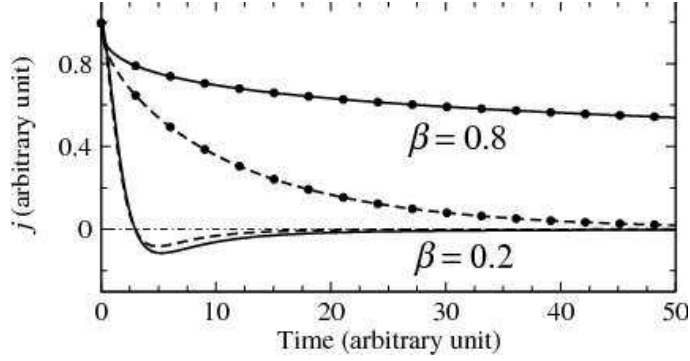}
\caption{Typical behaviour of the response $j(t)$ for a quasiperiodic system.  Without disorder (solid and solid dotted lines) and with disorder (dashed and dashed dotted  lines). For $\beta >0.5$ the response $j(t)$ is positive but for $\beta < 0.5$ the response is positive at small times and negative at long times (backscattering).
\label{BackscatteringQC}}
\end{center}
\end{figure}

For $\beta <0.5$ the behavior of conductivity with frequency and disorder is not that of a metal. This can be attributed to the phenomenon of backscattering. Indeed after the relation (\ref{RelXCp}) between quantum diffusion and velocity correlation one gets
\ben
C_{0}(E,t)=\frac{\d^2\Delta X_{0}^2(E,t)}{\d t^2} \simeq A2\beta (2\beta -1) t^{2\beta-2}
\label{CBackscattering}
\een

if the quantum diffusion law is $\Delta X_{0}^2(E,t)\simeq A t^{2\beta}$ in the perfect quasiperiodic system. From (\ref{CBackscattering}) it appears that the velocity correlation function is negative at large time if $\beta<0.5$. The response $j(t)$ is then also negative. The phenomenon of backscattering then implies that the conductivity increases with disorder or with frequency, as discussed for weak-localization regime or for the Thouless regime 
(see figure \ref{BackscatteringQC}).

{\it Metal-Insulator transition}

Let us discuss now the role of quantum interferences according to the scaling theory of localization. As explained in part 3.1  the central quantity is the conductance of a cube with a size equal to the elastic mean free path $L(E_{\mathrm{F}},\tau)$. 

\ben
g \simeq e^2 n(E_{\mathrm{F}}) D(E_{\mathrm{F}},\tau) L(E_{\mathrm{F}},\tau) 
\een

Since the quantum interferences effect have not the possibility to operate at smaller length scale than $L(E_{\mathrm{F}},\tau)$ then this quantity can be computed with the RTA. 

We assume that $\Delta X_{0}^2(E_{\mathrm{F}},t)$ is nearly equal to the asymptotic form $A t^{2\beta}$ then the typical propagation length $L(E_{\mathrm{F}},\tau)$ on a time scale $\tau$ is of the order of:
\ben 
L^2(E_{\mathrm{F}},\tau) \simeq A\tau^{2\beta}
\een
and the diffusivity is after (\ref{D(X)p}) of the order of:
\ben 
D(E_{\mathrm{F}}, \tau) \simeq  \frac{A}{2} \tau^{2\beta-1}
\een
From this one deduces that the conductance of a cube of size $L(E_{\mathrm{F}},\tau) $ is given approximately  by
\ben
g \simeq e^2 \frac{n(E_{\mathrm{F}})}{2} A^{{3}/{2}} \tau^{3\beta-1}
\een

From this expression one concludes that if $\beta<1/3$  the conductance tends to zero at large $\tau$. This means that the system becomes insulating when the disorder {\it decreases}. This is what happens in the case of crystals if the Non Boltzmann contribution to transport dominates (see part 3.3). 

Yet one must note that due to the Guarneri inequality \cite{MayouPRL00}  the spectrum becomes singular continuous for a three dimensional system with $\beta <1/3$. In that case the density of states cannot be considered as a constant in the perfect system. Thus we have to assume that the density of states is sufficiently smooth when averaged on the energy scale given by the inverse scattering time $1/\tau$.

\section{Evidence of anomalous quantum diffusion in quasicrystals and approximants }
\label{SecCondQC_Approx}
In this part we present briefly  the experimental transport properties of phases such as AlMnSi, AlPdMn and AlCuFe or AlPdRe. These experimental transport properties indicate a conduction mode which is neither metallic nor semiconducting. For the $\alpha$-AlMnSi phase, recent ab-initio computations  are presented, which confirm the existence of an anomalous diffusion and allow for a semi-quantitative ab-initio computation of conductivity. Concerning AlCuFe and related quasiperiodic phases, which cannot be addressed by band structure calculations, we present a phenomenological model. This model  based on  anomalous quantum diffusion provides a coherent interpretation of the strange electronic transport of these systems.  
\subsection{Experimental transport properties of icosahedral and related approximant phases}
\label{ExperienceElectTranspPropert}

\begin{figure}[]
\begin{center}
\includegraphics[width=12cm]{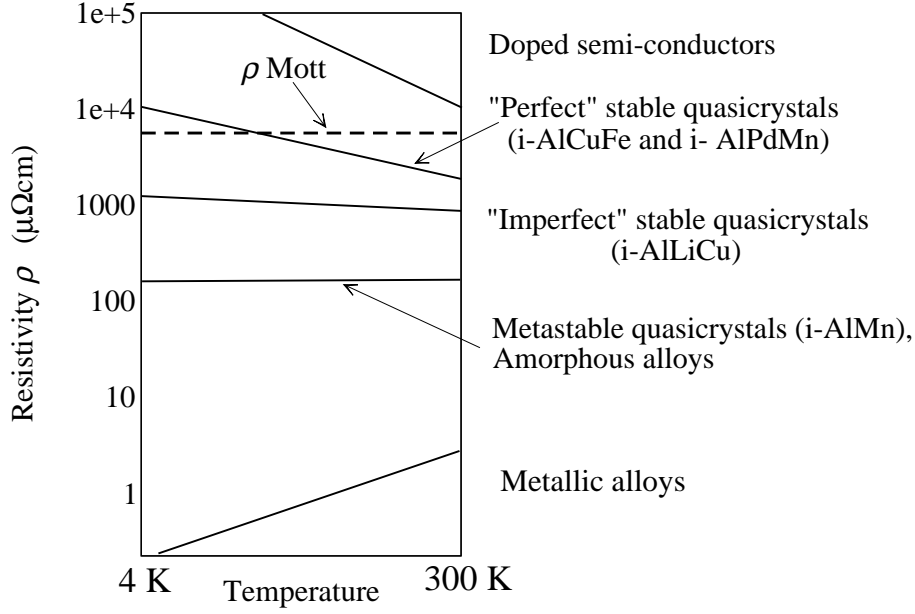}
\caption{Order of magnitude and schematic temperature dependencies of 
the resistivity of icosahedral quasicrystals compared to amorphous 
and metallic crystals. From \cite{Grenet00_Aussois}.
\label{Fig_Resistivite_schema}}
\end{center}
\end{figure}

Quasicrystals of high structural quality reveal unusual transport
properties \cite{Klein91,Poon92,Berger94,Grenet00_Aussois}
(figure \ref{Fig_Resistivite_schema}).
For instance, one of the main features is the low conductivity
$\sigma_{4K}=100-200$~$\rm \Omega cm^{-1}$
for icosahedral AlPdMn and AlCuFe and
$\sigma_{4K}<1$~$\rm \Omega cm^{-1}$ for
AlPdRe
\cite{Pierce93_science,Berger93,Akiyama93,Delahaye99,Delahaye01,Delahaye03_JPCM,Rosenbaum04},
although the DOS still has a metallic character. This
means that the high resistivity is due mainly to a small
diffusivity of electrons.

Experimental measurements show that {\it approximants phases}
like $\alpha$-AlMnSi~\cite{Berger94},
1/1 AlCuFeSi~\cite{Quivy96},
R-AlCuFe~\cite{Berger94},
1/1 AlReSi~\cite{Tamura01,Takeuchi04}
etc.,
have transport properties similar to
those of quasicrystals AlPdMn and AlCuFe.
This suggests that the local atomic order
on the length scale of the unit cell,
{\it i.e.} $\rm 10-30~\AA$, determines their transport properties.
As atomic medium-range order of quasicrystals and
approximants are similar, it should also be important
in the understanding of transport properties of
quasicrystals.
This remark is confirmed by the fact that AlTM crystals
with a small unit cell
(typically less than 50 atoms in a unit cell)
does not exhibit such particular transport properties.

The resistivity, $\rho=1/\sigma$, of crystals with
a small unit cell increases as temperature $T$ increases 
and generally follows  the Mathiessen rule:
\begin{equation}
\rho(T)=\rho_0 + \Delta{\rho}(T),
\end{equation}

By contrast, the resistivity of some quasicrystals and approximants
(AlPdMn, AlCuFe)
decreases when temperature increases , 
and their conductivity follows approximatively 
an \textit{``inverse Mathiessen rule''} \cite{Mayou93,Berger94}:
\begin{equation}
\sigma(T)=\sigma_0 + \Delta{\sigma}(T).
\end{equation} 

Besides, after annealing sample, with a consequent reduction of
the structural defects, the resistivity of quasicrystals and
approximants increases.
The relation between the particular transport 
properties of these phases
and their structure is still debated.
For AlPdMn quasicrystals,
J.J. Pr\'ejean {\it et al.}~\cite{Prejean02}
found that local defects might be related
with the occurrence of Mn atoms with localized magnetic moment.
Thus, magnetic properties, transport properties and structural quality
are intimately linked for those complex phases.

Another remarkable experimental result is 
the linear energy dependence of the optical conductivity 
of AlCuFe and the absence of  Drude peak \cite{Homes91,Burkov94_JPCM}.

The icosahedral AlPdRe is the most resistive known quasicrystalline material  
\cite{Delahaye01,Delahaye03_JPCM}.
This material displays a strong decrease of the conductivity when the temperature is reduced 
and the conductivity value can fall below 1\,$\rm (\Omega cm)^{-1}$ at 4\,K.
Although the behavior depends strongly on the composition and the preparation of the sample,
many authors \cite{Pierce93_science,Berger93,Akiyama93,Delahaye99,Delahaye01,Delahaye03_JPCM,Rosenbaum04} 
have reported that AlPdRe quasicrystal are very close 
to the metal-insulator transition.
Three successive regimes are revealed \cite{Delahaye03_JPCM} as the temperature is increased 
to room temperature: a low temperature variable range hopping-like behavior, 
followed by a
Thouless regime and a high temperature critical regime.

Experimentally a low density of states (DOS) at the Fermi
energy $E_{\rm F}$ is usually measured in
quasicrystals and their crystalline approximants.
For instance,
a density of states at
$E_{\rm F}$ reduced by $\sim 1/3$ from
its free electrons value is measured in
i-AlCuLi and R-AlLiCu approximant
\cite{Poon92,Wagner88,Wang88,Kimura89}.
The presence of the pseudogap in these phases
is confirmed by
photo-emission measurements~\cite{Belin93}
and NMR experiments~\cite{Hippert92}.

For icosahedral phases containing transition metal (TM) elements,
specific heat measurement indicate a DOS at
$E_{\rm F}$
of $\sim 1/3$ of the free electron value for
i-AlCuFe
and $\sim 1/10$ for i-AlCuRu~\cite{Biggs90}
and  i-AlPdRe~\cite{Pierce94}.
From photo-emission spectroscopy the pseudogap
in the DOS is confirmed for many icosahedral quasicrystals in
the systems:
AlMn (metastable)~\cite{Belin92a},
AlMnSi~\cite{Belin92a,Belin94_MSEA},
AlCuFe~\cite{Mori91,Matsubara91,Mori92,Belin92b,Belin94EuroPhys,Belin00},
AlCuFeCr~\cite{Belin92b},
AlPdMn~\cite{Zhang94,Belin94_JPCM,Stadnik97,Fournee02},
AlCuRu~\cite{Stadnik94},
AlPdRe~\cite{Stadnik97}.
The pseudogap has been also measured in many approximants of
quasicrystals. For instance
R-AlCuFe~\cite{Hippert92,Berger94},
1/1 AlCuFeSi~\cite{Quivy96}
$\alpha$-AlMnSi~\cite{Berger94,Belin94_MSEA},
1/1 AlCuRuSi~\cite{Mizutani01,Mizutani04},
1/1 AlReSi~\cite{Takeuchi04}
have a DOS at $E_{\rm F}$ reduced by a similar factor
as in i-AlCuTM and i-AlPdMn.
A pseudogap near $E_{\rm F}$ has been also confirmed by
ab-initio calculations of the electronic
structure in many icosahedral approximants
(figure \ref{Fig_DOS_Al6Mn_sugi}, 
see below and see also the chapter by Ishii and  Fujiwara in this book).

It has also been shown experimentally that transition metal elements have a 
important role on the unusual transport properties of quasicrystals and 
related phases 
\cite{BelinMayou93,Berger93b,Berger93c,Mayou93b,Dankhazi93,GuyPRB95_2}.

\subsection{Ab-initio electronic structure and quantum diffusion 
in perfect approximants}

{\it Density of states}

Electronic structure determinations have been
performed in the frame-work of 
density functional formalism
in the local density approximation (LDA)  by using the
self-consistent Tight-Binding (TB)
Linear Muffin Tin Orbital (LMTO) method
in the Atomic Sphere Approximation (ASA) \cite{Andersen75}.

\begin{figure}[]
\begin{center}
\includegraphics[width=8cm]{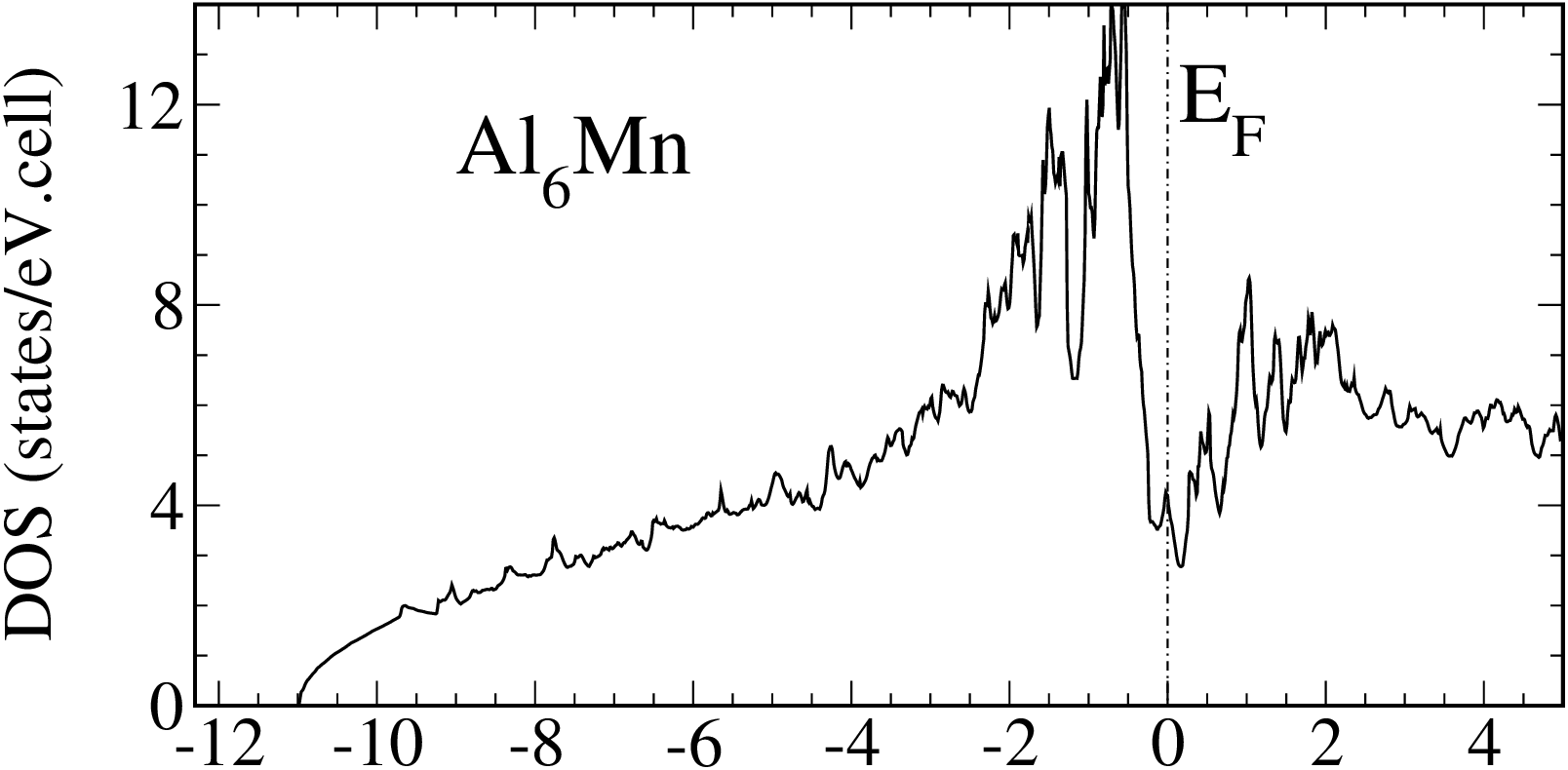}

\vskip 0.2cm
\includegraphics[width=8cm]{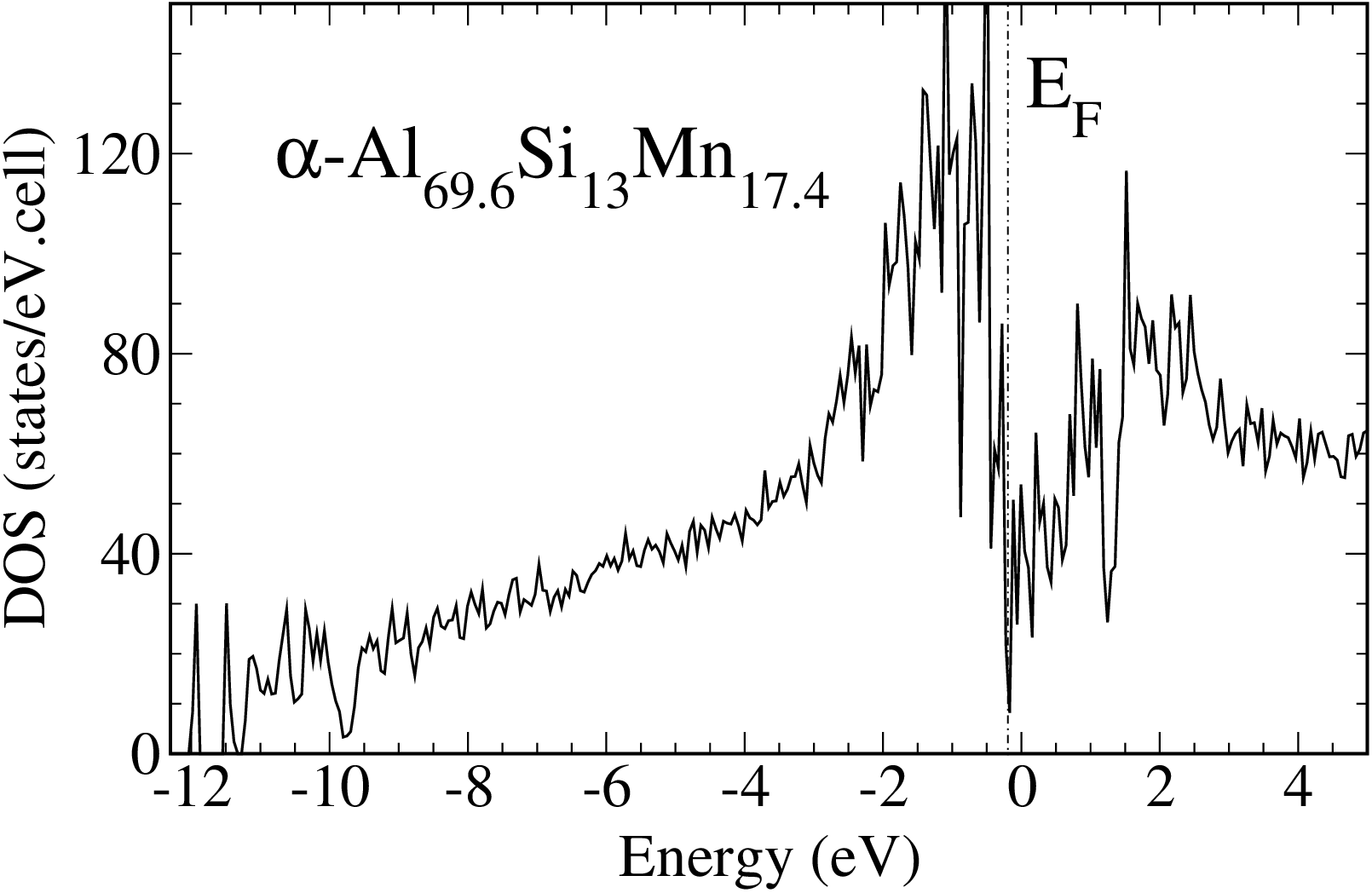}
\caption{LMTO total DOS of $\rm Al_6Mn$ \cite{Dankhazi93}
and $\alpha$-Al$_{69.6}$Si$_{13.0}$Mn$_{17.4}$.
\label{Fig_DOS_Al6Mn_sugi}}
\end{center}
\end{figure}

The LMTO DOS of an $\alpha$-AlMn idealized approximant
(Elser-Henley model \cite{ElserH85,Guyot85})
has been first calculated by T. Fujiwara \cite{Fujiwara89,Fujiwara93}.
This original work shows the presence of a Hume-Rothery pseudo-gap
near the Fermi energy, $\ef$, in agreement with experimental results
\cite{Poon92,Berger94} (see also figure \ref{Fig_DOS_Al6Mn_sugi}).
The role of the transition metal (TM) element in the pseudo-gap formation
has  been shown from  ab-initio calculations
\cite{PMS05} and experiments \cite{Dankhazi93}.
Indeed the formation of the pseudo-gap  results from
a strong sp--d coupling associated to an ordered sub-lattice
of TM atoms.
Just as for Hume-Rothery phases a description of the band energy
can be made in terms of pair interactions.
It has been shown that
a medium-range Mn--Mn interaction mediated by the
sp(Al)--d(Mn) hybridization plays a determinant role
in the occurrence of the pseudo-gap 
\cite{Friedel87,Fujiwara89,Fujiwara93,ZouPRL93,GuyEuro93,GuyPRB95,PMS05,Guy03,GuyPRL00,Guy04_ICQ8}.
It is thus essential to take into account the chemical nature of
the elements to analyze the electronic properties of approximants. The electronic structures of simpler crystals such as
orthorhombic $\rm Al_6Mn$,
cubic $\rm Al_{12}Mn$, present \cite{PMS05} also a pseudo-gap near $\ef$
which is less pronounced than in complex approximants phases. 
E.S. Zijlstra and S.K. Bose \cite{Zijlstra03}
show that Si atoms are in substitution with some Al atoms in the
$\alpha$-phase.
The main effect of Si is to shift $\ef$ in the
pseudo-gap in agreement with the Hume-Rothery mechanism that
minimizes the band energy.

{\it Role of clusters}

As for the local atomic order, 
one of the characteristics of the quasicrystals and approximants, 
is the occurrence of atomic clusters on a scale of 10--30 $\rm \AA$ \cite{Gratias00}.  
Nevertheless the clusters are not well defined because some of them overlap,  
and in addition there are a lot of glue atoms. 
The role of clusters has been much debated in particular by 
C. Janot \cite{Janot94}
and G. Trambly de Laissardi\`ere \cite{GuyPRB97}. 
C. Janot considers as a reference clusters that are isolated in vacuum but it is more realistic to consider a model of  clusters that are not isolated 
but  are embedded in metallic medium.   
The model \cite{GuyPRB97,GuyICQ6}  
is based on a standard description of intermetallic alloys. 
Considering the cluster embedded in a metallic medium,  the variation $\Delta n(E)$ 
of the DOS due to the cluster is calculated.
For electrons, which have energy in the vicinity of the Fermi level, 
transition atoms (such as Mn and Fe) are strong scatters whereas 
Al atoms are weak scatters. 
Then, following a standard approximation, the potential of Al atoms was neglected in reference 
\cite{GuyPRB97}.

\begin{figure}
\begin{center}
\includegraphics[width=8cm]{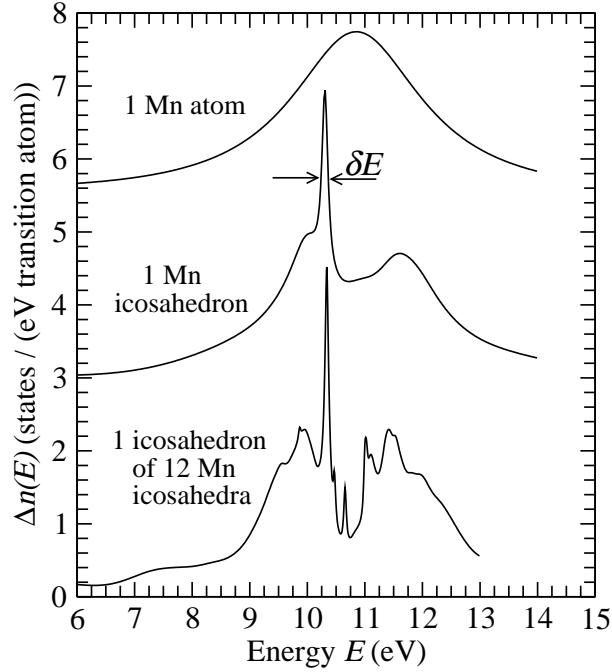}
\end{center}
\caption{\label{DOS_Clusters}Variation $\Delta n(E)$ of the DOS due 
to 1 Mn atom
(Virtual Bound State), 
1 Mn icosahedron, 
and 1 icosahedron of Mn icosahedra obtained after an inflation by a 
factor $\varphi^2$ 
of an initial Mn icosahedron. $\ef \simeq E_d = 10.88$\,eV. 
From \cite{GuyPRB97}.}
\end{figure}

In the figure \ref{DOS_Clusters}, $\Delta n(E)$  due to different clusters
 are shown. The Mn icosahedron is the actual Mn icosahedron  of the $\alpha$-AlMnSi approximant.
As an example of a larger cluster, we consider one icosahedron of Mn icosahedra, 
which appeared in the structural model proposed by C. Janot \cite{Janot94}.

$\Delta n(E)$ of clusters exhibits strong deviations from the 
Virtual Bound States (1 Mn atom)
\cite{Friedel56,Anderson61}.
Indeed several peaks and shoulders appear. The width of the most narrow peaks 
($50 - 100$\,meV)
are comparable to the fine peaks of the calculated DOS in the approximants. 
Each peak indicates a resonance due to the scattering by the cluster. 
These peaks correspond to states ``localized''  
by the icosahedron or the icosahedron of icosahedra. 
They are not eigenstate, they have finite lifetime of the order of $\hbar / \delta E$, 
where $\delta E$ is the width of the peak. 
Therefore, the stronger the effect of the localization by cluster is, the narrower is the peak. 
A large lifetime is the proof of a localization, 
but in the real space these states have a quite large extension on length scale of the cluster 
or the cluster of clusters. 

The physical origin of these states can be understood as follows. 
Let us consider 
incident electrons, with energy $E$ closed to $\ef$, scattered by the cluster. 
In AlMn alloys $\ef \simeq E_d$, where $E_d$ is the energy of the d-orbital. 
In this energy range, the potential of the Mn atom is strong and the Mn atoms 
can roughly be considered as hard spheres with radius of the order of the d-orbital size 
($\sim 0.5$\,$\rm \AA$). By an effect similar to that of a Faraday cage, 
electrons can by confined 
by the cluster provided that their wavelength $\lambda$ satisfies 
$\lambda \gtrsim l$, 
where $l$ is the distance between two hard spheres. 
In the case of $\alpha$-AlMnSi approximant, $\lambda \simeq 3.6$\,$\rm \AA$ 
(if we assume a free electron band and $\ef = 10.33$\, eV) and the distance $l$ is about 
$3.8$\,$\rm \AA$. 
Consequently, we expect to observe such a confinement. 
This effect is a multiple scattering effect, and it is not due to an overlap between d-orbitals 
because Mn atoms are not first neighbor.
We have also shown that these resonances are very sensitive to the geometry of the cluster 
\cite{GuyICQ6}. 
For instance, they disappear quickly when the radius of the Mn icosahedron increases. 

{\it Quantum diffusion in perfect crystals}

\begin{figure}[]
\begin{flushright}

\includegraphics[width=9cm]{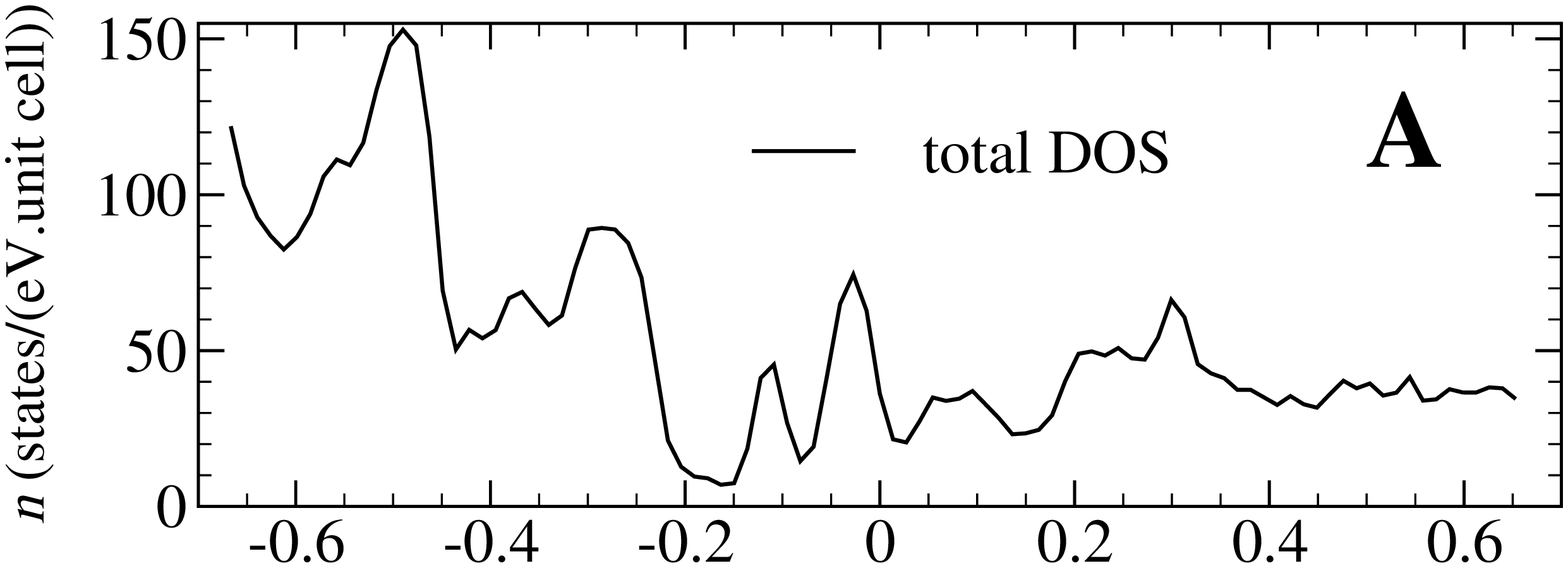}\hspace{2cm}

\vspace{.2cm}
\includegraphics[width=9.17cm]{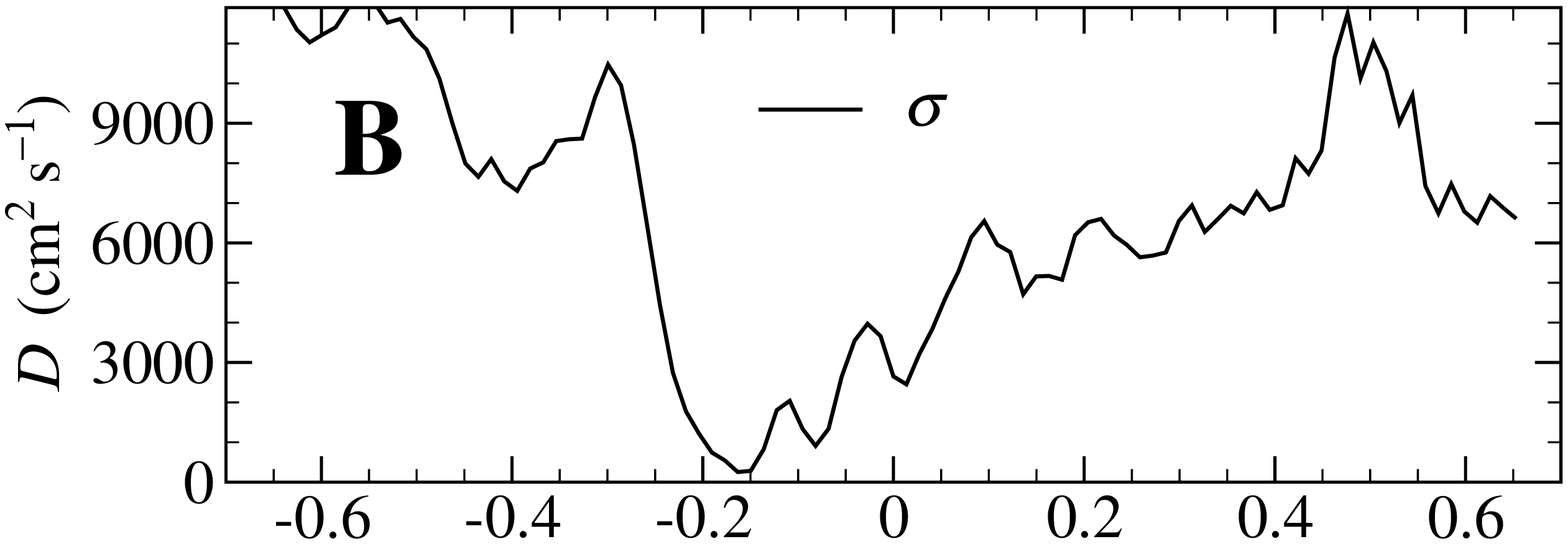}\hspace{2cm}

\vspace{.2cm}
\includegraphics[width=9.09cm]{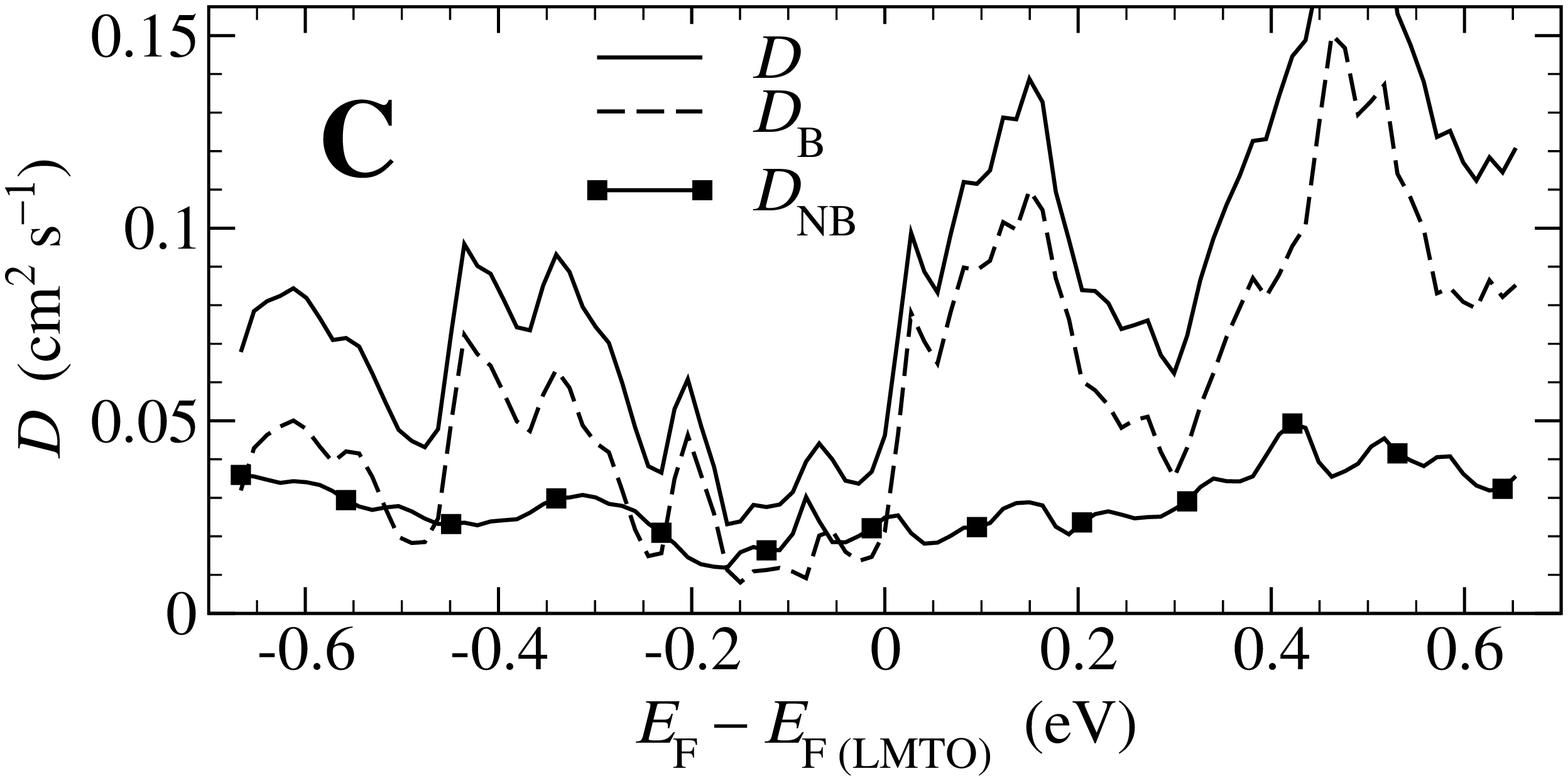}\hspace{2cm}
 
\end{flushright}

\caption{ \label{Fig_DOS_sig_dif}
({\bf A}) LMTO total DOS $n$, ({\bf B}) conductivity $\sigma$, 
and ({\bf C}) diffusivity $D$, in the
cubic approximant $\alpha$-Al$_{69.6}$Si$_{13.0}$Mn$_{17.4}$.
From \cite{PRL06}.
}
\end{figure}

In the following we present calculation of the quantum diffusion in perfect crystalline systems. 
Some works have already been done from ab-initio calculation and give indication 
of none ballistic diffusion \cite{Fujiwara96,Roche98,Krajci02}.
We consider the $\alpha$-AlMnSi approximant and compare it  
with  simpler crystals orthorhombic $\rm Al_6Mn$,
and cubic $\rm Al_{12}Mn$ \cite{PRL06,ICQ9,Julien06}.
For the  $\alpha$-AlMnSi phase, 
we use the experimental atomic structure
\cite{Sugiyama98}
and the Si positions proposed by  Ref.
\cite{Zijlstra03}
with composition:
$\alpha$-Al$_{69.6}$Si$_{13.0}$Mn$_{17.4}$.
In figure \ref{Fig_DOS_sig_dif}, the total DOS $n$ of the 
$\alpha$-AlMnSi phase
is presented versus the energy. The total density of
states is characterized \cite{Fujiwara89}
by a pseudogap near the  Fermi energy
$E_{\mathrm{F}}$.
Following the Hume-Rothery condition, 
it is expected that the most realistic value of
$E_{\mathrm{F}}$ in the actual $\alpha$-phase corresponds to
the minimum of the pseudo-gap, i.e.
$E_{\mathrm{F}}-E_{\mathrm{F(LMTO)}} = -0.163$\,eV for
our calculation.

\begin{figure}[]
\begin{center}
\includegraphics[width=9cm]{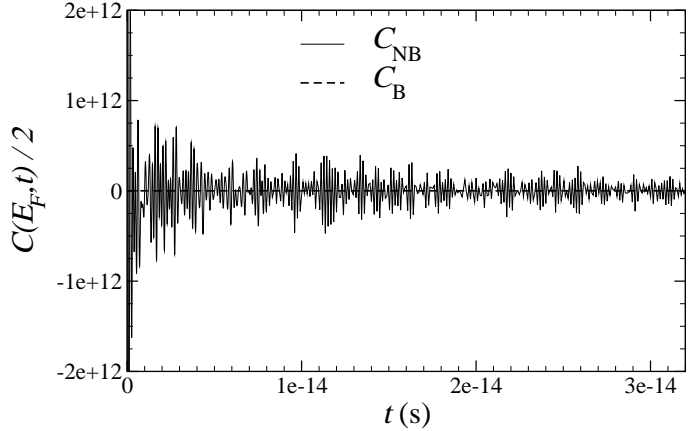}
\caption{Velocity correlation function $C(\ef,t)$ ($\rm
m^2\,s^{-2}$) in  $\alpha$-Al$_{69.6}$Si$_{13.0}$Mn$_{17.4}$
versus large time $t$
The dashed lines are the corresponding Boltzmann velocity
correlation function $C_{\rm B}(\ef,t)=2v_{\rm F}^2$.
From \cite{ICQ9}.
\label{Fig_vx2}}
\end{center}
\end{figure}

We compute the velocity correlation function  
$C(E,t)$ for crystals (complex approximants and simple
crystals).
In equations ($\ref{EqAutocorVit}$), ($\ref{eq_Cnk_Vx2}$), the average
is obtained by taking the eigenstates for each
$\vec{k}$ vector with and energy $E_n(\vec{k})$ such as
\begin{eqnarray}
E-\frac{1}{2}\Delta E<E_n(\vec{k})<E+\frac{1}{2}\Delta E.
\end{eqnarray}
$\Delta E$ is the energy resolution of the calculation. The calculated
$C(E,t)$ is sensitive to the  number $N_k$ of $\vec{k}$
vectors in the first Brillouin zone
when $N_k$ is too small. 
Therefore $N_k$ is increased
until $C(E,t)$ does not depend significantly on $N_k$.
For Al, $\rm
Al_{12}Mn$ and  $\alpha$-Al$_{69.6}$Si$_{13.0}$Mn$_{17.4}$,
$\Delta E$ is equal to $0.272$, $0.272$, and $0.0272$\,eV, respectively,
and  $N_k$ is equal to $120^3$, $40^3$ and $32^3$, respectively.
$C(\ef,t)$ for  the cubic approximant $\alpha$-$\rm AlMnSi$ is shown in
figure \ref{Fig_vx2}. 

After ($\ref{EqAutocorVit}$) and ($\ref{eq_Cnk_Vx2}$), 
$C(\ef,t)$ is the sum of a constant Boltzmann term $C_{\B}(E,t)$  
and  a non Boltzmann term containing  oscillating terms that average to zero 
on long time scale:
\ben
C(E,t)=C_{\B}(E,t)+C_{\NB}(E,t)
\een
\ben
C_{\B}(E,t)=2~ \Big\langle V_{x}^2 \Big\rangle_E
\een
\ben
\lim_{\tau \to \infty}\int_{0}^{\infty}C_{\NB}(E,t) \e^{-t/\tau}{\rm d}t = 0
\een
where $V_{x}^2$ is the square of Boltzmann velocity (intra-band velocity) 
along the $X$ direction at the
Fermi energy: $v_{\rm F}= 9.4\,10^{7}$, $3.6\,10^{7}$, and
$2.7\,10^6$\,cm.s$^{-1}$, for Al,  $\rm Al_{12}Mn$ and
$\alpha$-Al$_{69.6}$Si$_{13.0}$Mn$_{17.4}$,
respectively. This last result is
very similar to the original work of T. Fujiwara et al.
for the $\alpha$-$\rm Al_{114}Mn_{24}$ (with the
atomic structure model of Elser-Henley) \cite{Fujiwara93},
for a model of icosahedral  approximant AlCuFe \cite{GuyPRB94_AlCuFe}.
The strong reduction of
$v_{\rm F}$ in the approximant phase with respect to simple
crystal phases shows the importance of a quasiperiodic
medium-range order (up to distances equal to 12--20\,$\rm \AA$). This
leads to a very small Boltzmann conductivity for approximants
\cite{Fujiwara93,GuyPRB94_AlCuFe}.
In the case of a decagonal approximant AlCuCo,
a strong anisotropy has been found between $v_{\rm F}$ in the ``pseudo''
quasiperiodic directions and $v_{\rm F}$ in the periodic direction
\cite{GuyPRBAlCuCo}.

\begin{figure}[]
\begin{center}
\includegraphics[width=9cm]{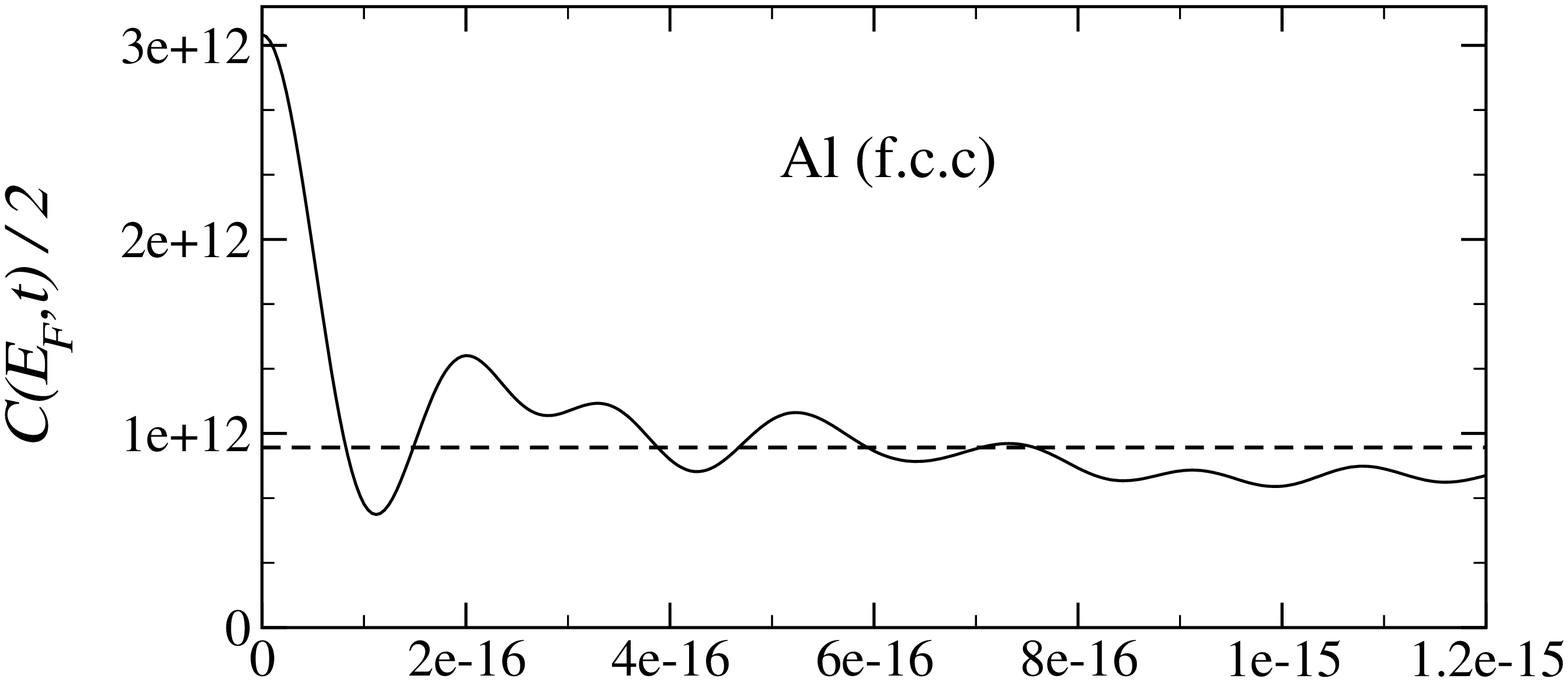}

\vspace{.1cm}
\includegraphics[width=9cm]{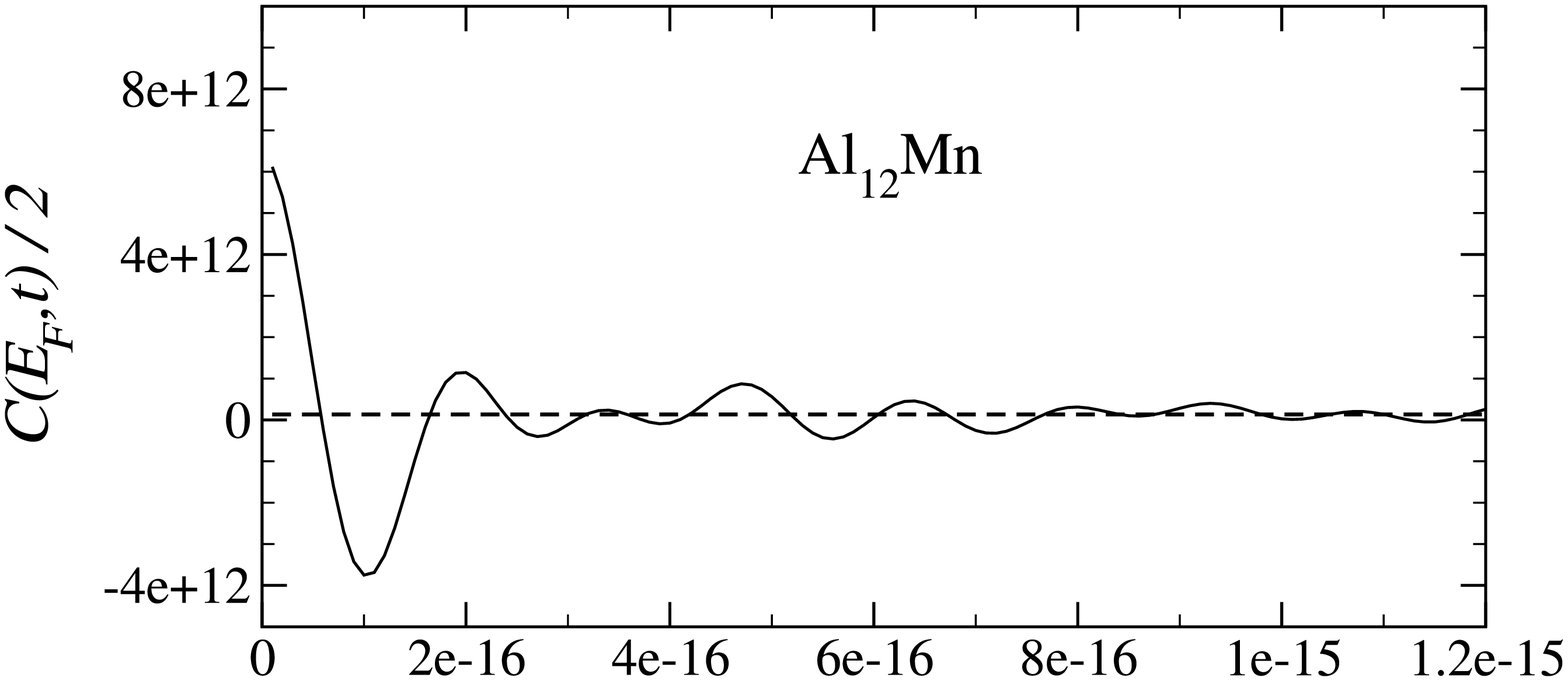}

\vspace{.1cm}
\includegraphics[width=9cm]{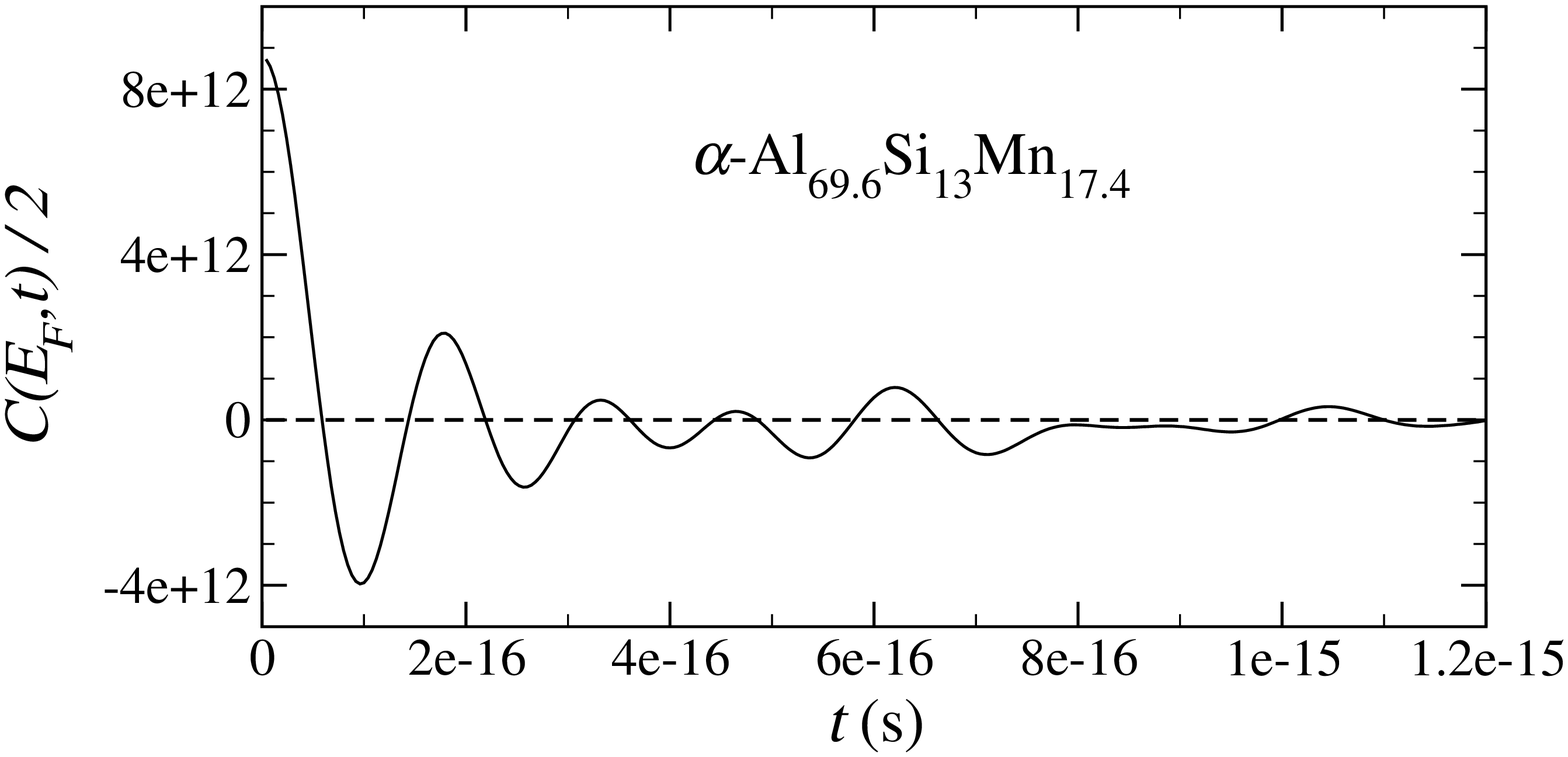}
\caption{Velocity correlation function $C(\ef,t)$ ($\rm
m^2\,s^{-2}$) versus small time $t$
The dashed lines are the corresponding Boltzmann velocity
correlation function $C_{\rm B}(\ef,t)=2v_{\rm F}^2$. From \cite{ICQ9}.
\label{Fig_vx2_small}}
\end{center}
\end{figure}

On small time scale $t$ (figure \ref{Fig_vx2_small}),
$C(\ef,t)$ and $C_{\B}(\ef,t)$ differ,
and there is a new difference
between approximant and simple crystal.
In the case of Al (f.c.c.) phase, $C(\ef,t)$ is always positive,
and the Boltzmann value is reached
rapidly when $t$ increases.
But for some $t$ values the velocity correlation function $C(\ef,t)$ is
negative for $\rm Al_{12}Mn$ and $\alpha$-$\rm Al_{114}Mn_{24}$.
That means that at these times the phenomenon of
{\it backscattering} occurs.

The transport properties
depend on the average value of $C(\ef,t)$ on a time scale equals to the
scattering time $\tau$ \cite{MayouPRL00,Triozon04}
(see for instance equation (\ref{DRTA})).
A realistic value of $\tau$ has been estimated to about
$10^{-14}$\,s \cite{Mayou93}.
For the simple crystals $\rm Al_{12}Mn$, $C(\ef,t)$ is
mainly positive when $t>2\,10^{-15}$\,s.
But for the complex approximant
$\alpha$-$\rm Al_{114}Mn_{24}$, a lot of $t$ values
correspond to $C(\ef,t)<0$,
even when $t$ is close to $\tau$
or larger (figure \ref{Fig_vx2}).
Therefore, in the case of $\rm Al_{12}Mn$, the backscattering
(negative value of $C(\ef,t)$) should have a negligible effect on
the transport properties, whereas this effect must be determinant
for the approximant.

\begin{figure}[]
\begin{center}
\includegraphics[width=9cm]{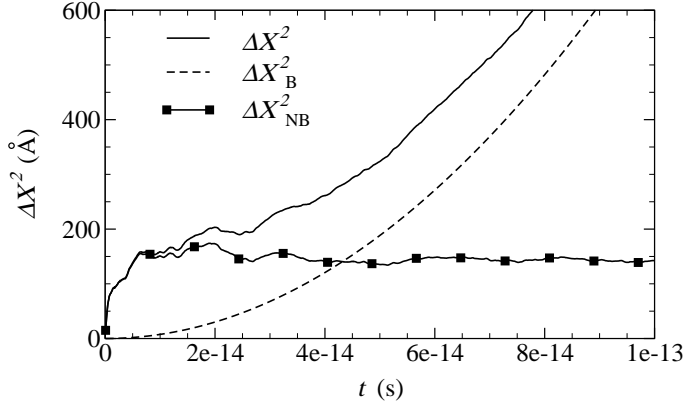}
\end{center}
\caption{ \label{Fig_X2}
Square spreading 
$\Delta X^2$
of electrons states with Fermi energy 
$E_{\mathrm{F}}$ versus time $t$, 
in the cubic approximant $\alpha$-Al$_{69.6}$Si$_{13.0}$Mn$_{17.4}$.
$\Delta X^2 =\Delta X^2_{\mathrm{B}} + \Delta X^2_{\mathrm{NB}} $
(see text). 
From \cite{PRL06}.
}
\end{figure}

As discussed in part 3 the phenomenon of
backscattering is associated to unusual quantum diffusion.
It is
illustrated on the plot of the average spreading of states
$\Delta X^2$ versus time $t$ (figure \ref{Fig_X2}).
After ($\ref{Eq_DeltaX2}$),  $\Delta X^2$ results in two term:
\begin{equation}
\Delta X^2(E,t) = V_{\mathrm{B}}(E)^2 t^2 + \Delta X_{\mathrm{NB}}^2(E,t),
\label{Eq_DeltaX3}
\end{equation}

A Boltzmann term $V_{\B}(E)^2 t^2$  and a non-Boltzmann term.
The non-Boltzmann contribution, $\Delta X^2_{\rm NB}$,
which comes from the non-diagonal matrix element
(\ref{Calcul_DeltaX2}),
increases very rapidly and saturates to a maximum value
of the order of the square size of the unit cell.
In the $\alpha$-approximant, at small time $t$,  
$\Delta X^2_{\rm B}$ is smaller than in Al due to a very small velocity
$V_{\rm F}$ of the electron with energy $\ef$.
The calculated $V_{\rm F}$ is equal to $2.7\,10^7$\,cm.s$^{-1}$, 
which is about 30 times smaller than aluminum values.
Thus $\alpha$-AlMnSi is a non-conventional metal at these
time scale i.e. when the scattering time is $\tau < \tau^*$
where $\tau^* \simeq 3\,10^{-14}\,$s.
In a normal crystal, the $\Delta X^2_{\rm NB}(t)$ 
term is negligible with respect
to the Boltzmann term $\Delta X^2_{\rm B}(t)$.
On the contrary, in the approximant both terms have
the same  magnitude at the  realistic scattering times scale,
typically a fews $10^{-14}$\,s.

\subsection{Ab-initio RTA  model for the  conductivity of approximants}

Within a relaxation time approximation the diffusivity
$D(E,\omega)$ is calculated. At low frequency one gets 
\ben
\RRe\, \sigma(E,\omega)=\e^2n(E)D(E,\omega)
\een
\ben
D(E,\omega)=D_{\B}(E ,\omega)+D_{\NB}(E, \omega)
\label{DDD1}
\een
\ben
D_{\B}(E,\omega) = \frac{V^2 \tau}{1+\omega^2\tau^2} 
\label{DB31}
\een
and 
\ben
D_{\NB}(E,\omega) = \frac{1}{2} \,  {\RRe}\,
\left\{ (\frac{1}{\tau} - i\omega)^2
\int_0^{+\infty} \e^{(i\omega -1/\tau)t} \Delta X_{\NB}^2(E,t) {\d}t
\right\}
\label{DNB21}
\een

The  $D_{\mathrm{B}}$ values for $\alpha$-Al$_{69.6}$Si$_{13.0}$Mn$_{17.4}$
(figure \ref{Fig_DOS_sig_dif})
are similar in magnitude to those
obtained by T. Fujiwara {\it et al.}
\cite{Fujiwara89}
for the idealized approximant $\alpha$-Al$_{114}$Mn$_{24}$ approximant.
$D_{\mathrm{NB}}$ is almost independent
on $E$, whereas the
$D_{\mathrm{B}}$ values depend strongly on $E$
and is particularly small in the pseudo-gap.

\begin{figure}[]
\begin{center}
\includegraphics[width=10cm]{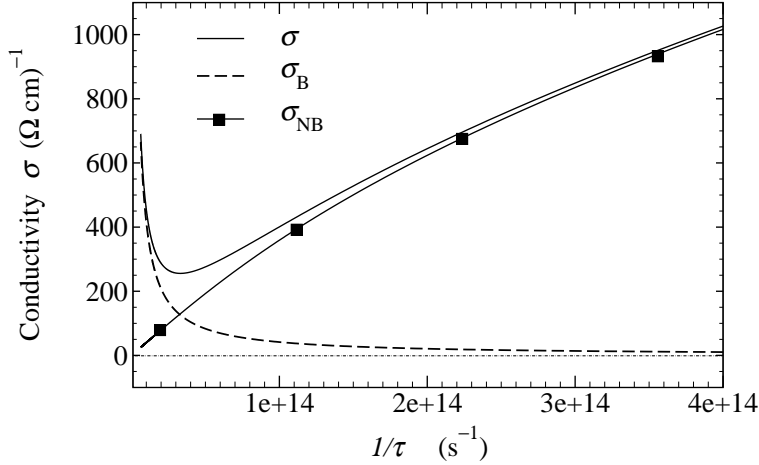}
\end{center}
\caption{ \label{Fig_Conductivity_t6_T_e.012D.001}
Ab-initio dc conductivity $\sigma$ in cubic approximant
$\alpha$-Al$_{69.6}$Si$_{13.0}$Mn$_{17.4}$ versus
inverse scattering time.
}
\end{figure}

\begin{figure}[]
\begin{center}

~~~~~~\includegraphics[width=8cm]{Conductivity_t6_logT_e-012D-001.eps}

\vskip .4cm
\includegraphics[width=9cm]{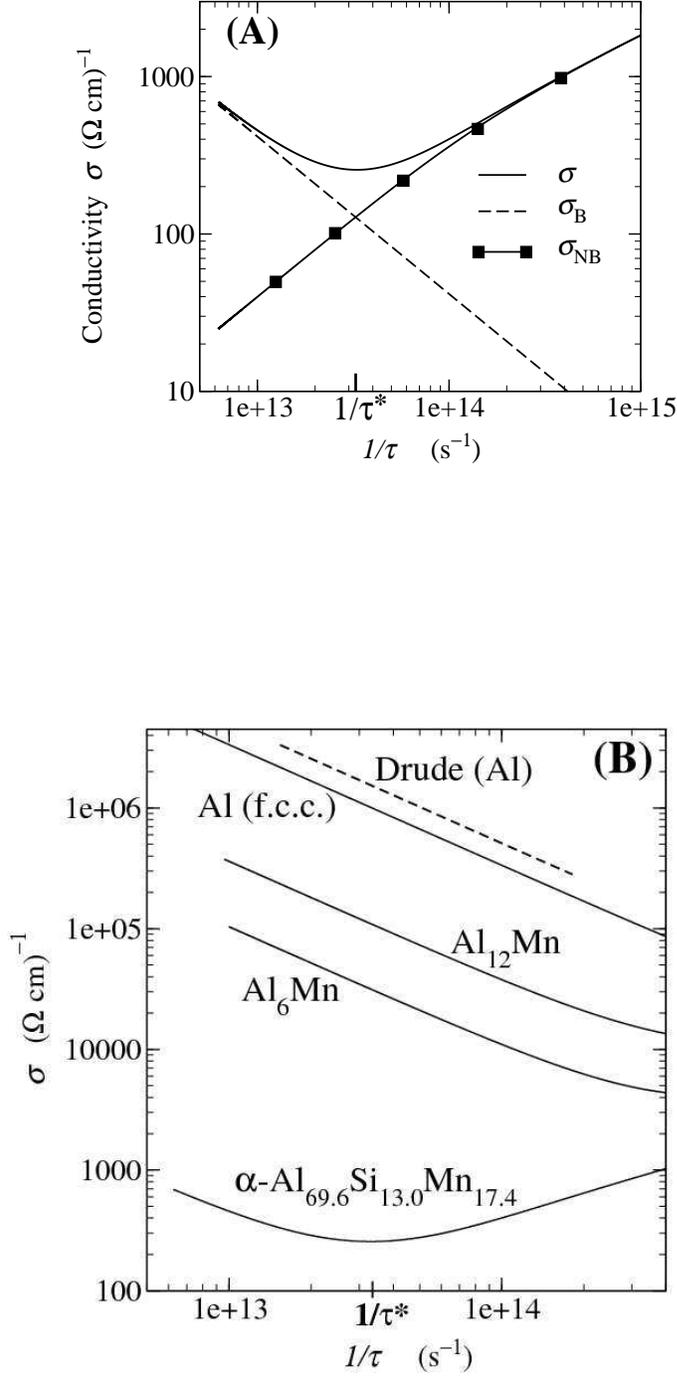}
\end{center}

\caption{ \label{Fig_sigma}
Ab-initio electrical conductivity $\sigma(E_{\mathrm{F}})$ 
(logarithmic scale)
versus inverse scattering time
$1/\tau$ (logarithmic scale). 
{\bf (A)} 
In cubic approximant $\alpha$-Al$_{69.6}$Si$_{13.0}$Mn$_{17.4}$:
$\sigma(E_{\mathrm{F}})$ is the sum of 
a ballistic term (Boltzmann term), 
$\sigma_{\mathrm{B}} = e^2 n(E_{\mathrm{F}}) V_{\mathrm{F}}^2 \tau$,
and a non ballistic term (non Boltzmann term),
$\sigma_{\mathrm{NB}}$. 
{\bf (B)} 
In
pure Al (f.c.c.),
the Boltzmann term dominates, and the model is compatible
with 
a simple Drude model (dashed line). 
In cubic Al$_{12}$Mn and  orthorhombic Al$_6$Mn crystal, 
the model predicts also a 
Boltzmann behavior as expected experimentally. From \cite{PRL06}. }
\end{figure}

The predicted static conductivity (dc conductivity) of the 
$\alpha$-AlMnSi phase,
assuming the value of the Fermi energy given above i.e. 
$E_{\mathrm{F}}-E_{\mathrm{F(LMTO)}} = -0.163$\,eV,
is shown  figures \ref{Fig_Conductivity_t6_T_e.012D.001} 
and \ref{Fig_sigma}
versus the inverse scattering time.
Two regimes appear clearly: 
the metallic regime (Boltzmann regime) at large scattering time,
$\tau >\tau^*$, 
and the insulating like regime (non Boltzmann regime)
at small scattering time, $\tau <\tau^*$. 
$\tau^*= 3.07 \, 10^{-14}$\,s is defined as the time for which 
the Boltzmann and non-Boltzmann contributions are equal.
As expected from our model, 
$\sigma_{\mathrm{NB}}$ is almost proportional to $1/\tau$. 
Therefore, in the non Boltzmann regime, the conductivity 
increases with disorder 
as observed experimentally.
For realistic $\tau$ values,
$\tau <  \tau^*$
\cite{Mayou93}, $\sigma_{\mathrm{NB}}$ dominates and
$\sigma$ increases when $1/\tau$ increases i.e. when
defects or temperature increases.
$\sigma$ varies from 250~($\Omega$\,cm)$^{-1}$ for
$\tau = 3.3\, 10^{-14}$\,s,
to 2000~($\Omega$\,cm)$^{-1}$ for
$\tau = 10^{-15}$\,s.
This is consistent with experimental results
in $\alpha$-AlMnSi:
$\sigma(4~K) \simeq 200$~($\Omega$\,cm)$^{-1}$
and $\sigma(900~K) \simeq 2000$~($\Omega$\,cm)$^{-1}$ 
and with standard estimates for the scattering time 
in these systems \cite{Berger94}.
Furthermore for $\tau$ equal to  a few~$10^{-14}$\,s,
i.e. when the Boltzmann term is negligible, the mean free path is given
by the square root of the saturation value of
$\Delta X^2_{\mathrm{NB}}$ 
and is of the order of 15\,$\rm \AA$. This is in agreement with estimates
in the literature
\cite{Berger94}. 
As discussed in
part 3
this means also that the systems is far from the
Anderson transition despite its low conductivity. 
From the ab-initio calculations the estimated value of the ratio  
$g_{0}/g_{c}$ for the $\alpha$-AlMnSi phase is about $2-3$. 
This means that this  system is always metallic as discussed in part 3.3.  
According to  figure \ref{g0_sur_gc_invX_alpha} 
the $\alpha$-AlMnSi phase is a phase of type (a).

{\it Optical conductivity}

\begin{figure}
\begin{center}
~~~~~~~~~\includegraphics[width=11cm]{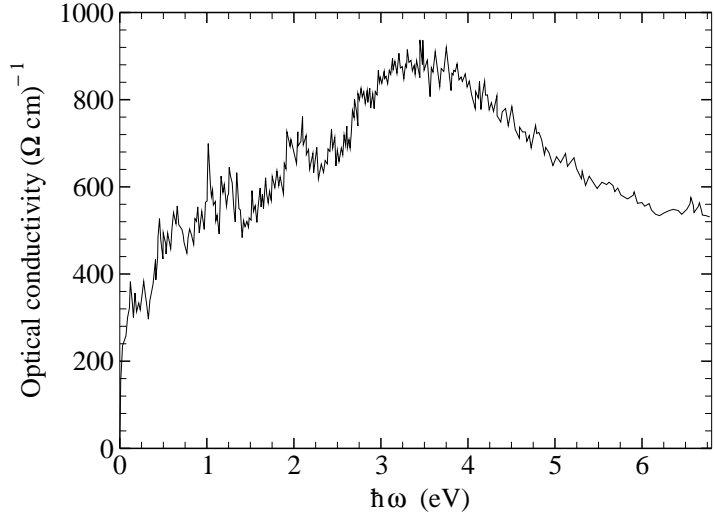}

\caption{Inter-band optical conductivity, $\sigma_{\mathrm{NB}}(\omega)$,
in
$\alpha$-$\rm Al_{114}Mn_{24}$ calculated from the
LMTO results, with a relaxation time $\tau$ equals to infinity.
From T. Fujiwara {\it et al.}~\cite{Fujiwara93}.
\label{Fig_sigma_omega_Fujiwara}}
\end{center}
\end{figure}

\begin{figure}[]
\begin{center}
\includegraphics[width=7cm]{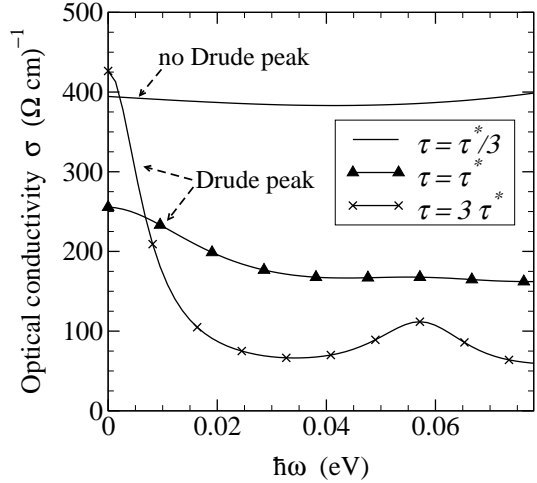}
\end{center}
\caption{ \label{Fig_sigma_omega}
Ab-initio optical conductivity $\sigma(\omega)$ 
in cubic approximant $\alpha$-Al$_{69.6}$Si$_{13.0}$Mn$_{17.4}$
for
three $\tau$ values.
$\omega$ is the pulsation.
For $\tau = 3 \tau^*$,
the non Boltzmann conductivity $\sigma_{\mathrm{NB}}$
is smaller than 
Lorentzian of the
Boltzmann conductivity $\sigma_{\mathrm{B}}$,
$\sigma_{\mathrm{B}}(\omega)=
\sigma_{\mathrm{B}}(0)/(1+\omega^2\tau^2)$.
For $\tau = \tau^*= 3.07 \times 10^{-14}$\,s,
$\sigma_{\mathrm{NB}}(0) = \sigma_{\mathrm{B}}(0)$.
For $\tau = \tau^*/3$, 
the non Boltzmann conductivity $\sigma_{\mathrm{NB}}(\omega)$
dominates.
}
\end{figure}

Within the relaxation time approximation used here, 
the optical conductivity  $\sigma(\omega)$ is
 the sum of two terms.
The Boltzmann contribution
($\sigma_{\rm B}(\omega)$, diagonal elements of
the velocity operator) gives rise to the so-called Drude peak
and the non Boltzmann conductivity gives rise to a nearly frequency independent contribution. This is a consequence of the fact that $\Delta X_{\NB}^2(E,t)$ is nearly constant on the time scale of $\tau$  (see part 3).

T. Fujiwara et al. \cite{Fujiwara93} 
has also estimated the optical conductivity from the LMTO 
band dispersion of a $\alpha$-$\rm Al_{114}Mn_{24}$ 
(figure \ref{Fig_sigma_omega_Fujiwara}). 
This calculation reproduces the linearity and the peak position observed 
experimentally. 
Our ab-initio calculation (figure \ref{Fig_sigma_omega}) confirms that  
a Drude peak can be identified in the Boltzmann  regime,
$\tau >\tau^*$, whereas in the non Boltzmann regime,
$\tau < \tau^*$, the Drude peak disappears.

The role of transition metal elements (TM\,= Fe, Mn, Co, Pd, Re) 
in the electronic structure
of quasicrystals and related phases as been often discussed in the literature 
\cite{Friedel87,Fujiwara89,Fujiwara93,ZouPRL93,GuyEuro93,GuyPRB95,PMS05,Guy03,GuyPRL00,Guy04_ICQ8,Virginie2,Hippert99}.
Because of their strong scattering potential with respect to Al(Si) atoms,
TM elements play a crucial role in the formation of the Hume-Rothery pseudogap
that contributes to the stability of these phases.
This effect is related to an effective
medium range interaction between TM atoms mediated by a strong 
sp(Al)-d(TM) hybridization
\cite{PMS05}.
TM elements have also a very important role on the transport properties.
As an example,
it is shown in the previous paragraph  how a Mn-cluster can ``localize'' 
electrons \cite{GuyPRB97,GuyICQ6}.

\begin{figure}
\begin{center}
\includegraphics[width=10cm]{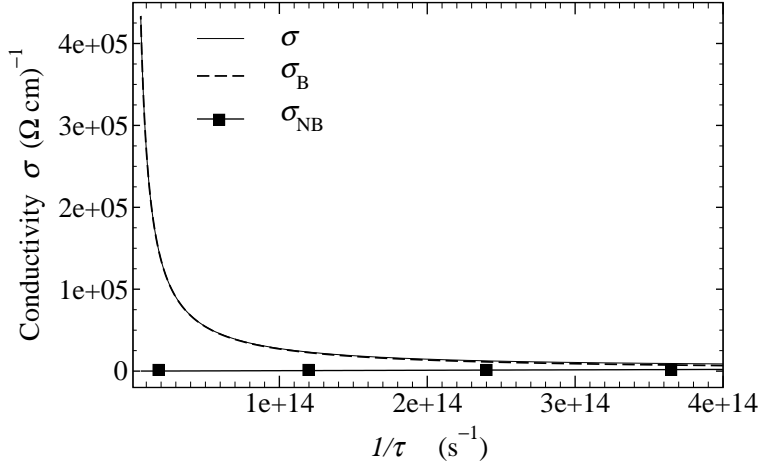}
\end{center}
\caption{\label{Fig_Sig_aAlCuSi} 
Ab-initio dc conductivity $\sigma$ in an hypothetical cubic approximant
$\alpha$-Al$_{69.6}$Si$_{13.0}$Cu$_{17.4}$ versus
inverse scattering time.
}
\end{figure}

To evaluate the effect of TM elements
on the conductivity calculation in the RTA, we have considered an
hypothetical   $\alpha$-Al$_{69.6}$Si$_{13.0}$Cu$_{17.4}$
constructed by
putting Cu atoms in place of Mn atoms in the actual
$\alpha$-Al$_{69.6}$Si$_{13.0}$Mn$_{17.4}$ structure.
Cu atoms have almost
the same number of sp electrons as Mn atoms,
but their d DOS is very small at $\ef$.
Therefore in $\alpha$-Al$_{69.6}$Si$_{13.0}$Cu$_{17.4}$,
the effect of  sp(Al)-d(TM) hybridization on electronic states
with energy near $\ef$ is very small.
As a result, the pseudogap disappears in total DOS,
and the dc-conductivity is now ballistic (metallic) 
as shown on figure~\ref{Fig_Sig_aAlCuSi}.

\subsection{Phenomenological model for the low frequency conductivity 
of AlCuFe quasicrystals}

The ab-initio calculations which rest on the Bloch theorem are applicable to approximants only.
Here  we present a  phenomenological model of the optical conductivity of AlCuFe QCs, which 
should be approximately valid also for the  related QC phases AlPdMn or AlFeCr 
\cite{BasICQ5,Demange}.  This anomalous diffusion model allows to derive  an analytical expression for the conductivity that fits experiments very well. In particular the model explains quantitatively the main experimental facts:

 -  the increase of conductivity with disorder
 
 - the {\it"inverse Mathiessen rule"} \cite{Berger94,Prejean02} that is the fact that  the increases of conductivity due to different sources of scattering are additive
 
 - the absence of the Drude peak 

\begin{figure}[]
\begin{center}
\includegraphics[width=10cm]{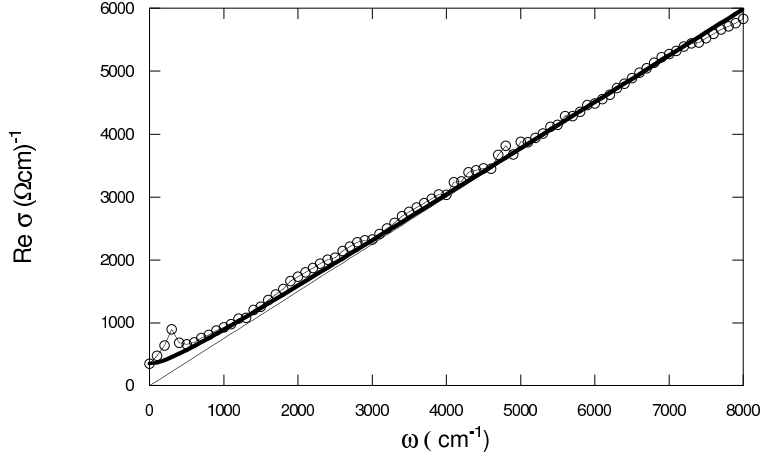}
\caption{Real part of the conductivity as a function of the frequency for different cases. 
Line with circles: experimental conductivity of an AlCuFe QC [9]. 
Thin line: conductivity of the model without
defect $\RRe\,\sigma_{0}(\omega) = \sigma_{1}({|\omega|}/{\omega_{1}})$. 
Thick line: conductivity of the
model with defect, for  $\tau = 3\,10^{-14}s$.
From \cite{Triozon04}.} 
\label{fig omega}
\end{center}
\end{figure}

One needs first a model of
conductivity of the {\it perfect} system at all frequencies. 
For low frequencies, according to the discussion in part 2  we assume:
\beq
\RRe\,\sigma_{0}(\omega) = \sigma_{0}\left(\frac{|\omega|}{\omega_{1}}\right)^{1-2\beta}   
\qquad\hbox{for $|\omega| < \omega_{1}$}
\label{Modelsigma}
\eeq
\noindent A small value of $\beta$ ($\beta \ll 1$ ) 
is imposed by the nearly linear experimental variation of 
$\RRe\,\sigma(\omega)$ at
$\omega < 8000 \,{\rm cm^{-1}}$ (see figure \ref{fig omega}). 
This means that the system without defects would be 
insulating. In (\ref{Modelsigma}) we take $\beta =0$,  
$\omega_{1}\simeq  8000 \,{\rm cm^{-1}}$ 
and $\sigma_{0}\simeq 6000 \,{\rm (\Omega cm)^{-1}}$. At higher frequencies 
one uses other analytical expressions. 
For  $8000 \,{\rm cm^{-1}} < \omega < {\rm 25\, 000 \,cm^{-1}}$ a polynomial of $\omega$  
reproduces the experimental value. For $\omega > {\rm 25\, 000\, cm^{-1}}$,  
we take the Drude expression according to \cite{Homes91}. Let us note that the experimental 
uncertainty on the high frequency conductivity  \cite{Homes91} has essentially no effect on 
the results presented here.

Within the RTA (\ref{sigma},\ref{Relaxation}) one has for 
the optical conductivity:
\beq
\RRe\,\sigma(\omega,\tau) = 
\displaystyle{\int\limits_{-\infty}^{+\infty}}  
\frac{\RRe\,\sigma_{0}(\omega -\omega ')}
{\pi\tau (\omega '^{2} + {1}/{\tau^{2}})} \d\omega ' 
\label{Convolution1}
\eeq
\noindent i.e. the real part $\RRe\,\sigma (\omega,\tau)$  
of the conductivity of the system with defects  is the convolution of  
$\RRe\,\sigma_{0} (\omega)$ of the perfect system and of a Lorentzian of width $1/\tau$. 

\begin{figure}[]
\begin{center}
\includegraphics[width=10cm]{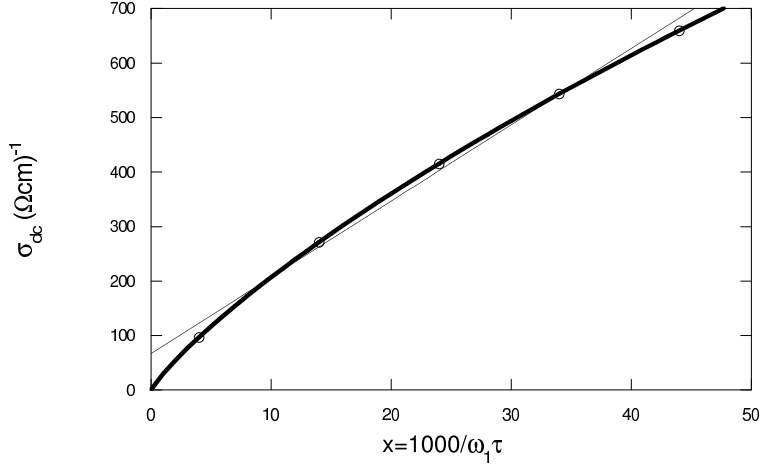}
\caption{Thick line: Variation of $\sigma_{dc}$ with $x=1000/\omega_{1}\tau$. 
$\tau$ is given by $\tau = (6.6/x) \,10^{-13}\,s $. 
The straight thin line shows that  $\sigma_{dc}$ varies nearly linearly 
with  $1/\tau$ in the range $\sigma_{dc} = 150--700  \,{\rm (\Omega cm)^{-1}}$.
From \cite{Triozon04}.
}
\label{fig tau}  
\end{center}
\end{figure}

As shown in part 2 for $\omega <\omega_{1}$ the conductivity is well represented by
\beq
{\RRe\,\sigma} \simeq \frac{A}{\tau} \left[\alpha + 
\log\left(\frac{\omega_{1} \tau}{\sqrt{1+(\omega \tau)^{2}}}\right)
+ \omega \tau {\rm Arctg}(\omega \tau)\right]\label{Resigma1}
\eeq
\noindent where 
$A=2\sigma_{0}/\pi \omega_{1}$The analytical expression (\ref{Resigma}) 
with $\alpha \simeq 0.7$,  
$\hbar\omega_{1}\simeq 1\,$eV and 
$\sigma_{1}\simeq 6000 \,{\rm (\Omega cm)^{-1}}$ describes well the electronic
conductivity in figures \ref{fig omega}, \ref{fig tau}.

Let us  focus on the low frequency conductivity 
($\omega <\omega_{1}$) which is the real test of the model
(\ref{Convolution1},\ref{Modelsigma},\ref{Resigma1}).
Figure \ref{fig omega} gives a  comparison of the experimental
$\RRe\,\sigma_{exp}(\omega)$ for AlCuFe \cite{Homes91} with the theoretical 
$\RRe\,\sigma (\omega,\tau)$.
The scattering time $\tau$ is chosen to reproduce the experimental
dc-conductivity $\sigma_{dc} \simeq 350 \,{\rm (\Omega cm)^{-1}}$. 
One finds $\tau \simeq 3\, 10^{-14}\,{\rm s}$ which is rather long, in agreement with 
the high structural quality of these systems. The fit is good except 
for the peak in the experimental curve around $200 \,{\rm cm^{-1}}$. 
This peak is attributed to the conductivity of phonons \cite{Homes91} 
which is not incorporated in the model. The mean-free path $\Lambda$ is related to the scattering time $\tau$ and to the diffusivity $D$ through $D=\Lambda ^{2}/3\tau$. 
One estimates \cite{Berger94,Prejean02}  
$D \geq 0.2 \,{\rm cm^{2}/s}$ and since $\tau \simeq 3\,  10^{-14}\,{\rm s}$, one has  
$\Lambda \geq 15-20$ $\rm \AA$.     From (\ref{Convolution1}) and (\ref{Resigma1}) one gets the dc-conductivity $\sigma_{dc}(\tau)$ as a function of the relaxation time $\tau$ (see figure \ref{fig tau}). $\sigma_{dc}(\tau)$ increases with $1/\tau$ and varies nearly linearly with $1/\tau$ on a large range of values of $\sigma_{dc}(\tau)$ i.e. $\sigma_{dc}\simeq A+B/\tau$. For two independent sources of scattering characterized by scattering times $\tau_{1}$ and $\tau_{2}$ it is common that the inverse relaxation times add. Then $1/\tau \simeq 1/\tau_{1} +1/\tau_{2}$ and   $\sigma_{dc}\simeq A+B/\tau \simeq A+ B /\tau_{1} + B/\tau_{2}$. Thus each source of disorder gives its contribution to the conductivity in agreement with the ``inverse Mathiessen rule'' \cite{Berger94}.     

\begin{figure}[]
\begin{center}
\includegraphics[width=8cm]{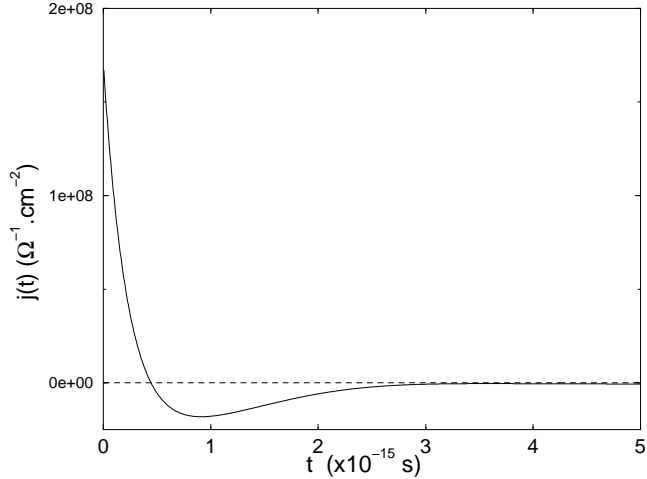}
\caption{Value of $j(t)$ deduced from the experimental conductivity. 
The negative value of j(t) at large times indicates backscattering.
From \cite{Triozon04}.
} 
\label{fig jt}  
\end{center}
\end{figure}

The present phenomenological model treats the disorder within the relaxation time approximation (RTA). 
Indeed, as shown now, the  RTA is applicable to AlCuFe QCs, at least for  
$T\leq 200-300\,$K. 
A first indication is that  quantum interferences have been found for 
$T\leq 200-300\,$K \cite{Berger94}. They indicate that the main scattering sources are elastic in this temperature range. Indeed, if the dominant scattering were inelastic the coherence of the electron wavefunction would be lost at each scattering event. In that case there would be no interferences in the diffusive regime. 
In addition  the experimental fits \cite{Berger94,Prejean02} show that the quantum interferences and the  electron-electron interaction  give only a correction to the conductivity. Therefore, as the elastic scattering dominates and as quantum interferences are weak, the RTA is a good approximation 
for the AlCuFe QC studied in \cite{Homes91} at least at $T\leq 200-300 K$. 
In particular a scenario of hopping between localized critical states, 
such as proposed by Janot \cite{Janot94} is not consistent 
with the present analysis.

Note that the model is consistent with the observed weak-localization effects.
Indeed in the context of the scaling theory of localization  \cite{TVR}  
the importance of quantum interferences depends on the ratio between the 
dc-conductivity of the system $\sigma_{dc}$ and the Mott value 
$\sigma_{\rm Mott} \simeq 600\,{\rm (\Omega cm)^{-1}}/ \Lambda $ 
where $\Lambda$ is the mean-free path expressed in Angstr\"oms. 
If $R= \sigma_{dc}/\sigma_{\rm Mott} \gg  1 $ the effect of the quantum 
interferences on $\sigma_{dc}$ is small. Here $R= \sigma_{dc}/\sigma_{\rm Mott}
\simeq 5-10$ and the localization effects are only corrections.

To conclude, the
dynamics of electrons in AlCuFe quaiscrystals 
and related  systems such as  
AlPdMn \cite{BasICQ5},  AlCrFe \cite{Demange} quasicrystals
is not free electron like. The minimum of optical conductivity 
at low frequency 
and  the increase of dc-conductivity with disorder 
(in the RTA scheme) are intimately related to  the backscattering 
(figure \ref{fig jt}) or equivalently to an anomalous diffusion law with 
$\beta < 0.5$ (Ref. \cite{MayouPRL00}). 
The low value of $\beta$ implies a $\it {slow}$ anomalous diffusion 
$L(t)\propto t^{\beta}$, which suggests the proximity to a localized state. 
This value $\beta \simeq 0$ is consistent with the results on approximant 
phases. Indeed the Non Boltzmann term in the quantum diffusion 
$\Delta X_{\NB}^2(t)$ which saturates very quickly is analogous to a quantum 
diffusion with $\beta\simeq 0$.

\section{Conclusion}
\label{SecConclusion}

In this chapter we concentrated  on quantum diffusion and  electronic conduction properties in quasiperiodic and periodic systems. We found that deviations from the standard ballistic propagation exist either in quasiperiodic or even in periodic systems. This anomalous diffusion mode has deep consequences on the conduction properties at zero and low frequency. 

The anomalous diffusion mode is related to a tendency to localization and to a phenomenon of backscattering which is well known in disordered systems. The phenomenon of backscattering is the fact that an impulse of electric field creates a current density $J(t)$ which is opposite to the electric field at large time. Backscattering is associated  with an increase of conductivity with frequency and disorder. 

The physics of phonons in quasicrystals could be affected by the anomalous diffusion phenomenon. In particular it has been argued that the heat conductivity could be sensitive to this effect \cite{Mayou00_Aussois}.

The concepts developed here open also a new  insight in the physics 
of correlated systems. Indeed recent studies of some heavy fermions 
or polaronic systems 
\cite{Vidhyadhiraja05,Fratini03,Fratini05},  
where charge carriers are also slow, show that their conduction properties 
present a deep analogy with those described here. 
In particular a transition from a metallic like regime 
at low temperature where scattering is weak  to an insulating 
like regime at higher temperature with a stronger scattering is observed.

%
%
%
%
\section*{Acknowledgements}

The works presented in the review paper have been done
since the 90's.
Our work owes much to the discussions with
Prof. T. Fujiwara,
Prof. J. Bellissard,
Prof. J. Friedel
and Prof. N.W. Ashcroft.
We are very grateful to
many colleagues with whom we had collaborations during this time:
C. Berger,
F. Cyrot--Lackmann,
J. Delahaye,
T. Grenet,
J.P. Julien,
T. Klein,
L. Magaud,
J.J. Pr\'ejean,
S. Roche
and
F. Triozon.
We also thanks
F. Hippert,
R. Mosseri,
J. Vidal
and 
C. Sire
for fruitful discussions.
%

%
%
%
\bibliographystyle{plain}

\begin{thebibliography}{00}






\bibitem{Shechtman84} Shechtman D, Blech I,
Gratias D, Cahn JW.
{Phys Rev Lett} 1984;53:1951.

\bibitem{Pierce93_science} Pierce FS, Poon SJ, Guo Q.
Science 1993;261:737;
Pierce FS, Poon SJ,  Biggs BD.
{Phys Rev Lett} 1993;70:3919.

\bibitem{Berger93} Berger C, Grenet T, Lindqvist P, Lanco P,
Grieco JC, Fourcaudot G, Cyrot--Lackmann F.
Solid State Commun 1993;87:977.

\bibitem{Akiyama93} Akiyama H, Honda Y, Hashimoto T,
Edagawa K, Takeuchi S.
Jpn J Appl Phys 1993;32:L1003.


\bibitem{Ashcroft76} Ashcroft NW, Mermin DE.
Solid State Physics. Sauders College Publishing,
1976.

\bibitem{MayouPRL00} Mayou D.
{Phys Rev Lett} 2000;{85}:1290.

\bibitem{Macia02} Maci\'{a} E.
Phys Rev 2002;B66:174203.

\bibitem{Macia04} Maci\'{a} E.
Phys Rev 2004;B69:132201.

\bibitem{Solbrig00} 
Solbrig H, Landauro CV, L\"{o}ser A.
Mat Sc Eng 2000:A294-296:596.
Landauro CV, Solbrig H.
Mat Sc Eng 2000:A294-296:600.
Physica 2001:B301:267.

\bibitem{Haussler02}
Ha\"ussler P, Haberkern R, Madel C, Barzola-Quiquia J, Lang M.
J Alloys and Comp 2002;342:228.

\bibitem{IntroQC} Janot C, Quasicrystals a Primer, Clarendon Press-Oxford (1992). 

\bibitem{Kohmoto83} Kohmoto M, Kadanoff LP, Tang C. Phys Rev Lett 1983;{50},1870. 
\bibitem{Tokihiro88} Tokihiro T, Fujiwara T, Arai M.
Phys. Rev 1988;B38:5981.

\bibitem{FujiwaraSofArt} Fujiwara T, Tsunetsugu H.
In: Di Vincenxo DP, Steinhart PJ, editors,
Quasicrystals: The states of the art,
Singapore: World Scientific, 1991. 

\bibitem{Roche96}
Roche S, Trambly de Laissardi\`ere G, Mayou D.
{J Math Phys} 1996;{38}:1794.

\bibitem{Bellissard03}
Bellissard J.
In: Garbaczeski P, Olkieicz R, editors.
{Dynamics of Dissipation},
Lecture Notes in Physics.
Berlin: Springer, 2003; p. 413.

\bibitem{Fujiwara96}   Fujiwara T,  Mitsui T, Yamamoto S.  
Phys Rev B 1996;{53},R2910.

\bibitem{Sire94} Sire C.
In: Hippert F, Gratias D, editors.
{Lecture on Quasicrystals}.
Les Ulis: Les Editions de Physique, 1994; p. 505.

\bibitem{Piechon96} Pi\'echon F.  Phys Rev Lett 1996;{76},4372. 

\bibitem{Triozon02} 
Triozon F, Vidal J, Mosseri R, Mayou D.
Phys Rev 2002;B65:220202.

\bibitem{Klein91} Klein T, Berger C, Mayou D, Cyrot--Lackmann F.
{Phys Rev Lett} 1991;{66}:2907.

\bibitem{Poon92} Poon SJ.
Adv Phys 1992;41:303.

\bibitem{Berger94} Berger C.
In: Hippert F, Gratias D, editors.
{Lecture on Quasicrystals}.
Les Ulis: Les Editions de Physique, 1994; p. 463.

\bibitem{Roche97} Roche S, Mayou D.
{Phys Rev Lett} 1997;{79}:2518.

\bibitem{Grenet00_Aussois} Grenet T.
In: Belin-Ferr\'e E, Berger C, Quiquandon M, Sadoc A, editors.
Quasicrystals: Current Topics.
Singapor: World Scientific, 2000; p. 455.

\bibitem{Delahaye99} Delahaye J, Frison JP, Berger C.
{Phys Rev Lett} 1999;{81}:4204.

\bibitem{Delahaye01} Delahaye J,  Berger C.  Phys Rev B 2001;{64},094203.  

\bibitem{Delahaye03_JPCM} Delahaye J, Berger C,
Fourcaudot G.
J Phys: Condens Matter 2003;15:8753.

\bibitem{Rosenbaum04} 
Rosenbaum R, Murphy T, Brandt B, Wang C-R, Zhong Y-L, Wu S-W, Lin S-T, Lin J-J,
J Phys: Condens Matter 2004;16:821;
Rosenbaum R, Grushko B, Przepiorzynski B,
J. of Low Temperature physics 2006;142:101.

\bibitem{Quivy96}
Quivy A, Quiquandon M, Calvayrac Y, Faudot F, Gratias D, Berger C,
Brand RA, Simonet V, Hippert.
J Phys Condens Matter 1996;8:4223.


\bibitem{Tamura01}
Tamura R, Asao T, Takeuchi S.
{Phys Rev Lett} 2001;86:3104.


\bibitem{Takeuchi04} 
Takeuchi T, Onogi T, Otagiri T, Mizutani U, Sato H,
Kato K, Kamiyama T.
Phys Rev 2004;B68:184203.

\bibitem{Mayou93}
Mayou D, Berger C, Cyrot--Lackmann F, Klein T, Lanco P.
{Phys Rev Lett} 1993;{70}:3915.


\bibitem{Prejean02} Pr\'ejean JJ, Berger C,
Sulpice A, Calvayrac Y.
Phys Rev 2002;B65:R140203.


\bibitem{Homes91} 
Homes CC, Timusk T, Wu X,
Altounian Z, Sahnoune A, Str\"{o}m--Olsen JO.
Phys Rev Lett 1991;67:2694.

\bibitem{Burkov94_JPCM} Burkov SE, Timusk T, Ashcroft NW.
J Phys: Condens Matter 1992;4:9447.

\bibitem{Wagner88}
Wagner JL, Biggs BD, Wong KM, Poon SJ.
Phys Rev 1988;B38:7436.

\bibitem{Wang88}
Wang K, Garoche P, Calvayrac Y.
J Phys Colloq France 1988;49:C8-237.

\bibitem{Kimura89} Kimura K, Iwahashi H, Hashimoto T,
Takeuchi S, Mizutani S, Ohashi S, Itoh G.
J Phys Soc Japan 1989;58:2472.

\bibitem{Belin93} Belin E, Danhkazi Z.
J Non Cryst Solids 1993;153-154:298.

\bibitem{Hippert92} Hippert F, Kandel L, Calvayrac Y,
Dubost B.
Phys Rev Lett 1992;69:2086.

\bibitem{Biggs90} Biggs BD, Poon SJ, Munirathnam NR.
Phys Rev Lett 1990:65:2700.

\bibitem{Pierce94} Pierce FS, Guo Q, Poon SJ.
Phys Rev Lett 1994:73:2220.


\bibitem{Belin92a}
Belin E, Kojnok J, sadoc A, Traverse A,
Harmelin M, Berger C, Dubois JM.
J Phys: Condens Matter 1992;4:1057.

\bibitem{Belin94_MSEA}
Belin E, Miyoshi  Y, Yamada Y, Ishikawa T, Matsuda T, Mizutani U.
Mat Sci Eng 1994;A181-182:730.

\bibitem{Mori91} 
Mori M, Matsuo S, Ishimasa T, Matsuura T, Kamiya K,
Inokuchi H, Matsukawa T.
J Phys: Condens Matter 1991;3:767.

\bibitem{Matsubara91}
Matsubara H, Ogawa S, Kinoshita T, Kishi K, Takeuchi S,
Kimura K, Suga S.
Jpn J Appl Phys 1991;30:L389.

\bibitem{Mori92} 
Mori M, Kamiya K, Matsuo S, Ishimasa T, Nakano H,
Fujimoto H, Inokuchi H.
J Phys: Condens Matter 1992;4:L157.

\bibitem{Belin92b} Belin E, Dankh\'azi Z, Sadoc A, Calvayrac Y,
Klein T, Dubois JM.
J Phys: Condens Matter 1992;4:4459.

\bibitem{Belin94EuroPhys} 
Belin E, Dankh\'azi Z, Sadoc A, Dubois JM, Calvayrac Y.
Europhys Lett 1994;26:677.

\bibitem{Belin00} 
Belin--Ferr\'e, Fourn\'ee V, Dubois JM.
J Phys: Condens Matter 2000;12:9159.

\bibitem{Zhang94} 
Zhang GW, Stadnik ZM, Tsai AP, Inoue A.
Phys Rev 1994;B50:6696.

\bibitem{Belin94_JPCM} 
Belin E, Dankh\'azi Z, Sadoc A, Dubois JM.
J Phys: Condens Matter 1994;6:8771

\bibitem{Stadnik97}
Stadnik Z M, Purdie D, Garnier M, Baer Y, Tsai AP,
Inoue A, Edagawa K, Takeuchi S, Buschow KHJ.
{ Phys Rev} 1997;B{55}:10938.

\bibitem{Fournee02} 
Fourn\'ee V, Belin--Ferr\'e E, P\^{e}cheur P,
Tobala J, Dankh\'azi Z, Sadoc A, M\"uller H.
J Phys: Condens Matter 2002;14:87.

\bibitem{Stadnik94}
Stadnik ZM, Zhang GW, Tsai AP, Inoue A.
J Phys: Condens Matt 1994;6:6885.

\bibitem{Mizutani01}
Muzitani U, Takeuchi T, Banno E, Fournee V,
Takana M, Sato H.
In: Belin--Ferr\'e E, Thiel PA, Tsai AP, Urban K, editors.
{Mat Res Soc Symp Soc Proc}, Vol. 643.
Warrendale: Materials Research Society, 2001;
p. K13.1.1.

\bibitem{Mizutani04}
Mizutani U, Takeuchi T, Sato H.
Prog Mat Sci 2004;49:227.

\bibitem{BelinMayou93} Belin E, Mayou D.
Physica scipta 1993;T49A:356.

\bibitem{Berger93b} Berger C, Cyrot--Lackmann F, Mayou D.
J Non-Cryst Solids 1993;153:412.

\bibitem{Berger93c} Berger C, Belin E, Mayou D.
Annales de Chimie-Science des Mat\'eriaux 1993;18:485.


\bibitem{Mayou93b} Mayou D, Cyrot--Lackmann F, Trambly de Laissardi\`ere G,
Klein T.
J Non-Cryst Solids 1993;154:430.

\bibitem{Dankhazi93}
Dankh\'azi Z, Trambly de Laissardi\`ere G, Nguyen--Manh D, Belin E,
Mayou D.
{ J Phys: Condens Matter} 1993;{5}:3339.

\bibitem{GuyPRB95_2} Trambly de Laissardi\`ere G,
Dankh\'azi Z, Belin E,  Sadoc A,
Nguyen--Manh D, Mayou D, Keegan MA, Papaconstantopoulos D.
{Phys Rev} 1995;B{51}:14035.

\bibitem{Andersen75} Andersen OK. 1975
{Phys Rev} 1975;B{12}:3060.

\bibitem{ElserH85}
Elser V, Henley C.
{Phys Rev Lett} 1985;55:2883.

\bibitem{Guyot85}
Guyot P, Audier M.
{Philos Mag} 1985;B52:L15.


\bibitem{Friedel87} Friedel J, D\'enoyer F.
{C R Acad Sci Paris, Ser II} 1987;{305}:171.

\bibitem{Fujiwara89} 
Fujiwara T.
{Phys Rev} 1989;B40:942.

\bibitem{Fujiwara93} Fujiwara T, Yamamoto S, Trambly de
Laissardi\`ere G.
{Phys Rev Lett} 1993;{71}:4166;\\
Mat Sci Forum 1994;150-151:387.

\bibitem{ZouPRL93} Zou J, Carlsson AE.
{Phys Rev Lett} 1993;{70}:3748.

\bibitem{GuyEuro93} Trambly de Laissardi\`ere G, Mayou D,
Nguyen Manh D.
{Europhys Lett} 1993;{21}:25;\\
{J Non-Cryst Solids} 1993;153-154:430.

\bibitem{GuyPRB95} Trambly de Laissardi\`ere G, Nguyen Manh D,
Magaud L, Julien JP,
Cyrot--Lackmann F, Mayou D. 1995
{Phys Rev} 1995;B{52}:7920.

\bibitem{PMS05}
Trambly de Laissardi\`ere G, Nguyen Manh D, Mayou D, 
{Prog  Mater Sci} 2005;{50}:679.

\bibitem{Guy03} Trambly de Laissardi\`ere G.
Phys Rev 2003;B68:045117.

\bibitem{GuyPRL00}
Trambly de Laissardi\`ere G, Mayou D.
{Phys Rev Lett} 2000;{85}:3273.

\bibitem{Guy04_ICQ8} Trambly de Laissardi\`ere G,
Nguyen Manh D, Mayou D.
J Non-Cryst Solids 2004;334-335:347.

\bibitem{Virginie2} Simonet V, Hippert F, Audier M, Trambly de Laissardi\`ere G.
{Phys Rev} 1998;B{58}:R8865.

\bibitem{Hippert99} Hippert F, Simonet V,
Trambly de Laissardi\`ere G, Audier M, Y. Calvayrac Y.
J Phys: Condens Mat 1999;{11}:10419.


\bibitem{Zijlstra03} 
Zijlstra ES, Bose SK.
Phys Rev 2003;B67:224204.

\bibitem{Gratias00} 
Gratias D, Puyraimond F, Quiquandon M, Katz A.
{Phys Rev} 2000;B63:24202.

\bibitem{GuyPRB94_AlCuFe}
Trambly de Laissardi\`ere G, Fujiwara T.
{Phys Rev} 1994;B{50}:5999.


\bibitem{GuyPRBAlCuCo} Trambly de Laissardi\`ere G, Fujiwara T.
{Phys Rev} 1994;B50:9843;\\
Mat Sci Eng 1994;A181-182:722.


\bibitem{Janot94} Janot C, de Boissieu M.
Phys Rev Lett 1994;72:1674.


\bibitem{GuyPRB97} 
Trambly de Laissardi\`ere G, Mayou M.
{Phys Rev} 1997;B{55}:2890.

\bibitem{GuyICQ6} Trambly de Laissardi\`ere G,
Roche S, Mayou D.
{Mat Sci Eng} 1997;A{226}-{228}:986.


\bibitem{Friedel56} Friedel J.
Can J Phys 1956;34:1190.

\bibitem{Anderson61} Anderson PW.
Phys Rev 1961;124:41.


\bibitem{Roche98}
Roche S, Fujiwara T.
{Phys Rev } 1998;{B58}:11338.

\bibitem{Krajci02} Kraj\v{c}\'{\i} M, Hafner J, Mihalkovic M.  
Phys Rev 2002;{B65}:024205.   

\bibitem{PRL06}
Trambly de Laissardi\`ere G, Julien JP, Mayou D.
Phys Rev Lett 2006;97:026601.

\bibitem{ICQ9} Trambly de Laissardi\`ere G, Julien JP, Mayou D.
Phys Mag 2006;86:663.

\bibitem{Julien06} Julien JP, Trambly de Laissardi\`ere G, Mayou D.
In: Julien JP, Maruani J, Mayou D, Wilson S, Delgado-Barrio G, editors.
Recent Advances in the Theory of Chemical and Physical Systems.
Progress in Theoretical Chemistry and Physics. Vol 15.
Dordrecht: Springer, 2006; p. 535.

\bibitem{Sugiyama98}
Sugiyama K, Kaji N, Hiraga K.
{Acta Cryst} 1998;C{54}:445.

\bibitem{BasICQ5} Basov DN, et al. 
In: Janot C, Mosseri R, editors.
Proceedings of the Fifth International Conference on Quasicrystals. 
Singapore: World Scientific, 1995; p. 564. 

\bibitem{Demange} Demange V, Milandri A, de Weerd MC,
Machizaud F, Jeandel G, Dubois JM.  
Phys Rev 2002;{B65},144205.


\bibitem{Triozon04} 
Triozon F, Mayou D.
J Non-Cryst Solids 2004;{334}-{ 335},376.

\bibitem{TVR}  Lee PA, Ramakrishnan TV.  Rev Mod Phys 1985;{57},287.    

\bibitem{Mayou00_Aussois} Mayou D.
In: Belin-Ferr\'e E, Berger C, Quiquandon M, Sadoc A, editors.
Quasicrystals: Current Topics.
Singapor: World Scientific, 2000; p. 412.

\bibitem{Vidhyadhiraja05}
Vidhyadhiraja NS, Logan DE.
{J Phys: Condens Matter} 2005;{17},2959.

\bibitem{Fratini03}
Fratini S, Ciuchi S.
{Phys Rev Lett} 2003;{91}:256403.

\bibitem{Fratini05}
Fratini S, Ciuchi  S.
Phys Rev 2006;B74:075101;
and private communication.



\end{thebibliography}

\end{document}